\documentclass[prb,aps,tightenlines,twocolumn,superscriptaddress,showpacs,preprintnumbers,10pt]{revtex4-1} 
\usepackage{amsmath}
\usepackage{amssymb}
\usepackage{mathrsfs}
\usepackage{bm}
\usepackage{graphicx}
\usepackage{xfrac}
\usepackage{url}
\usepackage{tikz}
\usepackage{multirow}
\usepackage{mathtools}
\usepackage{braket}
\usepackage{courier}
\usepackage[version=4]{mhchem}
\usepackage[T1]{fontenc}
\usepackage{scalerel}
\usepackage{soul}
\usepackage{algpseudocode}
\usepackage[ruled]{algorithm2e}
\usepackage{tabularx}
\usepackage{xspace}

\usepackage{array}
\newcolumntype{L}[1]{>{\raggedright\let\newline\\\arraybackslash\hspace{0pt}}m{#1}}
\newcolumntype{C}[1]{>{\centering\let\newline\\\arraybackslash\hspace{0pt}}m{#1}}
\newcolumntype{R}[1]{>{\raggedleft\let\newline\\\arraybackslash\hspace{0pt}}m{#1}}

\AtBeginDocument{%
    \newwrite\bibnotes
    \def\bibnotesext{Notes.bib}
    \immediate\openout\bibnotes=\jobname\bibnotesext
    \immediate\write\bibnotes{@CONTROL{REVTEX41Control}}
    \immediate\write\bibnotes{@CONTROL{%
    apsrev41Control,author="08",editor="1",pages="1",title="0",year="1"}}
     \if@filesw
     \immediate\write\@auxout{\string\citation{apsrev41Control}}%
    \fi
}%

\algnewcommand{\Or}{\textbf{ or }}
\algnewcommand{\And}{\textbf{ and }}
\algnewcommand{\Break}{\textbf{break}}
\algnewcommand{\Continue}{\textbf{continue}}
\algrenewcommand\algorithmicforall{\textbf{foreach}}

\newcommand{\RN}[1]{\textup{\uppercase\expandafter{\romannumeral#1}}}
\newcommand{\BRN}[1]{\textbf{\textup{\uppercase\expandafter{\romannumeral#1}}}}
\definecolor{dartmouthgreen}{rgb}{0.05, 0.5, 0.06}

\newcommand{\exx}{E_{\rm xx}}

\newcommand{\mpi}[0]{\texttt{MPI}}
\newcommand{\omp}[0]{\texttt{OpenMP}}

\newcommand{\cf}[0]{\textit{cf}. }

\newcommand*\diff{\mathop{}\!\mathrm{d}}
\def\dd{\diff}
\newcommand{\appropto}{\mathrel{\vcenter{
  \offinterlineskip\halign{\hfil$##$\cr
    \propto\cr\noalign{\kern2pt}\sim\cr\noalign{\kern-2pt}}}}}

\newcommand{\DefineAuthor}[2]{%
  \expandafter\newcommand\csname #1note\endcsname[1]{%
    \textbf{\textcolor{#2}{\textbf{#1:} ##1}}}%
  \expandafter\newcommand\csname #1\endcsname[1]{
    \textbf{\textcolor{#2}{##1}}}
  \expandafter\newcommand\csname #1cancel\endcsname[1]{%
    \textbf{\textcolor{#2}{\sout{##1}}}}%
  \expandafter\newcommand\csname #1change\endcsname[2]{%
    \textbf{\textcolor{#2}{\sout{##1} ##2}}}%
  \newenvironment{#1text}{\color{#2}}{\color{black}}
}

\definecolor{dartmouthgreen}{rgb}{0.05, 0.5, 0.06}
\DefineAuthor{red}{red}
\DefineAuthor{blue}{blue}
\DefineAuthor{RAD}{red}
\DefineAuthor{HK}{blue}
\DefineAuthor{ZMS}{magenta}
\DefineAuthor{MCA}{dartmouthgreen}
\DefineAuthor{FIG}{dartmouthgreen}
\DefineAuthor{GEN}{dartmouthgreen}
\DefineAuthor{COMMENT}{red}

\begin{document} 

\title{High-Throughput Condensed-Phase Hybrid Density Functional Theory \\ for Large-Scale Finite-Gap Systems: The \texttt{SeA} Approach}
\author{Hsin-Yu Ko}
\affiliation{Department of Chemistry and Chemical Biology, Cornell University, Ithaca, NY 14853, USA}
\author{Marcos F. Calegari Andrade}
\affiliation{Department of Chemistry, Princeton University, Princeton, NJ 08544, USA}
\affiliation{Quantum Simulations Group, Materials Science Division, Lawrence Livermore National Laboratory, Livermore, CA 94550, USA}
\author{Zachary M. Sparrow}
\affiliation{Department of Chemistry and Chemical Biology, Cornell University, Ithaca, NY 14853, USA}
\author{Ju-an Zhang}
\affiliation{Department of Chemistry and Chemical Biology, Cornell University, Ithaca, NY 14853, USA}
\author{Robert A. DiStasio Jr.}
\email{distasio@cornell.edu}
\affiliation{Department of Chemistry and Chemical Biology, Cornell University, Ithaca, NY 14853, USA}

\date{\today}


\begin{abstract}
High-throughput electronic structure calculations (often performed using density functional theory (DFT)) play a central role in screening existing and novel materials, sampling potential energy surfaces, and generating data for machine learning applications. 
By including a fraction of exact exchange (EXX), hybrid functionals reduce the self-interaction error in semi-local DFT and furnish a more accurate description of the underlying electronic structure, albeit at a computational cost that often prohibits such high-throughput applications.
To address this challenge, we have constructed a robust, accurate, and computationally efficient framework for high-throughput condensed-phase hybrid DFT and implemented this approach in the \texttt{PWSCF} module of \texttt{Quantum ESPRESSO} (\texttt{QE}).
The resulting \texttt{SeA} approach (\texttt{SeA} = SCDM+\texttt{exx}+ACE) combines and seamlessly integrates: (\textit{i}) the selected columns of the density matrix method (SCDM, a robust non-iterative orbital localization scheme that sidesteps system-dependent optimization protocols), (\textit{ii}) a recently extended version of \texttt{exx} (a black-box linear-scaling EXX algorithm that exploits sparsity between localized orbitals in real space when evaluating the action of the standard/full-rank $\hat{V}_{\rm xx}$ operator), and (\textit{iii}) adaptively compressed exchange (ACE, a low-rank $\hat{V}_{\rm xx}$ approximation).
In doing so, \texttt{SeA} harnesses three levels of computational savings: \textit{pair selection} and \textit{domain truncation} from SCDM+\texttt{exx} (which only considers spatially overlapping orbitals on orbital-pair-specific and system-size-independent domains) and \textit{low-rank $\hat{V}_{\rm xx}$ approximation} from ACE (which reduces the number of calls to SCDM+\texttt{exx} during the self-consistent field (SCF) procedure).
Across a diverse set of $200$ non-equilibrium \ce{(H2O)64} configurations (with densities spanning $0.4$~g/cm$^3\mathrm{-}1.7$~g/cm$^3$), \texttt{SeA} provides a one--two order-of-magnitude speedup in the overall time-to-solution, i.e., $\approx 8\times$$-$$26\times$ compared to the convolution-based \texttt{PWSCF(ACE)} implementation in \texttt{QE} and $\approx 78\times$$-$$247\times$ compared to the conventional \texttt{PWSCF(Full)} approach, and yields energies, ionic forces, and other properties with high fidelity.
As a proof-of-principle high-throughput application, we trained a deep neural network (DNN) potential for ambient liquid water at the hybrid DFT level using \texttt{SeA} via an actively learned data set with $\approx 8{,}700$ \ce{(H2O)64} configurations.
Using an out-of-sample set of \ce{(H2O)512} configurations (at non-ambient conditions), we confirmed the accuracy of this \texttt{SeA}-trained potential and showcased the capabilities of \texttt{SeA} by computing the ground-truth ionic forces in this challenging system containing $>1{,}500$ atoms.
\end{abstract}

\maketitle

\section{Introduction \label{sec:intro}}

High-throughput \textit{ab initio} electronic structure calculations play a key role in the computational screening and design of novel materials,~\cite{curtarolo_high-throughput_2013,von_lilienfeld_towards_2014} exploring and sampling potential energy surfaces (PES), as well as generating quantum mechanical data needed for machine learning (ML) applications.~\cite{unke_machine_2021,meuwly_machine_2021,huang_ab_2021}
Since the accuracy is largely governed by the underlying electronic structure method, it is crucial to perform such high-throughput calculations at an appropriate level of theory.
Among the available \textit{ab initio} electronic structure methods, Kohn-Sham (KS) density functional theory (DFT)~\cite{hohenberg_inhomogeneous_1964,kohn_self-consistent_1965} has emerged as the computational workhorse for simulating large molecules and complex condensed-phase systems.
Despite being exact in theory, the exchange-correlation (xc) functional encoding the non-trivial many-body interactions between electrons remains unknown to date; hence, the accuracy of DFT calculations in practice largely relies on functional approximations.~\cite{parr_density-functional_1989,fiolhais_primer_2003,becke_perspective:_2014,mardirossian_thirty_2017,medvedev_density_2017,kepp_comment_2017,hammes-schiffer_conundrum_2017,medvedev_response_2017,lehtola_recent_2018}
For condensed-phase systems, the xc functional is predominantly computed within the (semi-local) generalized gradient approximation (GGA), which includes a local dependence on the electron density, $\rho(\bm r)$, and its gradient, $\nabla \rho(\bm r)$.
To reduce the deleterious self-interaction error (SIE)~\cite{perdew_self-interaction_1981,cohen_insights_2008} at the GGA level (in which each electron spuriously interacts with itself), hybrid functionals~\cite{becke_densityfunctional_1993} include a fraction of exact exchange (EXX) in the xc contribution to the DFT energy.
In doing so, hybrid functionals tend to furnish a more accurate description of the underlying electronic structure, albeit at a high computational cost that often limits their use in high-throughput applications (particularly for large-scale condensed-phase systems).

Since the cost of performing condensed-phase hybrid DFT calculations using the conventional convolution-based EXX algorithm~\cite{gygi_self-consistent_1986} often significantly exceeds that of GGA-based KS-DFT, methods for efficiently performing such challenging calculations are key to the routine use of hybrid functionals in high-throughput applications and have therefore received considerable attention from the community.~\cite{chawla_exact_1998,izmaylov_efficient_2006,sorouri_accurate_2006,guidon_ab_2008,guidon_robust_2009,wu_order-n_2009,gygi_compact_2009,guidon_auxiliary_2010,duchemin_scalable_2010,bylaska_parallel_2011,varini_enhancement_2013,gygi_efficient_2013,distasio_jr._individual_2014,dawson_performance_2015,damle_compressed_2015,lin_adaptively_2016,boffi_efficient_2016,barnes_improved_2017,hu_interpolative_2017,hu_projected_2017,mountjoy_exact_2017,dong_interpolative_2018,carnimeo_fast_2018,mandal_enhanced_2018,mandal_speeding-up_2019,mandal_efficient_2020,paper1,mandal_achieving_2021,paper2}
For large-scale condensed-phase systems with finite gaps, a numerically accurate evaluation of the EXX energy ($\exx$) can be accomplished with linear-scaling cost using \texttt{exx},~\cite{paper1,paper2} a real-space algorithm that exploits the sparsity in the exchange interaction provided by a localized representation of the occupied orbitals (e.g., maximally localized Wannier functions, MLWFs~\cite{marzari_maximally_1997,marzari_maximally_2012}).
To facilitate a linear-scaling (or order-${N}$) evaluation of $\exx$ (as well as other important EXX-related quantities, \textit{vide infra}), the \texttt{exx} algorithm exploits the following two levels of sparsity (for more details, see Sec.~\ref{method:exx} and Refs.~\onlinecite{paper1,paper2}): (\textit{i}) pair selection---\texttt{exx} only considers spatially overlapping orbital pairs that will have a non-vanishing EXX interaction; and (\textit{ii}) domain truncation---\texttt{exx} only evaluates the corresponding EXX interactions on orbital-pair-specific and system-size-independent spatial domains (instead of the entire real-space mesh).
With a massively parallel hybrid \texttt{MPI}/\texttt{OpenMP} implementation in the \texttt{CP} module of the open-source \texttt{Quantum ESPRESSO} (\texttt{QE}) package~\cite{giannozzi_advanced_2017} in conjunction with on-the-fly MLWF localization,~\cite{sharma_ab_2003} \texttt{exx} has enabled hybrid DFT based \textit{ab initio} molecular dynamics (AIMD) of large-scale condensed-phase systems containing $500\mathrm{-}1000$~atoms in the microcanonical/canonical ($NVE$/$NVT$) and isobaric-isoenthalpic/isobaric-isothermal ($NpH$/$NpT$) ensembles with a wall time cost that is comparable to GGA-based KS-DFT.~\cite{paper1,paper2} 
In doing so, current (and pilot) implementations of \texttt{exx} have been used to conduct several challenging theoretical investigations at the hybrid DFT level regarding the electronic structure of semi-conducting solids,~\cite{wu_hybrid_2009,chen_electronic_2011} the structure and local order of ambient liquid water,~\cite{distasio_jr._individual_2014,santra_local_2015} the structural and dynamical properties of aqueous ionic solutions,~\cite{bankura_systematic_2015,chen_hydroxide_2018} the thermal properties of the pyridine-I molecular crystal,~\cite{ko_thermal_2018} as well as isotope effects on the structure of liquid water.~\cite{ko_isotope_2019}

While the current implementation of \texttt{exx}~\cite{paper1,paper2} was designed for large-scale AIMD simulations (particularly the Car--Parrinello~\cite{car_unified_1985} or CPMD variant), the computational efficiency of the core \texttt{exx} algorithm also makes it well-suited for accurately and rapidly evaluating the action of the EXX operator ($\hat{V}_{\rm xx}$) on the (proto-)KS orbitals ($\{ \phi_i \}$) during high-throughput self-consistent field (SCF) calculations at the hybrid DFT level (i.e., $\{ \hat{V}_{\rm xx} \phi_i \}$, which is equivalent (to within a sign) to the EXX contribution to the wavefunction forces needed to propagate the CPMD equations of motion, see Sec.~\ref{method:exx}).
However, direct use of this version of \texttt{exx} would not meet the \textit{implicit} robustness requirements of a high-throughput framework (i.e., black-box and automatable algorithms), since this implementation contains system-dependent parameters in: (\textit{i}) the optimization protocol used to iteratively obtain the MLWFs needed by \texttt{exx} (e.g., initial guesses, convergence criteria, choice of optimization algorithm), and (\textit{ii}) the pair-selection and domain-truncation protocols used inside \texttt{exx}~\cite{paper1,paper2} to harness the aforementioned two levels of computational savings (i.e., the pair distance cutoff, Poisson equation radii, and multipole expansion radii).
In addition to these robustness requirements, there is potentially a third level of computational savings that could be harnessed to increase the efficiency of high-throughput SCF calculations at the hybrid DFT level.
As pointed out by Lin,~\cite{lin_adaptively_2016} these savings result from replacing unnecessary evaluations of $\{ \hat{V}_{\rm xx} \phi_i \}$ using the standard/full-rank EXX operator---the typical bottleneck during SCF calculations at the hybrid DFT level---with more computationally efficient evaluations using a low-rank approximation for $\hat{V}_{\rm xx}$ (i.e., the adaptively compressed exchange (ACE) operator).

%
%
\begin{figure}
    \centering
    \includegraphics[width=\linewidth]{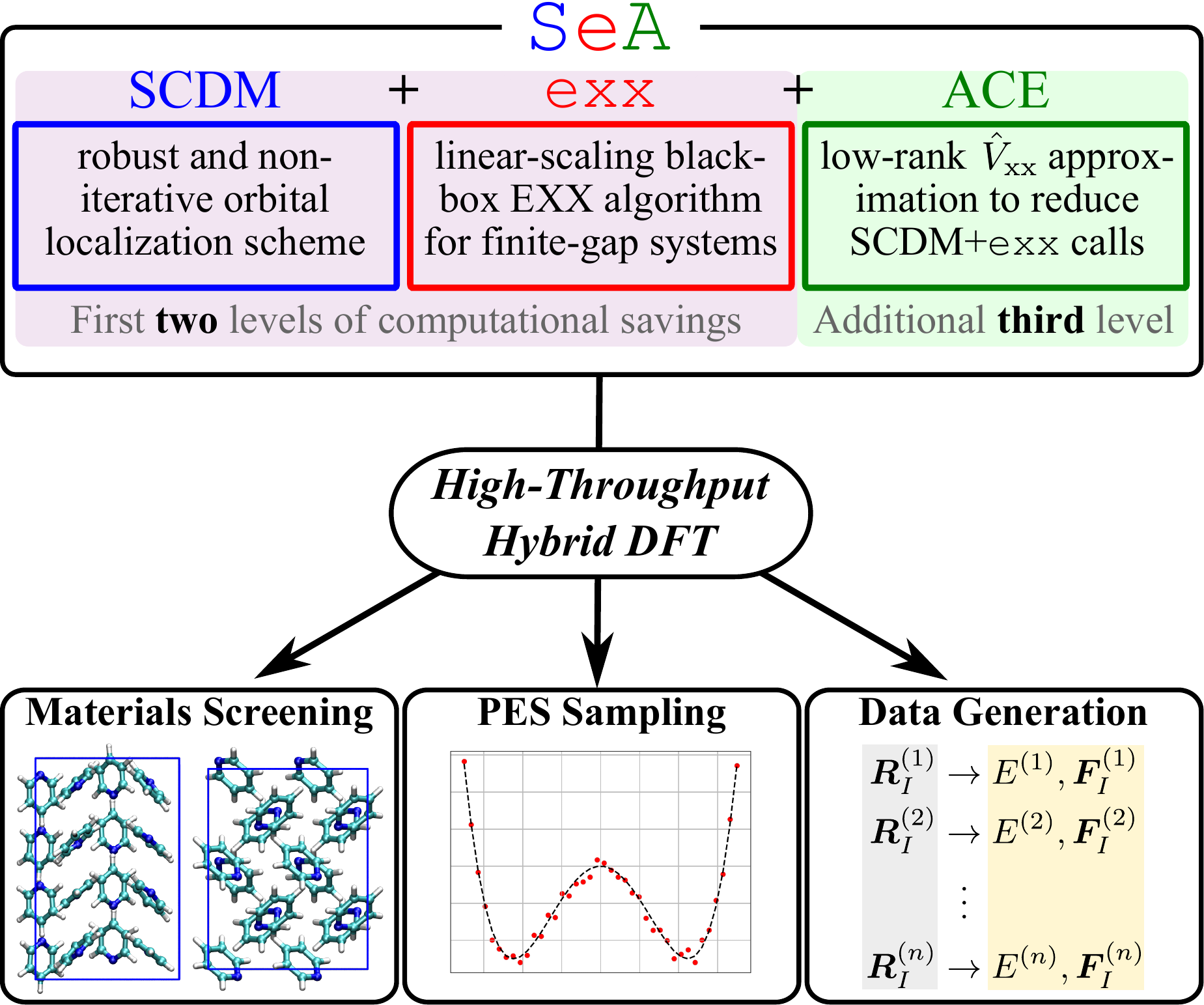}
    \caption{
    Schematic illustration of the \texttt{SeA} approach, which seamlessly integrates three theoretical and algorithmic advances (the non-iterative selected columns of the density matrix (SCDM) orbital localization scheme,~\cite{damle_compressed_2015} the linear-scaling \texttt{exx} algorithm,~\cite{paper1,paper2,paper3} and the adaptively compressed exchange (ACE) formalism~\cite{lin_adaptively_2016}) into a robust, accurate, and computationally efficient framework for performing high-throughput condensed-phase hybrid DFT calculations on large-scale finite-gap systems.
    By harnessing three levels of computational savings, \texttt{SeA} paves the way towards the routine use of hybrid DFT in high-throughput applications such as materials screening and discovery, potential energy surface (PES) sampling, and data generation for machine learning (ML).
    }
    \label{fig:SeA_roadmap}
\end{figure}
%
%
In this work, we directly address these issues by combining and seamlessly integrating three theoretical and algorithmic advances into a robust, accurate, and computationally efficient framework for performing high-throughput condensed-phase hybrid DFT calculations on large-scale finite-gap systems.
As depicted in Fig.~\ref{fig:SeA_roadmap}, the resulting \texttt{SeA} framework (\texttt{SeA} = SCDM+\texttt{exx}+ACE) includes the selected columns of the density matrix (SCDM) orbital localization scheme of Damle, Lin, and Ying,~\cite{damle_compressed_2015} a recent black-box extension~\cite{paper3} of our linear-scaling \texttt{exx} algorithm,~\cite{paper1,paper2} and the ACE formalism of Lin.~\cite{lin_adaptively_2016}
For an accurate and efficient evaluation of the action using the standard/full-rank EXX operator---the typical bottleneck during SCF calculations at the hybrid DFT level---\texttt{SeA} employs the non-iterative SCDM procedure for localized orbital generation (thereby eliminating the need for system-dependent optimization protocols) in conjunction with a recently extended black-box version~\cite{paper3} of \texttt{exx} that replaces all system-dependent parameters with a single system-\textit{independent} orbital coverage threshold.
As described in Ref.~\onlinecite{paper3}, this version of \texttt{exx} is quite robust and able to treat both homogeneous and heterogeneous (i.e., multi-phase and/or multi-component) systems with controllable accuracy (i.e., with an \textit{a priori} estimated error) and markedly improved computational efficiency.
Hence, the combined use of SCDM with this version of \texttt{exx} (SCDM+\texttt{exx}) forms an effectively black-box EXX engine that harnesses two levels of computational savings (i.e., pair selection and domain truncation) when evaluating the most computationally demanding step in hybrid DFT (i.e., evaluation of $\{ \hat{V}_{\rm xx} \phi_i \}$).
To further improve the overall efficiency, \texttt{SeA} harnesses a third level of computational savings by employing the ACE formalism to reduce the number of $\{ \hat{V}_{\rm xx} \, \phi_i \}$ evaluations during the iterative SCF procedure.
In doing so, \texttt{SeA} not only reduces the number of calls to the most computational demanding step in hybrid DFT calculations (via ACE), but also drastically reduces the cost of evaluating each of the remaining calls (via SCDM+\texttt{exx}).
The development of \texttt{SeA} was inspired by the following pioneering works that improved the efficiency of hybrid DFT by harnessing either one or two levels of computational savings based on orbital locality: Gygi and co-workers (who used pair selection~\cite{gygi_efficient_2013,dawson_performance_2015} via recursive subspace bisection (RSB)~\cite{gygi_compact_2009}), Giannozzi and co-workers (who used pair selection in conjunction with ACE to form the localized ACE (L-ACE) method~\cite{carnimeo_fast_2018}), and Nair and co-workers (who used pair selection~\cite{mandal_enhanced_2018} in conjunction with the ACE operator and multiple time scale approach~\cite{tuckerman_reversible_1992} for efficient hybrid DFT based AIMD~\cite{mandal_speeding-up_2019,mandal_achieving_2021}).
\texttt{SeA} differs from these previous developments by providing a single framework that harnesses \textit{three} levels of savings (pair selection, domain truncation, and ACE) to further enable the routine use of hybrid DFT in high-throughput applications involving large-scale condensed-phase systems.

The remainder of this manuscript is organized as follows.
In Sec.~\ref{sec:method}, we provide a brief review of the SCDM, \texttt{exx}, and ACE approaches as well as their seamless combination/integration into the \texttt{SeA} framework implemented in the \texttt{PWSCF} module of \texttt{QE}.
In Sec.~\ref{sec:acc_prf}, we assess the accuracy and performance of \texttt{SeA} on a diverse set of $200$ non-equilibrium \ce{(H2O)64} configurations (which include both intact and auto-ionized water molecules, with system densities spanning $0.4$~g/cm$^3\mathrm{-}1.7$~g/cm$^3$), and demonstrate that \texttt{SeA} yields a one--two order-of-magnitude ($\approx 11\times$$-$$116\times$) speedup in the computational bottleneck in the convolution-based \texttt{PWSCF(ACE)} implementation in \texttt{QE}, while providing energies, ionic forces, and other properties with high fidelity.
%
In doing so, \texttt{SeA} enables the routine use of hybrid DFT in high-throughput applications for systems with sizes similar to (and beyond) \ce{(H2O)64}, and delivers single-point energy and ionic force evaluations for such systems with an $\approx 8\times$$-$$26\times$ speedup in the overall time-to-solution compared to \texttt{PWSCF(ACE)} and an $\approx 78\times$$-$$247\times$ speedup compared to the conventional (non-ACE) \texttt{PWSCF(Full)} EXX implementation in \texttt{QE}.
As a proof-of-principle high-throughput application, we used \texttt{SeA} in Sec.~\ref{sec:app} to train a deep neural network (DNN) potential for ambient ($T = 300$~K, $p = 1$~Bar) liquid water at the PBE0~\cite{perdew_rationale_1996,adamo_toward_1999} level based on an actively learned data set containing $\approx 8,700$ \ce{(H2O)64} configurations.
We then assessed the accuracy of this DNN potential on an out-of-sample test set (\ce{(H2O)512} at $T = 330$~K and $p = 1$~Bar), and showcased the capabilities of \texttt{SeA} by directly computing the (ground-truth) ionic forces in these challenging condensed-phase systems containing $>1{,}500$ atoms.
This manuscript then is ended with a brief summary of our findings in Sec.~\ref{sec:conclusion} and some potential future research directions in Sec.~\ref{sec:future_outlook}.

\section{Theory \label{sec:method}}

In this section, we begin with a brief review of the non-iterative SCDM orbital localization scheme,~\cite{damle_compressed_2015} which will be used to generate localized occupied orbitals without the need for system-dependent optimization protocols (Sec.~\ref{method:scdm}).
Based on the set of localized orbitals computed with this scheme, we then describe a recent extension~\cite{paper3} to the linear-scaling \texttt{exx} algorithm~\cite{paper1,paper2} that replaces all system-dependent parameters with a single system-\textit{independent} orbital coverage threshold (Sec.~\ref{method:exx}); this algorithm harnesses two levels of computational savings (i.e., pair selection and domain truncation) and will be used to provide an accurate and efficient black-box evaluation of the action using the standard/full-rank EXX operator ($\hat{V}_{\rm xx}$).
We then briefly summarize the ACE method, which provides a third level of computational savings by furnishing an efficient low-rank approximation to $\hat{V}_{\rm xx}$ that eliminates unnecessary full-rank evaluations of the action during the iterative SCF procedure (Sec.~\ref{method:ace}).
This is followed by a detailed description of how these three theoretical and algorithmic advances are combined and seamlessly integrated into \texttt{SeA} (SCDM+\texttt{exx}+ACE)---a robust, accurate, and computationally efficient framework for high-throughput condensed-phase hybrid DFT for large-scale finite-gap systems (Sec.~\ref{method:SeA}).

\subsection{SCDM: A Robust and Non-Iterative Orbital Localization Scheme \label{method:scdm}}

To generate the localized orbitals needed for \texttt{exx}, \texttt{SeA} employs the selected columns of the density matrix (SCDM) approach,~\cite{damle_compressed_2015} a robust and non-iterative alternative to the well-known MLWF procedure~\cite{marzari_maximally_1997,marzari_maximally_2012} used throughout the development of \texttt{exx}.~\cite{wu_order-n_2009,distasio_jr._individual_2014,paper1,paper2}
Since this work focuses on EXX calculations involving large-scale finite-gap systems, we will center our discussion around the $\Gamma$-point specific SCDM algorithm.~\cite{damle_compressed_2015}
However, this choice is not an intrinsic limitation of \texttt{SeA}, as \texttt{exx} could (in principle) utilize localized orbitals obtained via Brillouin zone sampling of the canonical KS orbitals with the SCDM-k approach.~\cite{damle_scdm-k:_2017}
The theoretical foundation underlying SCDM is the ``nearsightedness'' principle popularized by Kohn,~\cite{kohn_density_1996,prodan_nearsightedness_2005} which states that the elements of the real-space one-particle density matrix ($\bm P$), i.e., $P(\bm r, \bm r') = \braket{\bm r | \hat P | \bm r'} = \sum_i^{N_{\rm occ}} \braket{\bm r \left| \phi_{i} \right.} \braket{\left. \phi_{i} \right| \bm r'}$, are exponentially decaying with respect to $|\bm r - \bm r'|$ in finite-gap systems.
On a real-space mesh with $N_{\rm grid}$ points, $\bm P = \bm{\Phi}\bm{\Phi}^{T}$ is an $N_{\rm grid} \times N_{\rm grid}$ matrix constructed from $\bm{\Phi}$, the corresponding $N_{\rm grid} \times N_{\rm occ}$ wavefunction matrix with columns given by a set of $N_{\rm occ}$ orbitals ($\{ \phi_i \}$) which spans the occupied space, e.g., the set of canonical KS occupied orbitals.
As such, $\bm P$ is rank-deficient with a column space dimension of $\mathrm{Rank}(\bm P) = N_{\rm occ}$, from which one can obtain a localized representation of the occupied space by selecting (and orthogonalizing) a set of $N_{\rm occ}$ well-conditioned columns of $\bm P$.

In the SCDM approach, this set of columns is obtained from a column-pivoted QR factorization of the $N_{\rm occ} \times N_{\rm grid}$ $\bm{\Phi}^T$ matrix (instead of the significantly larger $\bm P$ matrix), as this alternative formulation yields equivalent column selection at significantly lower computational cost.~\cite{damle_compressed_2015}
Performing this factorization yields:
\begin{align}
    \bm{\Phi}^{T} \bm{\Pi} = \bm{Q} \bm{R} ,
    \label{eq:phi_qr}
\end{align}
in which $\bm{\Pi}$ is the $N_{\rm grid} \times N_{\rm grid}$ permutation matrix (i.e., a dense representation of the selected column indices), $\bm{Q}$ is an $N_{\rm occ} \times N_{\rm occ}$ orthogonal matrix, and $\bm{R}$ is an $N_{\rm occ} \times N_{\rm grid}$ upper triangular matrix.
The proto-SCDM orbitals (i.e., a well-conditioned set of localized but non-orthogonal orbitals that spans the occupied space) are then given by the following $N_{\rm grid} \times N_{\rm occ}$ matrix:
\begin{align}
    \bm{X} \equiv \bm{P}_{:,C} = \bm{\Phi} (\bm{\Phi}^{T})_{:,C} ,
    \label{eq:nonortho_scdm}
\end{align}
in which ``$:$'' denotes all row indices, $C$ is the set of indices corresponding to the $N_{\rm occ}$ selected columns, and $(\bm{\Phi}^{T})_{:,C}$ is an $N_{\rm occ} \times N_{\rm occ}$ matrix containing the first $N_{\rm occ}$ columns of $\bm{\Phi}^{T} \bm{\Pi}$.
The final SCDM orbitals are then obtained via symmetric orthogonalization of the proto-SCDM orbitals, namely,
\begin{align}
    \bm{\widetilde{\Phi}} = \bm{X} (\bm{P}_{C,C})^{-1/2} ,
    \label{eq:ortho_scdm}
\end{align}
in which $\bm{P}_{C,C} = \bm{X}^T \bm{X}$ is the $N_{\rm occ}\times N_{\rm occ}$ overlap matrix in the proto-SCDM basis (which is also equivalent to the density matrix in the basis selected by $C$).
In this expression, we follow the convention used during the development of \texttt{exx}:~\cite{paper1,paper2} all quantities that depend on the choice of localized orbitals (i.e., MLWFs in previous work and SCDM in this work) are dressed with tildes, while all quantities invariant to the underlying orbital representation (e.g., $\exx$) are left unmodified.
By plugging Eq.~\eqref{eq:nonortho_scdm} into Eq.~\eqref{eq:ortho_scdm}, one can see that the input orbitals ($\bm \Phi$) are connected to the localized SCDM orbitals ($\bm{\widetilde{\Phi}}$) via:
\begin{align}
  \bm{\widetilde{\Phi}} = \bm{\Phi} (\bm{\Phi}^{T})_{:,C} (\bm{P}_{C,C})^{-1/2} \equiv \bm{\Phi} \bm U ,
  \label{eq:fwd_rot_phi}
\end{align}
in which
\begin{align}
    \bm U \equiv (\bm{\Phi}^{T})_{:,C} (\bm{P}_{C,C})^{-1/2}
    \label{eq:unitary_scdm}
\end{align}
is an $N_{\rm occ} \times N_{\rm occ}$ unitary matrix that allows one to transform between these orbital representations via a single matrix multiplication.
Hence, the non-iterative SCDM orbital localization scheme is able to furnish $\bm{U}$ without the need for system-dependent optimization protocols (e.g., initial guesses, convergence criteria, choice of optimization algorithm), which makes it well-suited to provide the localized orbitals required by \texttt{exx} in the high-throughput \texttt{SeA} framework.

\subsection{\texttt{exx}: A Black-Box Linear-Scaling Exact- \\Exchange Algorithm for Finite-Gap Systems \label{method:exx}}

With the completion of the non-iterative SCDM localization procedure, one can now perform the following unitary transformation (using $\bm U$ from Eq.~\eqref{eq:unitary_scdm}),
\begin{align}
  \widetilde{\phi}_{i}(\bm r) = \sum_{j} \phi_{j}(\bm r) (\bm U)_{ji} ,
  \label{eq:rot_phi}
\end{align}
to obtain a set of localized orbitals that spans the occupied space and forms a basis for a computationally efficient linear-scaling evaluation of all EXX-related quantities needed during hybrid DFT calculations on large-scale finite-gap systems at the $\Gamma$-point.
This is most easily illustrated by considering the canonical expression for $\exx$ (shown here for a closed-shell system for simplicity), 
\begin{align}
  \exx = - \sum_{ij} \int_{\Omega} \rho_{ij}(\bm r) v_{ij}(\bm r) \dd \bm{r} ,
  \label{eq:exxGen}
\end{align}
in which the sum includes all pairs of occupied orbitals, the integral is over the entire real-space domain  ($\Omega$) in the periodic unit cell, $\rho_{ij}(\bm{r}) \equiv \phi_{i}(\bm{r}) \phi_{j}(\bm{r})$ is the orbital-product density, and $v_{ij}(\bm r)$ is the corresponding orbital-product potential (i.e., the solution to Poisson's equation, $\nabla^2 v_{ij}(\bm r) = -4\pi \rho_{ij}(\bm{r})$).
%
In typical planewave/pseudopotential codes, this expression is evaluated with cubic-scaling cost using the conventional convolution-based EXX algorithm.~\cite{gygi_self-consistent_1986}
The fact that Eq.~\eqref{eq:exxGen} is invariant to the unitary transformation in Eq.~\eqref{eq:rot_phi} forms the theoretical foundation for the \texttt{exx} algorithm~\cite{paper1,paper2} in \texttt{QE}, which utilizes a basis of localized orbitals (MLWFs in previous work and SCDM in the current work) to exploit the following two levels of sparsity during the real-space evaluation of $\exx$:
\begin{itemize}
    \item \textit{Pair selection}: only spatially overlapping orbitals $\braket{ij}$ need to be included when evaluating Eq.~\eqref{eq:exxGen} in a localized basis; since each localized orbital $\widetilde{\phi}_{i}(\bm{r})$ has compact support and will only overlap with a limited and constant-scaling number of neighboring orbitals, exploiting this level of sparsity reduces the total number of orbital pairs from quadratic to linear ($\sum_{ij} \rightarrow \sum_{\braket{ij}}$) in Eq.~\eqref{eq:exxGen}.
    \item \textit{Domain truncation}: for each overlapping $\braket{ij}$ pair, the spatial integral in Eq.~\eqref{eq:exxGen} only needs to be evaluated on a domain which encompasses $\widetilde{\rho}_{ij}(\bm{r}) \equiv \widetilde{\phi}_{i}(\bm{r}) \widetilde{\phi}_{j}(\bm{r})$; hence, this integral can be performed on a series of orbital-pair-specific domains $\Omega_{ij}$ that are independent of the system size ($\int_{\Omega} \rightarrow \int_{\Omega_{ij}}$).
\end{itemize}
By harnessing these two levels of computational savings, \texttt{exx} enables an efficient linear-scaling evaluation of Eq.~\eqref{eq:exxGen}, which can now be rewritten in the following working form:
\begin{align}
  \exx = - \sum_{\braket{ij}} \int_{\Omega_{ij}} \widetilde{\rho}_{ij}(\bm r) \widetilde{v}_{ij}(\bm r) \dd \bm{r},
  \label{eq:exxGen_mlwf}
\end{align}
in which $\widetilde{\rho}_{ij}(\bm r)$ and $\widetilde{v}_{ij}(\bm r)$ are the orbital-product density and corresponding orbital-product potential in the localized representation.
After determining the set of overlapping pairs, \texttt{exx} computes $\widetilde{v}_{ij}(\bm r)$ for a given $\braket{ij}$ pair via the iterative (conjugate gradient) solution to Poisson's equation (PE),
\begin{align}
    \nabla^2 \widetilde{v}_{ij}(\bm r) = -4\pi\widetilde{\rho}_{ij}(\bm r) \qquad \bm r \in \Omega_{ij} ,
    \label{eq:pe}
\end{align}
in the near-field region (i.e., for $\bm r \in \Omega_{ij}$), in conjunction with boundary conditions provided by a sufficiently converged multipole expansion (ME):
\begin{align}
    \widetilde{v}_{ij}(\bm r) = 4\pi \sum_{lm} \frac{Q_{lm}}{(2l+1)} \frac{Y_{lm}(\theta,\varphi)}{r^{l+1}} \qquad \bm r \in \partial \, \Omega_{ij} .
    \label{eq:me}
\end{align}
In this expression, $Y_{lm}(\theta,\varphi)$ are the spherical harmonics and $Q_{lm}$ are the multipole moments associated with $\widetilde{\rho}_{ij}(\bm r)$, i.e., 
\begin{align}
    Q_{lm} = \int_{\Omega_{ij}} Y_{lm}^{*}(\theta,\varphi) \, r^l \widetilde{\rho}_{ij}(\bm r) \dd\bm{r} .
    \label{eq:mepole}
\end{align}

In addition to the EXX contribution to the energy, hybrid DFT calculations also require evaluation of: 
\begin{align}
    D_{\rm xx}^{i}(\bm r) \equiv -\left(\frac{\delta \exx}{\delta \phi_{i}^*(\bm r)}\right) = -\hat{V}_{\rm xx} \phi_i (\bm r) ,
    \label{eq:DxxVxx}
\end{align}
a quantity that is often referred to as the EXX contribution to the wavefunction forces in the AIMD community, as it is needed to propagate the CPMD equations of motion.
Of particular relevance to this work is the fact that $D_{\rm xx}^{i}(\bm r)$ is also equivalent (to within a sign) to the action of the EXX operator ($\hat{V}_{\rm xx}$) on the (proto-)KS orbitals, a quantity that is needed to construct the xc potential during SCF calculations at the hybrid DFT level.
While the evaluation of the canonical expression for $D_{\rm xx}^{i}(\bm r)$ (shown here for a closed-shell system, \cf Eq.~\eqref{eq:exxGen}), 
\begin{align}
    D_{\rm xx}^{i}(\bm r) = \sum_{j} v_{ij}(\bm r) \phi_{j}(\bm r) \qquad \bm r \in \Omega ,
    \label{eq:Dxx}
\end{align}
is typically the computational bottleneck during the iterative SCF procedure, this quantity can also be efficiently computed in \texttt{exx} by exploiting the two levels of sparsity inherent to a localized basis (i.e., pair selection and domain truncation).
Since each localized orbital has compact support in real space, the sum over all orbitals in Eq.~\eqref{eq:Dxx} can again be replaced by a sum over overlapping orbitals only ($\sum_{j} \rightarrow \sum_{j \in \braket{ij}}$), with each contribution only needing to be evaluated on an orbital-specific and system-size-independent domain ($\Omega \rightarrow \Omega_{ij}^{D}$, \textit{vide infra}) that encompasses $\widetilde{\phi}_{j}(\bm{r})$.
Hence, \texttt{exx} also enables an efficient linear-scaling evaluation of Eq.~\eqref{eq:Dxx}, which can now be rewritten in the following working form:
\begin{align}
    \widetilde{D}_{\rm xx}^{i}(\bm r) = \sum_{j\in \braket{ij}} \widetilde{v}_{ij}(\bm r) \widetilde{\phi}_{j}(\bm r) \qquad \bm r \in \Omega_{ij}^{D} ,
    \label{eq:tDxx}
\end{align}
in which $\widetilde{v}_{ij}(\bm r)$ in the near-field region (for $\bm r \in \Omega_{ij}$) is provided by the solution to the PE in Eq.~\eqref{eq:pe}, while  $\widetilde{v}_{ij}(\bm r)$ in the far-field region (for $\bm r \in \Omega_{ij}^{D}\setminus\Omega_{ij}$) is provided by the ME in Eq.~\eqref{eq:me}.
Unlike $E_{\rm xx}$, $\widetilde{D}_{\rm xx}^{i}(\bm r)$ is not invariant to the underlying orbital representation and is therefore dressed with a tilde (following the convention used for non-invariant quantities in Refs.~\onlinecite{paper1,paper2}).
However, the canonical form of this quantity in Eq.~\eqref{eq:Dxx} is straightforwardly obtained from the local form in Eq.~\eqref{eq:tDxx} via the following unitary transformation:
\begin{align}
  D_{\rm xx}^{i}(\bm r) = \sum_{j} \widetilde{D}_{\rm xx}^{j}(\bm r) (\bm U^{-1})_{ji} ,
  \label{eq:rotDr}
\end{align}
in conjunction with the $\bm U$ provided by the orbital localization scheme (MLWFs in previous work and SCDM in the current work, see Eq.~\eqref{eq:rot_phi}).
Here, we note that this transformation is of particular relevance to this work as it allows the linear-scaling \texttt{exx} algorithm to directly attack the computational bottleneck (i.e., the evaluation of $\{ D_{\rm xx}^{i}(\bm r) \}$ or $\{ \hat{V}_{\rm xx}\phi_i(\bm r) \}$ in Eq.~\eqref{eq:DxxVxx}) that prohibits the routine use of hybrid DFT in applications involving large-scale condensed-phase systems---the central idea in the high-throughput \texttt{SeA} framework developed herein (see Sec.~\ref{method:SeA}).

With a massively parallel hybrid \texttt{MPI}/\texttt{OpenMP} implementation~\cite{paper1,paper2} in the \texttt{CP} module of \texttt{QE}~\cite{giannozzi_advanced_2017} (in conjunction with on-the-fly MLWF localization~\cite{sharma_ab_2003}), \texttt{exx} has enabled hybrid DFT based AIMD/CPMD simulations of large-scale condensed-phase systems containing $500\mathrm{-}1000$~atoms in the $NVE$/$NVT$ and $NpH$/$NpT$ ensembles with a wall time cost that is comparable to GGA-based KS-DFT,~\cite{paper1,paper2} and has been used to perform a number of challenging theoretical applications to date.~\cite{wu_hybrid_2009,chen_electronic_2011,distasio_jr._individual_2014,santra_local_2015,bankura_systematic_2015,chen_hydroxide_2018,ko_thermal_2018,ko_isotope_2019}
However, the direct use of this version of \texttt{exx} would not meet the \textit{implicit} robustness requirements of a high-throughput hybrid DFT framework (i.e., black-box/automatable algorithms), since this implementation relies on a set of system-dependent parameters to harness the two levels of computational savings described above (i.e., a pair distance cutoff ($R_{\rm pair}$) for pair selection, PE radii ($R_{\rm PE}^{\rm s}$, $R_{\rm PE}^{\rm ns}$) and ME radii ($R_{\rm ME}^{\rm s}$, $R_{\rm ME}^{\rm ns}$) for domain truncation; see Refs.~\onlinecite{paper1,paper2} for more details).

To address this shortcoming, we now briefly describe a recent extension~\cite{paper3} of \texttt{exx} that replaces all five of these parameters with a single system-\textit{independent} orbital coverage threshold ($\epsilon$).
More specifically, $\epsilon$ forms the basis for the \textit{domain truncation} protocol in \texttt{exx}, in which an orbital-specific and system-size-independent domain ($\Omega_{ii} \subset \Omega$) is first determined for each $\widetilde{\phi}_{i}(\bm r)$ according to the following orbital normalization condition:
\begin{align}
    1 = \int_{\Omega} \left|\widetilde{\phi}_{i} (\bm r)\right|^2 \dd \bm{r} \geq \int_{\Omega_{ii}} \left|\widetilde{\phi}_{i} (\bm r)\right|^2 \dd \bm{r} \geq 1 - \epsilon .
    \label{eq:oii}
\end{align}
In other words, $\epsilon$ specifies the level of (numerical) compliance in the orbital normalization condition (i.e., an upper bound in the fractional particle loss) when determining the domain $\Omega_{ii}$ that encompasses each $\widetilde{\phi}_i (\bm r)$.
%
In practice, $\epsilon$ is a small ($\epsilon \ll 1$) dimensionless parameter, which is systematically improvable, i.e., as $\epsilon\rightarrow 0$, $\Omega_{ii} \rightarrow \Omega \,\, \forall \,\, \widetilde{\phi}_{i}$ and the error due to domain truncation vanishes.
As such, $\epsilon$ provides an \textit{a priori} estimation~\cite{dawson_performance_2015,paper3} of the accuracy in all EXX-related quantities computed using this approach (\textit{vide infra}).

In this recently extended version of \texttt{exx},~\cite{paper3} each $\Omega_{ii}$ is represented by a tight axis-aligned bounding box (AABB) with respect to the underlying real-space grid---a parallelepiped that accounts for the anisotropy in the shape and extent of each localized orbital by construction, and provides a natural framework for \textit{simultaneously} and \textit{consistently} exploiting the two levels of sparsity described above.
To proceed, \texttt{exx} completes the domain truncation process by finding the set of orbital-pair-specific domains $\Omega_{ij} \equiv \Omega_{ii} \cap \Omega_{jj}$. In doing so, \texttt{exx} simultaneously initiates the \textit{pair selection} process by identifying a set of proto-$\braket{ij}$ pairs with $\Omega_{ij} \neq \text{\O}$.
To complete the pair selection process and arrive at the final (and significantly smaller) set of $\braket{ij}$ pairs, \texttt{exx} then further screens the proto-$\braket{ij}$ pairs according to the following absolute orbital overlap criterion:
\begin{equation}
    S_{|i,j|} \equiv \int_{\Omega_{ij}} \left|\widetilde{\phi}_{i} (\bm r)\widetilde{\phi}_{j} (\bm r)\right| \dd \bm r = \int_{\Omega_{ij}} \left|\widetilde{\rho}_{ij} (\bm r)\right| \dd \bm r \geq \delta .
    \label{eq:sij}
\end{equation}
In this expression, $\delta = \delta(\epsilon) > 0$ is automatically determined in \texttt{exx} to maximize transferability across systems with varying degrees of heterogeneity.
As described in Ref.~\onlinecite{paper3}, this is accomplished by ensuring that the pair selection error (governed by $\delta$ in Eq.~\eqref{eq:sij}) is smaller than (but comparable to) the domain truncation error (governed by $\epsilon$ in Eq.~\eqref{eq:oii}).
Here, we note in passing that the use of an absolute orbital overlap criterion during pair selection in \texttt{exx} was inspired by Refs.~\onlinecite{mandal_enhanced_2018,carnimeo_fast_2018}; however, the integral in Eq.~\eqref{eq:sij} is restricted to $\Omega_{ij}$ in \texttt{exx} instead of the entire real-space domain $\Omega$, which forms the basis for a linear-scaling pair-selection protocol.

Following the detailed derivation provided in Ref.~\onlinecite{paper3}, $\Omega_{ij}^{D}$ was set to $\mathrm{scale}_{2\times} \, \Omega_{ii}$ (for $j = i$) and $\Omega_{jj}$ (for $j \neq i$) when computing $\{ \widetilde{D}_{\rm xx}^{i}(\bm r) \}$ via Eq.~\eqref{eq:tDxx}, which enables an accurate and efficient linear-scaling evaluation of these mission-critical terms (\textit{vide infra}).
As described in Ref.~\onlinecite{paper3}, this theoretical extension of \texttt{exx} is quite robust and able to treat both homogeneous and heterogeneous (i.e., multi-phase and/or multi-component) systems with controllable accuracy, i.e., with an \textit{a priori} estimated error.
We have also completely overhauled the \texttt{exx} codebase with a comprehensive three-pronged algorithmic strategy designed to increase computational efficiency, decrease communication overhead, and minimize processor idling, which makes this version well-suited to be the core EXX engine in the high-throughput \texttt{SeA} framework.
To maximize the impact of these developments, we also plan to release this EXX engine as a free software library (\texttt{exxl}). 

\subsection{ACE: An Efficient Low-Rank Approximation to the Exact-Exchange Operator \label{method:ace}}

While the combined use of SCDM and this recently extended version of \texttt{exx} already provides a robust, accurate, and efficient core EXX engine for high-throughput hybrid DFT calculations of large-scale finite-gap systems, there is a third level of computational savings (independent from the two levels provided by SCDM+\texttt{exx}) that could be harnessed to further increase the efficiency of the SCF procedure for hybrid functionals.
As pointed out by Lin,~\cite{lin_adaptively_2016} the most computationally demanding steps during the iterative solution to the KS-DFT equations at the hybrid level---the repeated evaluation of the action using the standard/full-rank EXX operator (e.g., $\{ \hat{V}_{\rm xx} \phi_i \}$ in Eqs.~\eqref{eq:DxxVxx}--\eqref{eq:Dxx})---can be replaced with more efficient evaluations of these terms using a low-rank approximation for $\hat{V}_{\rm xx}$ (i.e., the adaptively compressed exchange (ACE) operator, $\hat{V}_{\rm xx}^{\rm ACE}$).
As outlined in Algorithm~\ref{alg:ace}, the ACE approach exploits the double-loop SCF structure commonly used when performing hybrid DFT calculations in condensed-phase electronic structure packages like \texttt{QE}. 
%
%
\begin{algorithm}
    \While{$\{\left| \phi_i \right> \}$ not converged 
    }{
      \tcp{outer loop}
      \texttt{ACE\_Construction}($\{\left| \phi_i \right>\}$; $\hat{V}_{\rm xx}^{\rm ACE}$)\;
      $\{\left|  \chi_i \right>\} \leftarrow \{ \left| \phi_i \right>\}$\;
      \While{$\{\left| \chi_i \right> \}$ not converged
      }{
        \tcp{inner loop}
        $\rho (\bm r) \leftarrow \sum_i |\chi_i (\bm r)|^2$\;
        \texttt{SCF\_Iteration}($\rho (\bm r)$, $\hat{V}_{\rm xx}^{\rm ACE}$; $\{\left| \chi_i \right>\}$)\;
      }
      $\{ \left| \phi_i \right>\} \leftarrow \{\left|  \chi_i \right>\}$\;
    }
    \caption{EXX-SCF procedure with ACE}
    \label{alg:ace}
\end{algorithm}
%
%
Although $\hat{V}_{\rm xx}^{\rm ACE}$ can in principle be constructed from (and applied to) both occupied and virtual/unoccupied (proto-)KS orbitals,~\cite{lin_adaptively_2016} we will limit our discussion below to the ACE procedure involving occupied (proto-)KS orbitals only.
We note in passing that this choice is not an intrinsic limitation of the \texttt{SeA} approach developed in this work, and an extension to include virtual (proto-)KS orbitals is underway in our group and will be discussed in future work.

During each outer-loop iteration in the EXX-SCF procedure with ACE (see Algorithm~\ref{alg:ace}), the \texttt{ACE\_Construction} step (the bottleneck for large-scale hybrid DFT calculations) takes the current set of proto-KS orbitals $\{\left| \phi_i \right>\}$ as input, and performs a full-rank evaluation of the action to obtain $\{ \hat{V}_{\rm xx} \left| \phi_{i}\right> \}$.
Since $\left| D_{\rm xx}^{i}\right> = -\hat{V}_{\rm xx} \left| \phi_{i}\right>$ via Eq.~\eqref{eq:DxxVxx}, the use of $\left| D_{\rm xx}^{i} \right>$ instead of $\hat{V}_{\rm xx}\left| \phi_{i} \right>$ will lead to a sign difference in certain quantities (e.g., $M_{ij}$ below) when compared to the original ACE formulation.~\cite{lin_adaptively_2016}
In this section, we use bra-ket notation to emphasize the fact that the ACE approach is independent of the underlying representation; in other words, one has the flexibility to work in real and/or reciprocal space (i.e., using $D_{\rm xx}^{i}(\bm r) = \left< \bm r \left| D_{\rm xx}^{i}\right>\right.$ and/or $D_{\rm xx}^{i}(\bm G) = \left< \bm G \left| D_{\rm xx}^{i}\right>\right.$).
The output of \texttt{ACE\_Construction} is the ACE operator:
\begin{align}
  \hat{V}_{\rm xx}^{\rm ACE} = - \sum_k \left| \xi_{k} \right> \left< \xi_{k} \right| ,
  \label{eq:Vace}
\end{align}
a low-rank approximation of $\hat{V}_{\rm xx}$ that is obtained via a Cholesky decomposition of the $N_{\rm occ} \times N_{\rm occ}$ symmetric positive semi-definite matrix $M_{ij} \equiv \left< \phi_i \left| D_{\rm xx}^{j} \right. \right> = - \left< \phi_i \right| \hat{V}_{\rm xx} \left| \phi_{j} \right> = \left(\bm{L}\bm{L}^T\right)_{ij}$, which yields $\left| \xi_{k} \right> \equiv -\sum_i \left| D_{\rm xx}^{i} \right> \left( {\bm L}^{-T} \right)_{ik}$ for $k = 1,\ldots,N_{\rm occ}$.

With the completion of the \texttt{ACE\_Construction} step, a set of auxiliary orbitals, $\{\left| \chi_i \right>\}$, is initialized to the current set of proto-KS orbitals ($\{\left|  \chi_i \right>\} \leftarrow \{ \left| \phi_i \right>\}$) as one enters the inner loop in Algorithm~\ref{alg:ace}.
Inside the inner loop, the KS-DFT equations are iteratively solved with a series of calls to the \texttt{SCF\_Iteration} step, which takes $\rho(\bm r) = \sum_i |\chi_i (\bm r)|^2$ (the charge density in real space) and $\hat{V}_{\rm xx}^{\rm ACE}$ as input.
During each \texttt{SCF\_Iteration} call, $\rho(\bm r)$ is used to construct the semi-local (non-EXX) contribution to the KS Hamiltonian and $\hat{V}_{\rm xx}^{\rm ACE} \left| \chi_i \right>$ (evaluated with the \textit{fixed} $\hat{V}_{\rm xx}^{\rm ACE}$ operator) is used to compute the EXX action.
Throughout the inner-loop iterations, $\{\left| \chi_i \right>\}$ and the semi-local contributions to the KS Hamiltonian are updated towards self-consistency, while $\hat{V}_{\rm xx}^{\rm ACE}$ remains fixed/unmodified.
Upon convergence of the auxiliary orbitals, the code exits the inner loop and the proto-KS orbitals are updated ($\{\left| \phi_i \right>\} \leftarrow \{ \left| \chi_i \right>\}$) for the next outer-loop iteration.

The ACE-based EXX-SCF procedure in Algorithm~\ref{alg:ace} is completed once the proto-KS orbitals in the outer loop reach self-consistency, which is equivalent to a fully self-consistent hybrid DFT calculation.
At this point, $\hat{V}_{\rm xx}^{\rm ACE}$ reproduces the action of $\hat{V}_{\rm xx}$ on the occupied KS orbitals, and the EXX energy can be conveniently evaluated via (shown here for a closed-shell system at the $\Gamma$-point for consistency with Eqs.~\eqref{eq:exxGen} and \eqref{eq:DxxVxx}):
\begin{align}
    E_{\rm xx} = \sum_{i} \left< \phi_i \right| \hat{V}_{\rm xx}^{\rm ACE} \left| \phi_i \right> = - \sum_{ik} \left|\left< \phi_i  \left| \xi_k \right>\right| \right.^2 ,
    \label{eq:exx_ace}
\end{align}
in which each sum is over the converged occupied KS orbitals. 
%
By replacing the more costly evaluations of the action using $\hat{V}_{\rm xx}$ with $\hat{V}_{\rm xx}^{\rm ACE}$, the ACE approach offers an additional level of computational savings (on top of the two provided by SCDM+\texttt{exx}) that can be harnessed to attack the most computationally demanding steps during the iterative solution to the KS equations at the hybrid DFT level, thereby making it well-suited to further improve the efficiency and throughput of the \texttt{SeA} framework.

\subsection{\texttt{SeA} (SCDM+\texttt{exx}+ACE): High-Throughput Hybrid DFT for Large-Scale Finite-Gap Systems \label{method:SeA}}

In this section, we describe how \texttt{SeA} combines and seamlessly integrates SCDM, \texttt{exx}, and ACE into a robust, accurate, and computationally efficient framework (which has been implemented in \texttt{PWSCF}) for performing high-throughput condensed-phase hybrid DFT calculations on large-scale finite-gap systems.
For a robust, accurate, and efficient evaluation of the action using the standard/full-rank EXX operator ($\{\hat{V}_{\rm xx} \left| \phi_{i}\right>\}$)---the typical bottleneck during SCF calculations at the hybrid DFT level---\texttt{SeA} employs the non-iterative SCDM procedure for localized orbital generation described in Sec.~\ref{method:scdm} in conjunction with the recently extended black-box version~\cite{paper3} of \texttt{exx} described in Sec.~\ref{method:exx}.
%
To further improve the overall efficiency and throughput, \texttt{SeA} also employs the ACE operator formalism described in Sec.~\ref{method:ace} to reduce the number of times that $\{ \hat{V}_{\rm xx} \left| \phi_{i}\right> \}$ (or $\{ \left| D_{\rm xx}^{i}\right> \}$) needs to be evaluated during the iterative solution to the KS-DFT equations.
%
In doing so, \texttt{SeA} harnesses three distinct levels of computational savings by reducing the number of calls to the most computationally demanding step in hybrid DFT calculations (ACE, one level of savings), and also drastically reducing the cost of evaluating each of the remaining calls (SCDM+\texttt{exx}, two levels of savings).
%
As mentioned above, \texttt{SeA} was inspired by the work of Gygi and co-workers,~\cite{gygi_efficient_2013,dawson_performance_2015} Giannozzi and co-workers,~\cite{carnimeo_fast_2018} and Nair and co-workers,~\cite{mandal_enhanced_2018,mandal_speeding-up_2019,mandal_achieving_2021} but differs from these pioneering works (which harnessed one or two levels of computational savings via pair selection and ACE) by harnessing three levels of savings (pair selection, domain truncation, and ACE) to further enable the routine use of hybrid DFT in high-throughput applications involving large-scale finite-gap systems.

%
%
\begin{figure*}
    \centering
    \includegraphics[width=0.6\linewidth]{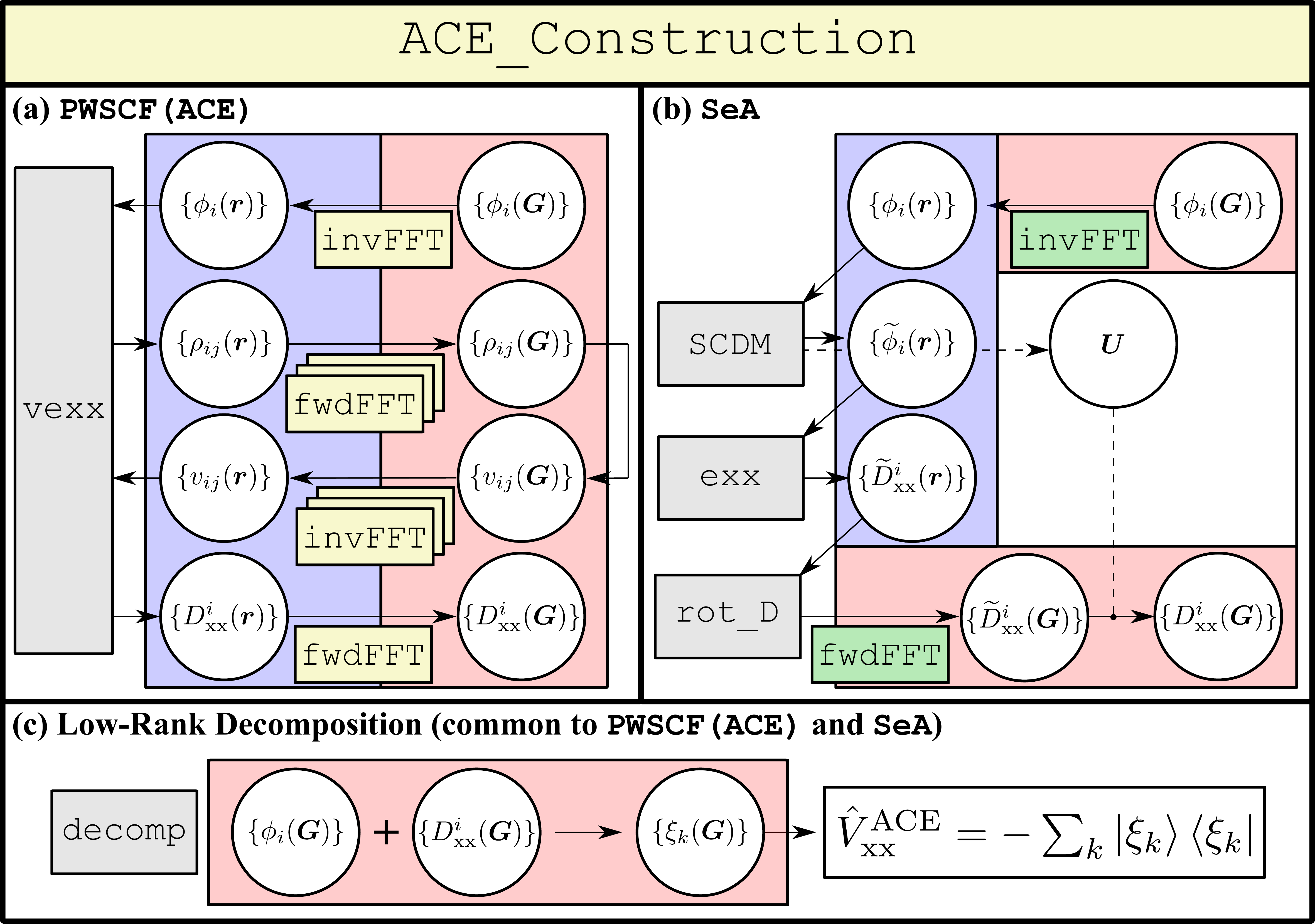}
    \caption{
    %
    Schematic illustration of the \texttt{ACE\_Construction} step in: (a) \texttt{PWSCF(ACE)}, the convolution-based ACE implementation in \texttt{PWSCF} and (b) \texttt{SeA}, the high-throughput hybrid DFT framework presented in this work.
    %
    Starting from the (proto-)KS orbitals (i.e., $\{ \phi_i (\bm G)\}$ in reciprocal space), both \texttt{PWSCF(ACE)} and \texttt{SeA} proceed to compute $D_{\rm xx}^i (\bm G) = -\hat{V}_{\rm xx} \phi_i (\bm G) \,\, \forall \,\, i$ (i.e., the action of the full-rank EXX operator needed during the iterative EXX-SCF procedure in hybrid DFT calculations, see Algorithm \ref{alg:ace}).
    %
    Then, both methods merge at: (c) a common low-rank decomposition function (\texttt{decomp}) that builds $\hat{V}_{\rm xx}^{\rm ACE}$ from $\{D_{\rm xx}^i(\bm G)\}$ and $\{\phi_i(\bm G)\}$.
    Pale blue (red) backgrounds indicate the real-space (reciprocal-space) aspects of each method.
    Single yellow and green boxes (with \texttt{fwdFFT} or \texttt{invFFT} enclosed) depict a linear ($\mathcal{O}(N_{\rm occ})$) number of FFT calls, while triple yellow boxes represent a quadratic ($\mathcal{O}(N_{\rm occ}^2)$) number of FFT calls.
    Each \texttt{fwdFFT}/\texttt{invFFT} call was performed at the \textit{appropriate} planewave resolution for each method---the density/potential FFT grid for \texttt{vexx} in \texttt{PWSCF(ACE)} ($N_{\rm FFT}$ points; yellow boxes) and the wavefunction FFT grid for \texttt{SeA} ($N_{\rm FFT}^{\rm wf} < N_{\rm FFT}$ points; green boxes).
    For more details, a pedagogical walk-through of the \texttt{ACE\_Construction} step is provided in Sec.~\ref{method:SeA}.
    }
    \label{fig:SeA_flow}
\end{figure*}
%
%
More specifically, \texttt{SeA} involves a key modification to the ACE-based EXX-SCF procedure, which already offers tremendous speedups during hybrid DFT calculations by replacing the repeated evaluation of $\{\hat{V}_{\rm xx} \left| \phi_{i}\right>\}$ in the conventional EXX-SCF procedure with a significantly more efficient evaluation of this term using $\hat{V}_{\rm xx}^{\rm ACE}$.
%
In the ACE-based EXX-SCF procedure depicted in Algorithm~\ref{alg:ace}, the construction of $\hat{V}_{\rm xx}^{\rm ACE}$ is the computational bottleneck (particularly for large-scale systems) and is accomplished via calls to the \texttt{ACE\_Construction} step during each outer-loop iteration.
To eliminate this step as the bottleneck during such large-scale hybrid DFT calculations, \texttt{SeA} directly attacks the cost of executing the \texttt{ACE\_Construction} step by using SCDM+\texttt{exx} to compute $\{ \hat{V}_{\rm xx} \left| \phi_{i} \right> \}$ (or $\{ \left| D_{\rm xx}^{i} \right> \}$).
To illustrate the combination and seamless integration of these methods in more detail, we will focus the remainder of this discussion on a $\Gamma$-point specific algorithm (i.e., as the target applications of \texttt{SeA} are high-throughput hybrid DFT calculations on large-scale systems); a more general extension of \texttt{SeA} based on Brillouin zone sampling will be discussed in future work.
Since $\left| D_{\rm xx}^{i} \right>$ is equivalent to $ \hat{V}_{\rm xx} \left| \phi_{i} \right>$ (to within a sign, cf. eq 12 and section IIB) and is naturally aligned with the \texttt{exx} framework,~\cite{paper1,paper2,paper3} we will set $\left| D_{\rm xx}^{i} \right>$ as the target quantity in the \texttt{ACE\_Construction} step.
As mentioned above in Sec.~\ref{method:ace}, we will also limit our discussion to the ACE procedure involving occupied (proto-)KS orbitals only; an extension of \texttt{SeA} to include virtual/unoccupied orbitals will also be discussed in future work.

To start, we first consider the \texttt{ACE\_Construction} step in the convolution-based ACE algorithm in the \texttt{PWSCF} module of \texttt{QE}, which will be referred to as \texttt{PWSCF(ACE)} throughout the remainder of this work. 
In \texttt{PWSCF(ACE)}, the \texttt{ACE\_Construction} step (as illustrated in Fig.~\ref{fig:SeA_flow}) includes a full-rank EXX evaluation to obtain $\{ D_{\rm xx}^i (\bm G) \}$ (via the \texttt{vexx} function, Fig.~\ref{fig:SeA_flow}(a)) followed by a low-rank decomposition to obtain $\hat{V}_{\rm xx}^{\rm ACE}$ (via the \texttt{decomp} function, Fig.~\ref{fig:SeA_flow}(c)).
During a $\Gamma$-point calculation, \texttt{vexx} first obtains the (proto-)KS orbitals in real space ($\{ \phi_i(\bm r) \}$) from their stored $N_{\rm pw}$ planewave coefficients ($\{ \phi_i(\bm G) \}$) via the inverse FFT (\texttt{invFFT}).
The complete set of orbital-product densities $\{\rho_{ij}(\bm r)\}$ is then constructed (as products over the $N_{\rm occ}(N_{\rm occ}+1)/2$ unique pairs of $\{\phi_i(\bm r)\}$) and converted to reciprocal space ($\{\rho_{ij}(\bm r)\} \longrightarrow \{\rho_{ij}(\bm G)\}$) using the forward FFT (\texttt{fwdFFT}).
Based on the convolution theorem, the corresponding orbital-product potentials are evaluated as $v_{ij}(\bm{G}) = 4 \pi \rho_{ij}(\bm{G})/\left|\bm{G}\right|^2$, which is again brought back to real space via the \texttt{invFFT} to form $\{ v_{ij}(\bm{r}) \}$.
Then, $\{ D_{\rm xx}^i (\bm r)\}$ is formed in real space via $D_{\rm xx}^i (\bm r) = \sum_{j} v_{ij}(\bm{r}) \phi_{j}(\bm{r})$ as described in Eq.~\eqref{eq:Dxx} and converted to the targeted $\{ D_{\rm xx}^i (\bm G)\}$ via the \texttt{fwdFFT}.
Hence, the cost of \texttt{vexx} is primarily governed by the two FFT operations involving $\{\rho_{ij}\}$ and $\{v_{ij}\}$ (each of which scales as $\mathcal{O}(N_{\rm FFT} \log N_{\rm FFT})$) for each of the $\mathcal{O}(N_{\rm occ}^2)$ unique pairs of occupied orbitals, which leads to an overall cubic-scaling algorithm (i.e., $\mathcal{O}( N_{\rm occ}^2 N_{\rm FFT} \log N_{\rm FFT})$) after neglecting the logarithmic term.
After obtaining the targeted $\{ D_{\rm xx}^i (\bm G) \}$, \texttt{PWSCF(ACE)} completes the \texttt{ACE\_Construction} step with a call to \texttt{decomp}, a function which constructs $\hat{V}_{\rm xx}^{\rm ACE}$ via Eq.~\eqref{eq:Vace} using the stored $\{ \phi_i(\bm G) \}$.
The procedure for constructing $\hat{V}_{\rm xx}^{\rm ACE}$ is schematically illustrated in Fig.~\ref{fig:SeA_flow}(c) (and covered in more detail in Sec.~\ref{method:ace}), and also requires cubic-scaling cost ($\mathcal{O}(N_{\rm pw} N_{\rm occ}^2 + N_{\rm occ}^3)$).
To ensure lossless \texttt{fwdFFT}/\texttt{invFFT} between $\rho_{ij}(\bm r) \longleftrightarrow \rho_{ij}(\bm G)$ and $v_{ij}(\bm r) \longleftrightarrow v_{ij}(\bm G)$ in \texttt{vexx} necessitates the use of the density/potential FFT grid (with $N_{\rm FFT}$ points).
Since $N_{\rm FFT} > N_{\rm pw}$, \texttt{vexx} generally dominates \texttt{decomp} and is almost always the computational bottleneck during the \texttt{ACE\_Construction} step in \texttt{PWSCF(ACE)}.

To greatly improve the computational efficiency of hybrid DFT calculations (particularly for large-scale systems), \texttt{SeA} directly attacks this computational bottleneck by replacing \texttt{vexx} with a procedure based on the combination of SCDM and \texttt{exx}.
As depicted in Fig.~\ref{fig:SeA_flow}(b), the \texttt{ACE\_Construction} step in \texttt{SeA} also starts from the stored $N_{\rm pw}$ planewave coefficients ($\{ \phi_i(\bm G) \}$), from which the $\{ \phi_i(\bm r) \}$ are obtained via \texttt{invFFT}.
However, this operation is performed at a planewave resolution that is sufficient to ensure lossless $\{\phi_i(\bm G)\} \xrightarrow[]{\texttt{invFFT}} \{\phi_i(\bm r)\}$ and $\{\widetilde{D}_{\rm xx}^i(\bm r)\} \xrightarrow[]{\texttt{fwdFFT}} \{\widetilde{D}_{\rm xx}^i(\bm G)\}$, namely, the wavefunction FFT grid with $N_{\rm FFT}^{\rm wf} < N_{\rm FFT}$ points.
As discussed in more detail below, this is the \textit{appropriate} planewave resolution for the \texttt{exx} function in Fig.~\ref{fig:SeA_flow}(b), since \texttt{SeA} directly forms $\{\widetilde{D}_{\rm xx}^i(\bm r)\}$ from $\{\phi_i(\bm r)\}$ in real space and therefore completely sidesteps the need for $\{ \rho_{ij}(\bm G) \}$ and $\{ v_{ij}(\bm G) \}$---the central quantities in \texttt{vexx} (cf. Fig.~\ref{fig:SeA_flow}(a)). 
With $N_{\rm FFT}^{\rm wf} < N_{\rm FFT}$ (i.e., $N_{\rm FFT}^{\rm wf} \approx N_{\rm FFT}/8$ when implemented with norm-conserving pseudopotentials), using the appropriate planewave resolution plays a significant role in increasing the computational efficiency (as well as decreasing the memory footprint, communication overhead, and processor idling) in \texttt{exx}; a more detailed discussion of this theoretical and algorithmic development will be provided in future work accompanying the public release of the \texttt{exxl} library.

With $\{ \phi_{i}(\bm r) \}$ on the wavefunction FFT grid, \texttt{SeA} applies the non-iterative SCDM algorithm (i.e., by calling the \texttt{SCDM} function) to obtain the localized orbitals in real space $\{\widetilde{\phi}_i(\bm r)\}$ and the corresponding unitary matrix $\bm U$ that allows one to transform between the local and canonical orbital representations (see Sec.~\ref{method:scdm}); with the use of the appropriate planewave resolution in \texttt{SeA}, execution of \texttt{SCDM} has a reduced cubic-scaling cost of $\mathcal{O}(N_{\rm FFT}^{\rm wf} N_{\rm occ}^2 + N_{\rm occ}^3)$.
With the localized SCDM orbitals in hand, \texttt{SeA} then calls the linear-scaling \texttt{exx} routine, which harnesses two levels of computational savings during the real-space evaluation of $\{\widetilde{D}_{\rm xx}^i(\bm r)\}$ via Eq.~\eqref{eq:tDxx} at $\mathcal{O}(N_{\rm occ})$ cost (see Sec.~\ref{method:exx}).
In doing so, \texttt{SeA} replaces the computationally dominant FFT operations involving $\{\rho_{ij}\}$ and $\{v_{ij}\}$ in \texttt{vexx} with a non-iterative cubic-scaling localization routine and a linear-scaling evaluation of $\{\widetilde{D}_{\rm xx}^i(\bm r)\}$.
Here, we remind the reader that the recently extended version~\cite{paper3} of \texttt{exx} included in \texttt{SeA} is based on a single system-independent orbital coverage threshold (instead of the set of system-dependent parameters used in previous \texttt{exx} versions~\cite{paper1,paper2}), which allows for a more straightforward (i.e., single-knob) tuning of the balance between accuracy and performance during this critical step in the \texttt{ACE\_Construction} step.
Then, \texttt{SeA} calls the \texttt{rot\_D} function, which first transforms $\{\widetilde{D}_{\rm xx}^i(\bm r)\}$ to $\{\widetilde{D}_{\rm xx}^i(\bm G)\}$ via \texttt{fwdFFT} with a quadratic-scaling cost of $\mathcal{O}(N_{\rm occ} N_{\rm FFT}^{\rm wf} \log N_{\rm FFT}^{\rm wf})$ after neglecting the logarithmic term; this step is the local representation analog of the $\{ D_{\rm xx}^i (\bm r)\} \xrightarrow{\texttt{fwdFFT}} \{ D_{\rm xx}^i (\bm G)\}$ transformation in \texttt{vexx} (albeit with reduced computational cost).
To form the targeted $\{D_{\rm xx}^i(\bm G)\}$, \texttt{rot\_D} then rotates $\{\widetilde{D}_{\rm xx}^i(\bm G)\}$ via:
\begin{align}
    D_{\rm xx}^{i}(\bm G) = \sum_{j} \widetilde{D}_{\rm xx}^{j}(\bm G) (\bm U^{-1})_{ji} ,
    \label{eq:rotD}
\end{align}
with a cubic-scaling associated cost of $\mathcal{O}(N_{\rm pw} N_{\rm occ}^2)$.
Here, we emphasize that the availability of the $\bm U$ matrix in \texttt{SeA} (provided by the earlier call to \texttt{SCDM}) facilitates this transformation, leveraging a strategy previously pointed out by Nair and co-workers.~\cite{mandal_enhanced_2018}  
While the order in which the \texttt{fwdFFT} and unitary rotation are applied leads to the same $\{D_{\rm xx}^i(\bm G)\}$, the order employed in \texttt{rot\_D} is the more computationally efficient of the two.
More specifically, the procedure in \texttt{rot\_D} ($\{ \widetilde{D}_{\rm xx}^i (\bm r)\} \xrightarrow{\texttt{fwdFFT}} \{\widetilde{D}_{\rm xx}^i (\bm G)\} \xrightarrow{U^{-1}} \{D_{\rm xx}^i (\bm G)\}$) has an associated computational cost of $\mathcal{O}(N_{\rm occ} N_{\rm FFT}^{\rm wf} \log N_{\rm FFT}^{\rm wf} + N_{\rm pw} N_{\rm occ}^2)$, while the alternative choice ($\{ \widetilde{D}_{\rm xx}^i (\bm r)\} \xrightarrow{U^{-1}} \{D_{\rm xx}^i (\bm r)\} \xrightarrow{\texttt{fwdFFT}} \{D_{\rm xx}^i (\bm G)\}$) scales as $\mathcal{O}(N_{\rm FFT}^{\rm wf} N_{\rm occ}^2 + N_{\rm occ} N_{\rm FFT}^{\rm wf} \log N_{\rm FFT}^{\rm wf})$.
As such, the \texttt{rot\_D} routine leverages the fact that $N_{\rm FFT}^{\rm wf} > N_{\rm pw}$ for lossless \texttt{fwdFFT} and \texttt{invFFT} to further improve the efficiency of \texttt{SeA}.

After obtaining the targeted $\{ D_{\rm xx}^i (\bm G) \}$, \texttt{SeA} merges with \texttt{PWSCF(ACE)} by calling the common \texttt{decomp} function described above (Fig.~\ref{fig:SeA_flow}(c)) to obtain $\hat{V}_{\rm xx}^{\rm ACE}$ and complete the \texttt{ACE\_Construction} step.
Here, we note that an approximate evaluation of $\{D_{\rm xx}^{i}\}$ can lead to a non-symmetric and/or non-positive-semi-definite $\bm M$ matrix (with elements $M_{ij} \equiv \left< \phi_i \left| D_{\rm xx}^{j} \right. \right>$), which will render the Cholesky decomposition procedure in \texttt{decomp} unstable (see Eq.~\eqref{eq:Vace} and surrounding discussion).
To address this potential stability issue, we have implemented an alternative/generalized low-rank decomposition scheme in the \texttt{decomp} function called by \texttt{SeA}; this eigensystem-based procedure is described in Appendix~\ref{app:es_comp_details_decomp} and comes also with a cubic-scaling cost of $\mathcal{O}(N_{\rm occ}^3 + N_{\rm occ}^2 N_{\rm pw})$.
In practice, this alternative decomposition scheme seems to be quite robust for \texttt{SeA} across a broad range of systems, and may also increase the robustness of other approximate ACE-based methods.~\cite{carnimeo_fast_2018,mandal_achieving_2021}

With $\hat{V}_{\rm xx}^{\rm ACE}$ in hand (at the end of a given outer-loop iteration), the current version of \texttt{SeA} computes $E_{\rm xx}$ via Eq.~\eqref{eq:exx_ace}.
Here, we note that Eq.~\eqref{eq:exxGen_mlwf} (i.e., the real-space energy expression in the \texttt{exx} approach) is not used in \texttt{SeA}, as the real-space evaluation of this quantity in a planewave code such as \texttt{QE} would be subject to small (but non-negligible) errors due to aliasing.
While \texttt{SeA} avoids such potential aliasing errors by using Eq.~\eqref{eq:exx_ace} to compute $E_{\rm xx}$, the approximate evaluation of $\{ D_{\rm xx}^{i} \}$ due to pair selection and domain truncation in \texttt{exx} can also lead to small numerical instabilities during the SCF procedure, e.g., minor oscillations in the total energy during the outer-loop iterations in \texttt{SeA} (see Algorithm~\ref{alg:ace}).
While pair selection in \texttt{exx} is a secondary source of error in $E_{\rm xx}$ by design (see Eq.~\eqref{eq:sij} and surrounding discussion in Sec.~\ref{method:exx}), pair selection is the primary cause of these numerical instabilities (which is consistent with the findings of Giannozzi and co-workers~\cite{carnimeo_fast_2018} for L-ACE).
In \texttt{SeA}, these oscillations are typically much smaller than the targeted accuracy in $E_{\rm xx}$, which is estimated \textit{a priori} for a given $\epsilon$ value.~\cite{paper3}
As such, we have constructed and implemented a convergence criterion based on successive idempotency in consecutive outer-loop iterations in \texttt{SeA}, wherein the EXX-SCF procedure converges after reaching this targeted level of accuracy (see Appendix~\ref{app:es_comp_details_conv}).

By constructing $\hat{V}_{\rm xx}^{\rm ACE}$ in the canonical orbital representation (via $\{ \phi_i(\bm G) \}$ and $\{ D_{\rm xx}^i (\bm G) \}$ in \texttt{decomp}), the current version of \texttt{SeA} sidesteps the need to work with localized orbitals during the inner-loop iterations in Algorithm~\ref{alg:ace}.
While the use of localized orbitals could provide additional computational savings during the inner-loop iterations, such savings would need to be balanced against the cost associated with multiple calls to \texttt{SCDM}; hence, this potential future research direction is beyond the scope of this work.
Since $\hat{V}_{\rm xx}^{\rm ACE}$ is independent of the underlying representation, it would also be possible to construct a fully real-space version of \texttt{SeA} (i.e., by eliminating the \texttt{fwdFFT} in \texttt{rot\_D} and using $\{\phi_i(\bm r)\}$ and $\{D_{\rm xx}^i(\bm r)\}$ to perform the low-rank decomposition in \texttt{decomp}).
While a na\"ive implementation of such an approach in a planewave code would be subject to aliasing errors (which can impede very tight SCF convergence, e.g., as needed for numerical phonon calculations), the incorporation of \texttt{SeA} into a real-space electronic structure code like \texttt{PARSEC}~\cite{kronik_parsec_2006} has the potential to enable hybrid DFT calculations across significantly larger length- and time-scales, and may therefore be a promising future research direction.

\section{Accuracy and Performance of \texttt{SeA} \label{sec:acc_prf}}

%
%
\begin{figure*}
    \centering
    \includegraphics[width=\linewidth]{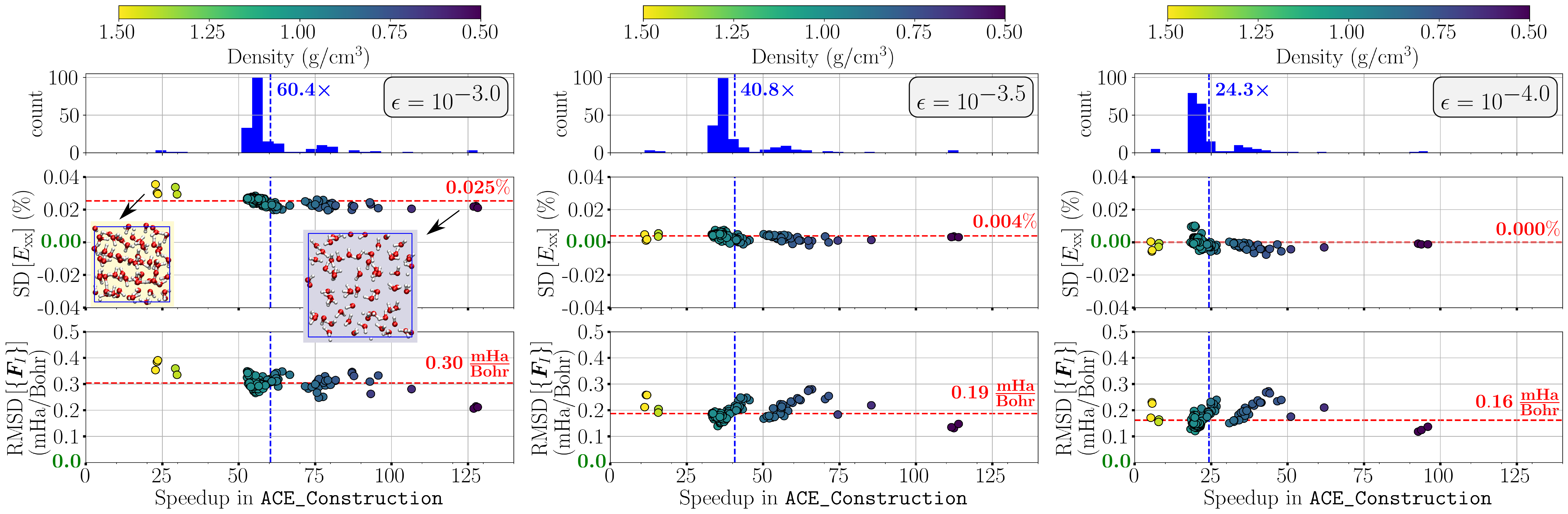}
    \caption{
    Accuracy and performance of \texttt{SeA} in the \texttt{PWSCF} module of \texttt{QE} (compared to the convolution-based \texttt{PWSCF(ACE)} implementation) when computing the EXX energy ($E_{\rm xx}$) and ionic forces ($\{\bm F_I\}$) at the PBE0 level for a diverse set of $200$ non-equilibrium \ce{(H2O)64} configurations (which include both intact and auto-ionized water molecules, with system densities $1.7$~g/cm$^3 > d > 0.4$~g/cm$^3$) using three typical $\epsilon$ settings: $\epsilon = 10^{-3.0}$ (\textit{left panel}), $\epsilon = 10^{-3.5}$ (\textit{middle panel}), and $\epsilon = 10^{-4.0}$ (\textit{right panel}).
    In each panel, the central (lower) scatter plot depicts the correlation between accuracy and performance when computing $E_{\rm xx}$ ($\{\bm F_I\}$) with \texttt{SeA}.
    In the central scatter plot, the accuracy in $E_{\rm xx}$ was quantified by the relative signed deviation: $\text{SD}[E_{\rm xx}] \equiv (E_{\rm xx}^{\texttt{SeA}}-E_{\rm xx}^{\texttt{PWSCF(ACE)}})/\left|E_{\rm xx}^{\texttt{PWSCF(ACE)}}\right|$ (red dashed line = mean $\text{SD}[E_{\rm xx}]$).
    In the lower scatter plot, the accuracy in $\{\bm F_I\}$ was quantified by the (component-wise) root-mean-square deviation: $\text{RMSD}[\{ \bm F_{I} \}] \equiv \sqrt{\sum_{I}\left|\bm F^{\texttt{SeA}}_{I} - \bm F^{\texttt{PWSCF(ACE)}}_{I} \right|^2/3N}$ wherein $N$ is the number of atoms (red dashed line = mean $\text{RMSD}[\{ \bm F_{I} \}]$).
    In both of these scatter plots, each point was colored according to the system density ($d$) and the performance of \texttt{SeA} was quantified by the speedup observed in the wall time cost of the \texttt{ACE\_Construction} step (i.e., the typical computational bottleneck in \texttt{PWSCF(ACE)}).
    In each panel, the top histogram depicts the distribution of speedups accomplished by \texttt{SeA} (blue dashed line = mean speedup) across these $200$ configurations.
    Each calculation was performed using a single Cori-Haswell node (with $16$ \texttt{MPI} processes and $4$ \texttt{OpenMP} threads per process).
    } 
    \label{fig:sea_acc_prf}
\end{figure*}
%
%

In this section, we assess the accuracy (Sec.~\ref{acc_prf:acc}) and performance (Sec.~\ref{acc_prf:prf}) of \texttt{SeA} by comparing this approach against the convolution-based ACE implementation in \texttt{PWSCF} (cf. the \texttt{SeA} and \texttt{PWSCF(ACE)} algorithms in Fig.~\ref{fig:SeA_flow}).
The computational details for all electronic structure calculations performed in this work can be found in Appendix~\ref{app:es_comp_details}.
To do so, we carried out a series of fully self-consistent PBE0~\cite{perdew_rationale_1996,adamo_toward_1999} calculations using \texttt{SeA} and \texttt{PWSCF(ACE)} for $200$ randomly selected \ce{(H2O)64} configurations (each with a cubic unit cell) that were collected during active learning~\cite{zhang_active_2019,zhang_dp-gen:_2020} of a deep neural network (DNN) potential for water at the SCAN meta-GGA level of theory,~\cite{sun_strongly_2015} which is known to perform quite well for aqueous systems.~\cite{sun_accurate_2016,chen_ab_2017,zheng_structural_2018,andreani_hydrogen_2020,calegari_andrade_structure_2018,calegari_andrade_free_2020,zhang_phase_2021}
This set contains a wide range of non-equilibrium configurations (including both intact and auto-ionized water molecules) with densities ranging from $\approx 0.4$~g/cm$^3$ to $\approx 1.7$~g/cm$^3$.
As such, this diverse set samples the configuration space of water across a wide range of system densities as well as sectors of configuration space that are relevant to auto-ionization (i.e., \ce{2H2O <=> H3O+ + OH-}).
While it is straightforward to apply \texttt{SeA} to significantly larger systems (\textit{vide infra}), our choice to consider \ce{(H2O)64} here allows for a systematic comparison with the more computationally demanding \texttt{PWSCF(ACE)} approach.

\subsection{Accuracy of \texttt{SeA} \label{acc_prf:acc}}

We begin by discussing the accuracy of \texttt{SeA} when computing $E_{\rm xx}$ for these $200$ non-equilibrium \ce{(H2O)64} configurations using typical settings for $\epsilon$ in \texttt{exx} (i.e., the system-independent orbital coverage threshold~\cite{paper3} in Eq.~\eqref{eq:oii}), namely: 
$10^{-3} \ge \epsilon \ge 10^{-4}$.
%
As depicted in Fig.~\ref{fig:sea_acc_prf} and Table~\ref{tab:acc_scalar_vector}, the use of successively smaller $\epsilon$ values consistently reduces the mean signed deviation (MSD) in $E_{\rm xx}$ between \texttt{SeA} and \texttt{PWSCF(ACE)} from $0.025\%$ ($\epsilon=10^{-3.0}$) to a visible plateau at $0.004\%$ ($\epsilon=10^{-3.5}$) and $0.000\%$ ($\epsilon=10^{-4.0}$).
While there is no formal variational principle in \texttt{SeA}, the $E_{\rm xx}$ values obtained for these aqueous systems tend to be variational, i.e., $E_{\rm xx}^{\texttt{SeA}} \ge E_{\rm xx}^{\texttt{PWSCF(ACE)}}$, for larger $\epsilon$ values ($\epsilon = 10^{-3.0}$).
Here, we would argue that the remaining (albeit small) systematic difference is largely due to pair selection (i.e., the exclusion of non-self pairs with small but finite $S_{|i,j|}$ values, cf.\ Eq.~\eqref{eq:sij}) and domain truncation (i.e., the use of system-size-independent domains $\Omega_{ii} \subset \Omega$ in Eq.~\eqref{eq:oii} to compute the $E_{\rm xx}$ contributions from both self- and non-self pairs).
To maximize transferability, the recently extended version~\cite{paper3} of \texttt{exx} incorporated into \texttt{SeA} ensures that the pair selection error (governed by $\delta$ in Eq.~\eqref{eq:sij}) is smaller than but comparable to the domain truncation error (governed by $\epsilon$ in Eq.~\eqref{eq:oii}).
For tighter $\epsilon$ settings ($\epsilon = 10^{-3.5}$ and $10^{-4.0}$), this residual difference is no longer systematic, which can be seen by comparing the MSD and mean absolute deviations (MAD) in Table~\ref{tab:acc_scalar_vector}.
Quite interestingly, we also note that the differences in $E_{\rm xx}$ are essentially constant across this diverse set of non-equilibrium aqueous configurations for all $\epsilon$ settings---this is reflected by the flat distributions in Fig.~\ref{fig:sea_acc_prf} as well as the nearly identical MAD and root-mean-square deviations (RMSD) in Table~\ref{tab:acc_scalar_vector}.
This provides strong evidence that the current \texttt{exx} approach~\cite{paper3} is quite robust and able to operate at an \textit{a priori} estimated level of accuracy (i.e., controllable error) set by $\epsilon$.
With $E_{\rm xx}$ deviations that are essentially flat across such a wide range of system densities, we also expect that \texttt{SeA} will be an accurate and reliable tool for screening materials, sampling PES, and generating high-quality data for ML applications (\textit{vide infra}).

As an additional assessment, we also considered the accuracy of the ionic forces (which are of central importance to MD simulations and ML data generation) obtained using \texttt{SeA} for this set of non-equilibrium aqueous configurations.
As depicted in Fig.~\ref{fig:sea_acc_prf} and Table~\ref{tab:acc_scalar_vector}, the use of successively smaller $\epsilon$ values again consistently reduces the RMSD in $\{\bm F_{I}\}$ between \texttt{SeA} and \texttt{PWSCF(ACE)} from $0.30$~mHa/Bohr ($\epsilon=10^{-3.0}$) to a visible plateau at $0.19$~mHa/Bohr ($\epsilon=10^{-3.5}$) and $0.16$~mHa/Bohr ($\epsilon=10^{-4.0}$).
For reference, these residual force differences are negligible when compared to the magnitude of $\{\bm F_{I}\}$ in these systems ($17.2 \pm 3.3$~mHa/Bohr, see Table~\ref{tab:acc_scalar_vector}) and small relative to the typical convergence criterion used during structural relaxations of condensed-phase systems ($\approx 1.0$~mHa/Bohr).
These small but observable residual differences between \texttt{SeA} ($\epsilon = 10^{-3.5}$ and $10^{-4.0}$) and \texttt{PWSCF(ACE)} are largely due to finite-size effects~\cite{nguyen_efficient_2009} in the evaluation of exact exchange in periodic systems,~\cite{chawla_exact_1998} which lead to errors in both of these approaches; the relative convergence of $E_{\rm xx}$ and $\{\bm F_{I}\}$ in \texttt{SeA} and \texttt{PWSCF(ACE)} as a function of system size will be discussed in future work.
With ionic force deviations that are essentially constant across such a wide range of system densities  for all $\epsilon$ settings, we expect that \texttt{SeA} will reproduce the structure and dynamics of large-scale finite-gap systems with high fidelity during MD simulations at the hybrid DFT level.
We also note in passing that the RMSD in $\{ \bm F_{I} \}$ between \texttt{SeA} and \texttt{PWSCF(ACE)} is an order-of-magnitude ($\approx 20\times$$-$$40\times$) smaller than the RMSD between PBE~\cite{perdew_generalized_1996} (GGA) and PBE0 (i.e., $6.63$~mHa/Bohr, see Table~\ref{tab:acc_scalar_vector}).
%
%
{
\renewcommand{\arraystretch}{1.2}
\begin{table*}[ht!]
    \caption{
    Accuracy of \texttt{SeA} in the \texttt{PWSCF} module of \texttt{QE} (compared to the convolution-based \texttt{PWSCF(ACE)} implementation) when computing scalar quantities (EXX energy ($E_{\rm xx}$), valence band width (VBW), and binding energy ($E_{\rm b}$)) and vector quantities (ionic forces ($\{\bm F_I\}$) and orbital eigenvalues ($\{\lambda_{i}\}$)) at the PBE0 level for a diverse set of $200$ non-equilibrium \ce{(H2O)64} configurations (which include both intact and auto-ionized water molecules, with system densities $1.7$~g/cm$^3 > d > 0.4$~g/cm$^3$) using three typical $\epsilon$ settings: $\epsilon = 10^{-3.0}$, $\epsilon = 10^{-3.5}$, and $\epsilon = 10^{-4.0}$.
    For each scalar quantity, we report the following deviation metrics computed from these $200$ configurations: mean signed deviation (MSD), mean absolute deviation (MAD), root-mean-square deviation (RMSD), and maximum absolute deviation (MAXD).
    For each vector quantity, we compute the component-wise MSD, MAD, RMSD, and MAXD \textit{for each configuration}, and report the corresponding population mean (denoted by $\braket{\cdots}_{200}$), i.e., the deviation metric averaged over the $200$ configurations.
    For a given scalar or vector quantity ($Q$), the above deviation metrics were computed on the difference between \texttt{SeA} and \texttt{PWSCF(ACE)}, i.e., $Q^{\texttt{SeA}} - Q^{\texttt{PWSCF(ACE)}}$; as such, a positive MSD corresponds to an overestimate of $E_{\rm xx}$, VBW, $E_{\rm b}$, or $\{ \lambda_i \}$ by \texttt{SeA}.
    As context for quantifying the reported deviations, quantity-specific measures for the range of PBE0 values obtained with \texttt{PWSCF(ACE)} for the $200$ aqueous configurations as well as the mean and standard deviation (in square brackets) are provided in the \texttt{PWSCF(ACE)} rows. 
    For an additional comparison, we also report property-specific deviation metrics between PBE and PBE0 (computed via $Q^{\rm PBE} - Q^{\texttt{PWSCF(ACE)}}$).
    }
    \label{tab:acc_scalar_vector}
    \begin{center}
    \begin{tabular}{ C{10mm} | C{19mm} | *{4}{C{11mm}} | *{4}{C{11mm}} | *{4}{C{11mm}} } 
    \hline\hline
    \multicolumn{14}{c}{{\textbf{\textit{Scalar Quantities}}}} \\
    \hline
    \multicolumn{2}{c|}{\textbf{Quantity} $\boldsymbol{\rightarrow}$}                    & \multicolumn{4}{c|}{\textbf{EXX Energy (\%)}} 
                       & \multicolumn{4}{c|}{\textbf{Valence Band Width (eV)}} 
                       & \multicolumn{4}{c}{\textbf{Binding Energy$^{a}$ (kcal/mol)}}  \\
    \hline
    Level & Method              & $\phantom{-}$MSD   & MAD     & RMSD    & MAXD    
                       & $\phantom{-}$MSD   & MAD     & RMSD    & MAXD
                       & $\phantom{-}$MSD   & MAD     & RMSD    & MAXD  \\
   \hline
   \multirow{3}{*}{PBE0} & \texttt{SeA} $\mathtt{(10^{-3.0})}$         & $0.025$            & $0.025$ & $0.025$ & $0.035$ 
                       & $\phantom{-}0.003$ & $0.011$ & $0.014$ & $0.035$
                        & $-0.28$            & $0.28$  & $0.28$  & $0.38$  \\
   & \texttt{SeA} $\mathtt{(10^{-3.5})}$         &           $0.004$  & $0.004$ & $0.004$ & $0.007$ 
                       & $\phantom{-}0.001$ & $0.005$ & $0.007$ & $0.020$
                       & $-0.12$            & $0.12$  & $0.12$  & $0.14$  \\
   & \texttt{SeA} $\mathtt{(10^{-4.0})}$         & $0.000$            & $0.002$ & $0.003$ & $0.010$ 
                       & $\phantom{-}0.000$ & $0.003$ & $0.004$ & $0.013$
                       & $-0.06$            & $0.06$  & $0.07$  & $0.08$  \\
   PBE            & \texttt{PWSCF}            & $\mathrm{-}$       & $\mathrm{-}$ & $\mathrm{-}$ & $\mathrm{-}$ 
                       & $-0.410$           & $0.410$ & $0.720$ & $1.390$
                       & $\phantom{-}0.54$  & $0.54$  & $0.55$  & $0.85$  \\
    \hline
  PBE0 & \texttt{PWSCF(ACE)} & \multicolumn{4}{c|}{$\mathrm{-}$} 
                       & \multicolumn{4}{c|}{$20.7 \le \text{VBW} \le 25.4 \;\: [ 21.9\pm 0.7 ]$}
                       & \multicolumn{4}{c}{$0.95 \le E_{\rm b} \le 10.91 \;\: [7.46 \pm 1.99]$}  \\
    \hline\hline
    \end{tabular}
    \end{center}
    \vspace{-1.95em}
    \begin{center}
    \begin{tabular}{ C{10mm} | C{19mm} | *{2}{C{15.45mm}} *{2}{C{19mm}} | *{2}{C{15.45mm}} *{2}{C{19mm}} } 
    \multicolumn{10}{c}{{\textbf{\textit{Vector Quantities}}}} \\
    \hline
    \multicolumn{2}{c|}{\textbf{Quantity} $\boldsymbol{\rightarrow}$}                  & \multicolumn{4}{c|}{\textbf{Ionic Forces (mHa/Bohr)}} & \multicolumn{4}{c}{\textbf{Orbital Eigenvalues (eV)}} \\
    \hline
   Level & Method          & $\braket{\text{MSD}}_{200}$ & $\braket{\text{MAD}}_{200}$ & $\braket{\text{RMSD}}_{200}$ & $\phantom{1}\braket{\text{MAXD}}_{200}$ & $\braket{\text{MSD}}_{200}$ & $\braket{\text{MAD}}_{200}$ & $\braket{\text{RMSD}}_{200}$ & $\braket{\text{MAXD}}_{200}$ \\
   \hline
   \multirow{3}{*}{PBE0} & \texttt{SeA} $\mathtt{(10^{-3.0})}$ & $\mathrm{-}^{b}$       & $0.25$  & $0.30$  & $\phantom{1}0.95$            & $0.014$& $0.015$& $0.018$ & $0.049$ \\
   & \texttt{SeA} $\mathtt{(10^{-3.5})}$         & $\mathrm{-}^{b}$       & $0.14$  & $0.19$  & $\phantom{1}0.61$            & $0.006$& $0.007$& $0.008$ & $0.023$ \\
   & \texttt{SeA} $\mathtt{(10^{-4.0})}$         & $\mathrm{-}^{b}$       & $0.10$  & $0.16$  & $\phantom{1}0.55$            & $0.003$& $0.003$& $0.004$ & $0.011$ \\
   PBE & \texttt{PWSCF}              & $\mathrm{-}^{b}$       & $5.70$  & $6.63$  &           $14.47$            & $\mathrm{-}^{d}$ & $0.521$& $0.604$ & $1.111$ \\
   \hline
   PBE0 & \texttt{PWSCF(ACE)} & \multicolumn{4}{c|}{$12.3 \le \text{RMSD}_{0}[\{ \bm{F}_I \}]^{c} \le 28.4 \quad [ 17.2 \pm 3.3 ]$}   & \multicolumn{4}{c}{$-6.2 \le \lambda_{\rm HOMO} \le -1.1 \quad [ -4.2 \pm 0.6 ]$}         \\
    \hline\hline
    \multicolumn{10}{l}{
    \parbox{0.99\linewidth}{\raggedright
    \sloppy\setlength\parfillskip{0pt}
    \footnotesize{\ \vspace{0.05em}\\
    \noindent \hspace{-0.2em}$^{a}$Binding energies were computed via $E_{\rm b} = E_{\ce{H2O(g)}} - \frac{1}{64}E_{\ce{(H2O)64}}$. \\
    \noindent \hspace{-0.2em}$^{b}$Removal of the net force acting on the center of mass ensures that $\text{MSD} = 0$ for each configuration; hence, $\braket{\text{MSD}}_{200}$ (the population mean) will vanish. \\
    \noindent \hspace{-0.2em}$^{c}$The magnitude of $\{\bm F_{I}\}$ in each configuration was quantified by $\text{RMSD}_0[\{\bm F_{I} \}] \equiv \sqrt{\sum_{I}\left|\bm F^{\texttt{PWSCF(ACE)}}_{I} \right|^2/3N}$, in which $N$ is the number of atoms. \\
    \noindent \hspace{-0.2em}$^{d}$To account for the vertical energy shift ($\approx 2.2$~eV) between the PBE and PBE0 levels of theory, the mean PBE and PBE0 eigenvalues were aligned for each configuration; hence, $\text{MSD} = 0$ for each configuration and $\braket{\text{MSD}}_{200}$ will vanish.
    }}
    }\\
    \end{tabular}
    \end{center}
\end{table*}
}
%
%

In Table~\ref{tab:acc_scalar_vector}, we also assessed the accuracy of \texttt{SeA} against \texttt{PWSCF(ACE)} when computing several other physical properties among this diverse set of aqueous configurations, namely: the valence band width (VBW), binding energy ($E_{\rm b}$), and orbital eigenvalues ($\{\lambda_{i}\}$).
As seen above for $E_{\rm xx}$ and $\{\bm F_{I}\}$, all of these properties rapidly converge as $\epsilon$ is reduced from $10^{-3.0}$ to $10^{-3.5}$ and $10^{-4.0}$.
For the VBW, the MAD were effectively negligible, and ranged from $3$~meV ($\epsilon = 10^{-4.0}$) to $11$~meV ($\epsilon = 10^{-3.0}$).
To put these results into context, such deviations are three orders of magnitude smaller than the typical VBW at the PBE0 level ($21.9 \pm 0.7$~eV) and one--two orders of magnitude smaller than the MAD between PBE and PBE0 ($0.410$~eV).
For the individual orbital eigenvalues, the $\braket{\text{MAD}}_{200}$ (i.e., the population mean corresponding to the $200$ MAD values) were also on the meV scale, and ranged from $3$~meV ($\epsilon = 10^{-4.0}$) to $15$~meV ($\epsilon = 10^{-3.0}$).
Similar to the VBW, these effectively negligible differences are two--three orders of magnitude smaller than the typical HOMO energy at the PBE0 level ($-4.2 \pm 0.6$~eV) and one--two orders of magnitude smaller than the MAD between PBE and PBE0 ($0.521$~eV).
When computing $E_{\rm b}$, the MAD between \texttt{SeA} and \texttt{PWSCF(ACE)} ranges from $0.06$~kcal/mol ($\epsilon=10^{-4}$) to $0.28$~kcal/mol ($\epsilon=10^{-3}$), and is quite flat across these $200$ snapshots.
While \texttt{SeA} already provides $E_{\rm b}$ with effectively sub-kJ/mol accuracy (i.e., well within ``chemical accuracy'' of $< 1$~kcal/mol), tighter $\epsilon$ values would be recommended when more stringent accuracy is needed (i.e., for very weakly bound systems).

\subsection{Performance of \texttt{SeA} \label{acc_prf:prf}}

\subsubsection{Attacking the Bottleneck of \texttt{PWSCF(ACE)} with \texttt{SeA} \label{acc_prf:prf_ace_cons}}

%
%
{
\renewcommand{\arraystretch}{1.3}
\begin{table*}
  \caption{
  Detailed wall time breakdown for EXX-SCF calculations at the PBE0 level using \texttt{PWSCF(ACE)} and the \texttt{SeA} ($\epsilon=10^{-3.5}$) implementation in the \texttt{PWSCF} module of \texttt{QE} for three selected \ce{(H2O)64} configurations (with system densities ranging from most dense to least dense) taken from the set of $200$ non-equilibrium aqueous configurations considered in Fig.~\ref{fig:sea_acc_prf}.
  Each calculation was performed using a single Cori-Haswell node (with $16$ \texttt{MPI} processes and $4$ \texttt{OpenMP} threads per process) starting from converged orbitals at the PBE level; the number of SCF iterations performed during these preliminary PBE calculations were excluded from the number of calls to \texttt{SCF\_Iteration}.
  }
  \label{tab:perf_breakdown_w064_ace_cons}
  \begin{tabular}{c *{3}{|c} *{3}{| C{13mm} | C{8mm} | C{13mm}}}
    \hline
    \hline
   \multicolumn{4}{c|}{\multirow{2}{*}{System Density $\rightarrow$}} & \multicolumn{3}{c|}{Most Dense}  & \multicolumn{3}{c|}{Ambient Density}  & \multicolumn{3}{c}{Least Dense}   \\
   \multicolumn{4}{c|}{}
   & \multicolumn{3}{c|}{($d=1.66$~g/cm$^3$)}
    &\multicolumn{3}{c|}{($d=0.99$~g/cm$^3$)}
    &\multicolumn{3}{c}{($d=0.42$~g/cm$^3$)} \\
    \hline
    \multirow{2}{*}{Level} & \multirow{2}{*}{Method} & \multicolumn{2}{c|}{\multirow{2}{*}{Function}} & Time & \multirow{2}{*}{Calls} & Relative 
    & Time & \multirow{2}{*}{Calls} & Relative
    & Time & \multirow{2}{*}{Calls} & Relative \\
    && \multicolumn{2}{c|}{\multirow{2}{*}{}} & (s/call) & & Cost 
    & (s/call) & & Cost
    & (s/call) & & Cost\\
\hline
\multirow{3}{*}{PBE0} &
    \multirow{3}{*}{\parbox{5.5em}{\texttt{PWSCF(ACE)}}}
    & \texttt{ACE\_Con-}        & \texttt{vexx}    & $227.5$ & $5$ & $96.8\%$
    & $478.3$ & $5$ & $97.5\%$
    & $1{,}798.0$ & $5$ & $97.9\%$ \\
    \cline{4-4}
   & &     \texttt{struction}    & \texttt{decomp}  & $\phantom{00}0.1$ & $5$ & $\phantom{0}0.1\%$
    & $\phantom{00}0.2$ & $5$ & $\phantom{0}0.0\%$
    & $\phantom{0,00}0.4$ & $5$ & $\phantom{0}0.0\%$ \\
    \cline{3-4}
  & & \multicolumn{2}{c|}{\texttt{SCF\_Iteration}} & $\phantom{00}2.9$ & $13$ & $\phantom{0}3.1\%$ 
    & $\phantom{00}4.7$ & $13$ & $\phantom{0}2.5\%$
    & $\phantom{0,0}13.8$ & $14$ & $\phantom{0}2.1\%$ \\
\hline
\multirow{6}{*}{PBE0} &
\multirow{6}{*}{\parbox{5.5em}{\texttt{SeA}\\ $\mathtt{(10^{-3.5})}$}}
    &                           & \texttt{SCDM}    & $\phantom{00}2.7$ & $5$ & $\phantom{0}9.7\%$ 
    & $\phantom{00}4.4$ & $5$ & $17.4\%$
    & $\phantom{0,0}10.4$ & $5$ & $19.2\%$ \\
    \cline{4-4}
   & &     \texttt{ACE\_Con-}    & \texttt{exx}     & $\phantom{0}17.2$ & $5$ & $62.3\%$
    & $\phantom{00}8.3$ & $5$ & $32.7\%$
    & $\phantom{0,00}4.1$ & $5$ & $\phantom{0}7.6\%$ \\
    \cline{4-4}
   & &    \texttt{struction}     & \texttt{rot\_D} & $\phantom{00}0.2$ & $5$ & $\phantom{0}0.6\%$
    & $\phantom{00}0.3$ & $5$ & $\phantom{0}1.1\%$
    & $\phantom{0,00}0.6$ & $5$ & $\phantom{0}1.1\%$ \\
    \cline{4-4}
   & &                          & \texttt{decomp} & $\phantom{00}0.1$ & $5$ & $\phantom{0}0.4\%$
   & $\phantom{00}0.2$ & $5$ & $\phantom{0}0.7\%$
   & $\phantom{0,00}0.4$ & $5$ & $\phantom{0}0.7\%$ \\
    \cline{3-4}
  & & \multicolumn{2}{c|}{\texttt{SCF\_Iteration}} & $\phantom{00}2.9$ & $13$ & $27.0\%$ 
    & $\phantom{00}4.7$ & $13$ & $48.1\%$
    & $\phantom{0,0}13.8$ & $14$ & $71.4\%$ \\
    \cline{3-13}
    & & \multicolumn{2}{c|}{Speedup$^{a}$} & \multicolumn{3}{c|}{$\mathbf{11.3}\boldsymbol{\times}$}  & \multicolumn{3}{c|}{$\mathbf{36.1}\boldsymbol{\times}$ }  & \multicolumn{3}{c}{$\mathbf{116.0}\boldsymbol{\times}$} \\
    \hline
    \hline
    \multicolumn{13}{l}{
    \parbox{0.9\linewidth}{
    \raggedright\sloppy\setlength\parfillskip{0pt}
    \footnotesize{\ \vspace{0.05em}\\
    \noindent \hspace{-0.2em}$^{a}$Following the convention used in Fig.~\ref{fig:sea_acc_prf}, the speedup was computed as the ratio of the total wall time spent in the \texttt{ACE\_Construction} step in \texttt{PWSCF(ACE)} vs.\ \texttt{SeA}.
    }}
    }\\
  \end{tabular}
\end{table*}
}
%
%

Having assessed the accuracy of \texttt{SeA} against \texttt{PWSCF(ACE)} for a number of different properties across a diverse set of $200$ non-equilibrium aqueous configurations, we now provide a preliminary analysis of the computational performance of the \texttt{SeA} implementation in the \texttt{PWSCF} module of \texttt{QE}.
We will focus our discussion on the wall time associated with the \texttt{ACE\_Construction} step (see Sec.~\ref{method:SeA} and Fig.~\ref{fig:SeA_flow}), i.e., the typical computational bottleneck during \texttt{PWSCF(ACE)} calculations of systems with sizes similar to \ce{(H2O)64} (\textit{vide infra}).
%
As shown in Fig.~\ref{fig:sea_acc_prf}, \texttt{SeA} provides an order-of-magnitude speedup of $\approx 60.4\times$ ($\epsilon=10^{-3.0}$), $\approx 40.8\times$ ($\epsilon=10^{-3.5}$), and $\approx 24.3\times$ ($\epsilon=10^{-4.0}$) on average in the \texttt{ACE\_Construction} step for these $200$ aqueous configurations.
%
%
As expected, the observed speedup decreases when performing higher-accuracy \texttt{SeA} calculations with tighter $\epsilon$ values.
As seen in Eq.~\eqref{eq:oii}, the use of tighter $\epsilon$ values corresponds to less aggressive domain truncation when determining the orbital-specific local domain ($\Omega_{ii} \subset \Omega$) that encompasses each $\widetilde{\phi}_{i}$. 
Hence, larger $\Omega_{ii}$ and $\Omega_{ij} = \Omega_{ii} \cap \Omega_{jj}$ domains will be used when computing the self ($\braket{ii}$) and non-self ($\braket{ij}$) contributions to $\exx$ with tighter $\epsilon$ values.
Tighter $\epsilon$ values also correspond to less aggressive pair selection; hence, the number of $\braket{ij}$ pairs will also be larger (see Eq.~\eqref{eq:sij} and surrounding discussion), which further increases the computational cost.

In general, the speedups observed in Fig.~\ref{fig:sea_acc_prf} are inversely correlated with the system density~\cite{footnote_sea_performance_density} for all $\epsilon$ settings; for simplicity, we will focus our discussion on the intermediate $\epsilon=10^{-3.5}$ setting (Fig.~\ref{fig:sea_acc_prf} middle panel) throughout the remainder of the manuscript.
While the mean speedup in the \texttt{PWSCF(ACE)} computational bottleneck is $\approx 40.8\times$ using this setting, we observed a significantly larger (i.e., a two order-of-magnitude) speedup of $\approx 116 \times$ for the least dense configuration ($d=0.42$~g/cm$^3$).
Since less dense systems will have a smaller number of overlapping $\braket{ij}$ pairs as well as smaller $\Omega_{ij}$ domains, this result is not surprising and provides further evidence that the recently extended version~\cite{paper3} of \texttt{exx} incorporated into \texttt{SeA} is able to automatically exploit the increased degree of sparsity present in such systems.
Even for the most dense configuration considered herein (which has significantly less sparsity and a system density of $d=1.66$~g/cm$^3$), \texttt{SeA} still provides an order-of-magnitude speedup ($\approx 11 \times$) in the most computationally demanding \texttt{ACE\_Construction} step.
By attacking the bottleneck of \texttt{PWSCF(ACE)}, \texttt{SeA} is therefore able to furnish highly accurate energies, ionic forces, and other physical properties at the hybrid DFT level at a significantly reduced computational cost for systems with sizes similar to (and beyond that of) \ce{(H2O)64}.

\subsubsection{Detailed Computational Analysis of \texttt{SeA} \label{acc_prf:prf_ace_cons_more_detail}}

%
%
{
\renewcommand{\arraystretch}{1.3}
\begin{table*}
  \caption{
  Overall time-to-solution for SCF calculations at the PBE and PBE0 levels for three selected \ce{(H2O)64} configurations (with system densities ranging from most dense to least dense) taken from the set of $200$ non-equilibrium aqueous configurations considered in Fig.~\ref{fig:sea_acc_prf}.
  Each calculation was performed using a single Cori-Haswell node (with $16$ \texttt{MPI} processes and $4$ \texttt{OpenMP} threads per process) with corresponding wall times for each method ($t_{\rm M}$) reported in seconds.
  PBE0 calculations were performed using the \texttt{SeA} ($\epsilon=10^{-3.5}$), \texttt{PWSCF(ACE)}, and \texttt{PWSCF(Full)} methods starting from converged orbitals at the PBE level; in all three cases, the reported $t_{\rm M}$ values include the initial wall time needed to obtain the PBE orbitals.
  To enable a straightforward comparison between methods, we also report the relative speedup of PBE ($t_{\rm M}/t_{\rm PBE}$) and \texttt{SeA} ($t_{\rm M}/t_{\texttt{SeA}}$) with respect to each method M.
 }
 \label{tab:perf_breakdown_w064_time_to_solution}
  \begin{tabular}{ c | c *{9}{| C{14mm}} } 
   \hline
   \hline
   \multicolumn{2}{c|}{\multirow{2}{*}{System Density $\rightarrow$}} & \multicolumn{3}{c|}{Most Dense}& \multicolumn{3}{c|}{Ambient Density}& \multicolumn{3}{c}{Least Dense}\\
   \multicolumn{2}{c|}{} & \multicolumn{3}{c|}{($d=1.66$~g/cm$^3$)} &\multicolumn{3}{c|}{($d=0.99$~g/cm$^3$)} &\multicolumn{3}{c}{($d=0.42$~g/cm$^3$)}\\
    \hline
   Level & Method (M) & $t_{\rm M}$\,(s) & $t_{\rm M}/t_{\rm PBE}$ & $t_{\rm M} / t_{\texttt{SeA}}$ 
  &  $t_{\rm M}$\,(s) & $t_{\rm M} / t_{\rm PBE}$ & $t_{\rm M} / t_{\texttt{SeA}}$ 
  &  $t_{\rm M}$\,(s) & $t_{\rm M} / t_{\rm PBE}$ & $t_{\rm M} / t_{\texttt{SeA}}$
  \\
    \hline
    PBE & \texttt{PWSCF}                     & $\phantom{00{,}0}16$ & $\phantom{00}1.0$ & $\phantom{0}0.1$
      & $\phantom{00{,}0}23$  & $\phantom{0{,}00}1.0$  & $\phantom{00}0.2$
      & $\phantom{00{,}0}89$  & $\phantom{0{,}00}1.0$  & $\phantom{00}0.3$
      \\
      \hline
  \multirow{3}{*}{PBE0} & \texttt{SeA} $\mathtt{(10^{-3.5})}$  & $\phantom{00{,}}154$ & $\phantom{00}9.6$ & $\phantom{0}1.0$
      & $\phantom{00{,}}151$  & $\phantom{0{,}00}6.6$  & $\phantom{00}1.0$
      & $\phantom{00{,}}360$  & $\phantom{0{,}00}4.0$  & $\phantom{00}1.0$
      \\
    & \texttt{PWSCF(ACE)}     & $\phantom{0}1{,}191$ & $\phantom{0}74.4$  & $\mathbf{\phantom{0}7.7}$
      & $\phantom{0}2{,}477$  & $\phantom{0{,}}107.7$ & $\mathbf{\phantom{0}16.4}$
      & $\phantom{0}9{,}274$  & $\phantom{0{,}}104.2$ & $\mathbf{\phantom{0}25.8}$
      \\
    & \texttt{PWSCF(Full)} & $12{,}012$           & $750.8$ & $\mathbf{78.0}$
      & $25{,}027$            & $1{,}088.1$ & $\mathbf{165.7}$
      & $89{,}050$            & $1{,}000.6$ & $\mathbf{247.4}$
      \\
    \hline
    \hline
  \end{tabular}
\end{table*}
}
%
%

Having assessed the accuracy and preliminary performance of \texttt{SeA} against \texttt{PWSCF(ACE)} on the $200$ non-equilibrium aqueous configurations in Fig.~\ref{fig:sea_acc_prf}, we now provide a more detailed analysis of the computational cost associated with each procedural component in these two approaches (i.e., each function call in Algorithm~\ref{alg:ace} and Fig.~\ref{fig:SeA_flow}).
To do so, we consider the wall time associated with EXX-SCF calculations at the PBE0 level with \texttt{SeA} and \texttt{PWSCF(ACE)} for three \ce{(H2O)64} configurations of varying system densities taken from this set: the most dense configuration ($d=1.66$~g/cm$^3$), a representative configuration with ambient density ($d=0.99$~g/cm$^3$), and the least dense configuration ($d=0.42$~g/cm$^3$).
Each EXX-SCF calculation was started from converged PBE orbitals that were stored on disk; the wall times associated with generating these initial orbitals were excluded from the current analysis and will be included when discussing the overall time-to-solution in Sec.~\ref{acc_prf:prf_tts}.
As reported in Table~\ref{tab:perf_breakdown_w064_ace_cons}, each PBE0 calculation required $5$ \texttt{ACE\_Construction} (outer-loop) steps and $13$$-$$14$ \texttt{SCF\_Iteration} (inner-loop) steps (cf. Algorithm~\ref{alg:ace}).
When using \texttt{PWSCF(ACE)} to perform these PBE0 calculations, a similar picture is observed for all three \ce{(H2O)64} configurations: the overall wall time in each case was dominated by the \texttt{ACE\_Construction} step ($96.9\%$$-$$97.9\%$ of the overall wall time) with the (negligible) remaining time spent in \texttt{SCF\_Iteration}.
During the \texttt{ACE\_Construction} step in \texttt{PWSCF(ACE)}---the computational bottleneck for systems with sizes similar to \ce{(H2O)64}---calculation of the targeted $\{D_{\rm xx}^{i} (\bm G)\}$ via the \texttt{vexx} function (Fig.~\ref{fig:SeA_flow}(a)) accounted for the lion's share of the cost ($96.8\%$$-$$97.9\%$ of the overall wall time) while the construction of $\hat{V}_{\rm xx}^{\rm ACE}$ via \texttt{decomp} (Fig.~\ref{fig:SeA_flow}(c)) was largely negligible ($\lesssim 0.1\%$).

As mentioned above, \texttt{SeA} harnesses computational savings from domain truncation and pair selection to provide a one--two order-of-magnitude speedup ($11.3\times$$-$$116.0\times$) in the bottleneck \texttt{ACE\_Construction} step for these three \ce{(H2O)64} configurations (see Fig.~\ref{fig:sea_acc_prf} and Table~\ref{tab:perf_breakdown_w064_ace_cons}).
Interestingly, this speedup is significant enough to \textit{displace} the computational bottleneck in ACE-based PBE0 calculations from the \texttt{ACE\_Construction} step to the \texttt{SCF\_Iteration} step.
This can be seen by considering the \texttt{ACE\_Construction}$\,{:}\,$\texttt{SCF\_Iteration} relative cost ratio (obtained from the data in Table~\ref{tab:perf_breakdown_w064_ace_cons}), which is: $73.0\%\,{:}\,27.0\%$ for the most dense case, $51.9\%\,{:}\,48.1\%$ for the ambient case, and $28.6\%\,{:}\,71.4\%$ for the least dense case.
In the least dense case, the increased cost of the inner-loop \texttt{SCF\_Iteration} step in \texttt{SeA} is largely due to the increased number of planewaves $N_{\rm pw}$ and grid points $N_{\rm FFT}$ as the simulation cell increases. 
Hence, the crossover from \texttt{ACE\_Construction} to \texttt{SCF\_Iteration} roughly occurs for \ce{(H2O)64} configurations with the ambient system density when performing hybrid DFT calculations using \texttt{SeA}.
This observation emphasizes the need for scalable and efficient GGA-based KS-DFT algorithms to reduce the cost of the \texttt{SCF\_Iteration} step in \texttt{SeA} and further extend the range of applicability of hybrid DFT. 

Within the \texttt{ACE\_Construction} step in \texttt{SeA}, the \texttt{SCDM} and \texttt{exx} functions account for most of the cost: $9.7\%+62.3\% = 72.0\%$ (out of $73.0\%$) for the most dense case, $17.4\%+32.7\% = 50.1\%$ (out of $51.9\%$) for the ambient case, and $19.2\%+7.6\% = 26.8\%$ (out of $28.6\%$) for the least dense case.
From this data, one can see that the wall times associated with these two functions have an opposite dependence on the system density, which leads to non-monotonic changes in the \texttt{ACE\_Construction} wall time: $20.2$~s/call for the most dense case, $13.2$~s/call for the ambient case, and $15.5$~s/call for the least dense case.
On the one hand, the wall time spent in \texttt{exx} decreases monotonically with decreasing density from $17.2$~s/call (most dense case) to $8.3$~s/call (ambient density) to $4.1$~s/call (least dense case), due to the increasing degree of sparsity as $d$ decreases (see Sec.~\ref{acc_prf:prf_ace_cons}).~\cite{footnote_sea_performance_density}
On the other hand, the $\mathcal{O}(N_{\rm FFT}^{\rm wf} N_{\rm occ}^2 + N_{\rm occ}^3)$ cost of \texttt{SCDM} in \texttt{SeA} (see Sec.~\ref{method:SeA}) increases monotonically with decreasing density from $2.7$~s/call (most dense case) to $4.4$~s/call (ambient density) to $10.4$~s/call (least dense case), due to the increasing number of real-space grid points ($N_{\rm FFT}^{\rm wf}$) as $d$ decreases.
While the cost of \texttt{SCDM} is comparable to \texttt{exx} for these \ce{(H2O)64} systems, the \texttt{SCDM} cost will grow more quickly (cubically) with system size than \texttt{exx} (linear); hence, a more efficient \texttt{SCDM} localization algorithm will become crucial when applying \texttt{SeA} to larger systems and is currently under active development in our group.
While the costs associated with \texttt{rot\_D} and \texttt{decomp} remain small compared to \texttt{exx} for \ce{(H2O)64}, both of these routines also scale cubically with system size (but with markedly smaller prefactors than \texttt{SCDM}) and will therefore become more important for substantially larger systems.

\subsubsection{High-Throughput Time-to-Solution with \texttt{SeA} \label{acc_prf:prf_tts}}

To showcase the throughput of \texttt{SeA}, we now compare the overall time-to-solution when performing EXX-SCF calculations at the PBE0 level using \texttt{SeA}, \texttt{PWSCF(ACE)}, and \texttt{PWSCF(Full)} (i.e., the conventional convolution-based (non-ACE) EXX approach) on the three \ce{(H2O)64} configurations of varying system densities discussed above in Sec.~\ref{acc_prf:prf_ace_cons_more_detail}.
Each of these PBE0 calculations was started from converged PBE orbitals stored on disk; the wall times associated with generating these initial orbitals are reported in Table~\ref{tab:perf_breakdown_w064_time_to_solution} and are also included in the overall time-to-solution for each of these methods.
As depicted in Table~\ref{tab:perf_breakdown_w064_time_to_solution}, \texttt{SeA} can complete each of these PBE0 calculations within a few minutes on a \textit{single} Cori-Haswell node: $154$~s (most dense case), $151$~s (ambient density), and $360$~s (least dense case)---timings which are only $4.0\times$$-$$9.6\times$ the cost of the analogous PBE calculation.
In contrast, \texttt{PWSCF(ACE)} takes approximately $0.3$$-$$2.6$~hours for the same set of calculations, while the conventional convolution-based (non-ACE) EXX implementation (\texttt{PWSCF(Full)}) spent a little more than a day ($89{,}050$~s) on the least dense \ce{(H2O)64} configuration.
Hence, \texttt{SeA} provides an order-of-magnitude speedup in the overall time-to-solution ($7.7\times$$-$$25.8\times$) compared to \texttt{PWSCF(ACE)} and a one--two order-of-magnitude speedup ($78.0\times$$-$$247.4\times$) when compared to \texttt{PWSCF(Full)}.

We also compared the performance of the \texttt{SeA} implementation in the \texttt{PWSCF} module of \texttt{QE} against the original \texttt{exx} implementation~\cite{paper1,paper2} in the \texttt{CP} module as well as the L-ACE approach~\cite{carnimeo_fast_2018} in the \texttt{PWSCF} module when performing the same PBE0 calculations.
When comparing against the MLWF-based \texttt{exx} implementation in \texttt{CP} (which harnesses only two levels of computational savings), we used the computational settings described in Ref.~\cite{paper1}.
In this case, we had to use six Cori-Haswell nodes due to memory requirements (which have been significantly reduced by using the appropriate planewave resolution in \texttt{SeA}, see Sec.~\ref{method:SeA}), and found that \texttt{SeA} provides an order-of-magnitude speedup ($\approx 20\times$$-$$40\times$) in the overall time-to-solution. 
%
We attribute this large speedup to: (\textit{i}) the extended version of \texttt{exx} used in \texttt{SeA} (which has been updated/optimized since Refs.~\onlinecite{paper1,paper2}), (\textit{ii}) the third level of computational savings provided by the ACE operator in \texttt{SeA}, and (\textit{iii}) the more efficient SCF mixing algorithm in \texttt{PWSCF} (vs.\ the damped dynamics scheme in \texttt{CP}).
When comparing against L-ACE, all calculations were performed within a single Cori-Haswell node; here, we followed Ref.~\cite{carnimeo_fast_2018} and used the $S_{\rm thr} = 0.004$ localization threshold (i.e., L-ACE(0.004)).
In doing so, we observed that \texttt{SeA} provides an $\approx 5$$-$$9\times$ speedup in the overall time-to-solution compared to L-ACE(0.004).
Although the following setting was not recommended by the developers of L-ACE, we also considered $S_{\rm thr}=0.007$$-$$0.010$ (which provides comparable accuracy to \texttt{SeA} with $\epsilon = 10^{-3.5}$) as another point for comparison. 
Here, we found that \texttt{SeA} provides an $\approx 4$$-$$7\times$ speedup, which is quite similar to that found when using L-ACE(0.004).
These speedups mostly originate from the additional level of computational savings (domain truncation) harnessed by \texttt{SeA}; as such, even greater speedups are expected for larger and/or more sparse systems.

\section{Proof-of-Principle High-Throughput Application: Training a Hybrid DFT Based DNN Model for Liquid Water with \texttt{SeA} \label{sec:app}}

As a proof-of-principle high-throughput application, we now demonstrate how \texttt{SeA} can be used to efficiently train a converged deep neural network (DNN) potential for ambient ($T = 300$~K, $p = 1$~Bar) liquid water at the PBE0~\cite{perdew_rationale_1996,adamo_toward_1999} level via the deep potential molecular dynamics (DPMD) active-learning protocol of Car, E, and co-workers~\cite{zhang_deep_2018,zhang_end--end_2018} (Sec.~\ref{app:liq_water}).
We then assess the accuracy of this \texttt{SeA}-trained DNN potential on an out-of-sample test set containing several elevated-temperature \ce{(H2O)512} snapshots in Sec.~\ref{app:water512}, and showcase the capabilities of \texttt{SeA} by directly computing the (ground-truth) ionic forces in these challenging systems containing $>1{,}500$ atoms.

\subsection{Convergence and Precision of the Model \label{app:liq_water}}

Having assessed the accuracy and performance of \texttt{SeA} on a diverse set of $200$ non-equilibrium \ce{(H2O)64} configurations (which include both intact and auto-ionized water molecules, and densities spanning $0.4$~g/cm$^3\mathrm{-}1.7$~g/cm$^3$) collected during the active learning of a DNN potential for water at the meta-GGA (SCAN) level,~\cite{zhang_active_2019,zhang_dp-gen:_2020} we now proceed to label (i.e., compute energies and ionic forces for) all $8{,}610$ configurations in this collection at the hybrid (PBE0) DFT level using \texttt{SeA} ($\epsilon = 10^{-3.5}$); see Appendix~\ref{app:es_comp_details} for computational details regarding the electronic structure calculations.
Based on this initial data set, we then trained a DNN for ambient liquid water ($T = 300$~K, $p = 1$~Bar) following the DPMD active-learning procedure outlined in Refs.~\cite{zhang_deep_2018,zhang_end--end_2018} (see Appendix~\ref{app:train_comp_detail} for training details).
Hence, the progress of our active-learning protocol was monitored by the model deviation ($\mathcal{E}$) in the ionic forces $\{ \bm F_I \}$:
\begin{align}
  \mathcal{E} \equiv \max_I \sqrt{\braket{ \left| \bm F_I - \braket{\bm F_I} \right|^2}} ,
  \label{eq:modeldev}
\end{align}
in which the $\max$ function was taken over all ions $I$, and the averages (denoted by $\braket{\cdots}$) were evaluated over an ensemble of four independently trained DNN models (i.e., trained using the same data but different initial random seeds) for each snapshot in a given trajectory.
From this expression, one can see that $\mathcal{E}$ is a precision-based criterion that is measured among equivalently trained DNN models; an accuracy-based assessment, in which the ionic forces from the final DNN model(s) are directly compared to the ground-truth PBE0 ionic forces, will be provided in Sec.~\ref{app:water512}.
For the trajectory needed in Eq.~\eqref{eq:modeldev}, we performed a DPMD simulation (i.e., a classical MD simulation propagated using one of the trained DNN models as the underlying force field) in the $NpT$ ensemble ($T = 300$~K, $p = 1$~Bar); see Appendix~\ref{app:train_comp_detail} for more details.
When configurations with large model deviations (i.e., $\mathcal{E} > 2.0$~mHa/Bohr) were encountered, these configurations were relabeled with \texttt{SeA} and added to the training set for the next round of active learning (i.e., DPMD exploration and configuration selection/relabeling).
The active-learning process was stopped when the $90$-th percentile of the $\mathcal{E}$ probability density function ($P(\mathcal{E})$) was below $1.0$~mHa/Bohr, a typical convergence criterion used during structural relaxation in condensed-phase systems.

\begin{figure}
    \centering
    \includegraphics[width=\linewidth]{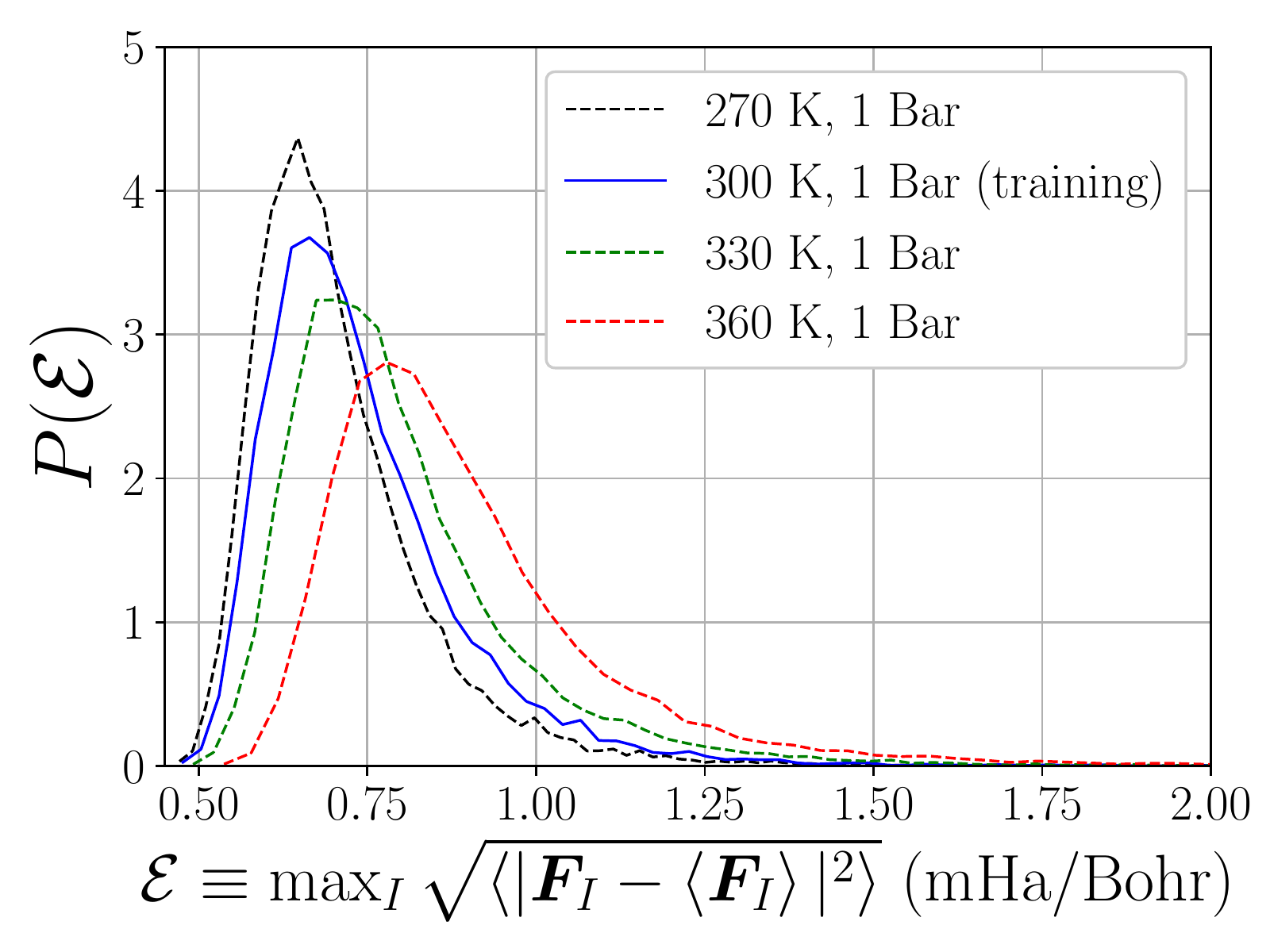}
    \caption{
    Normalized probability density functions of the model deviation ($P(\mathcal{E})$) in the ionic forces ($\{\bm{F}_{I}\}$) across four independently trained DNN models of liquid water at $T=300$~K and $p = 1$~Bar (solid blue line, training thermodynamic conditions); training protocol = active learning of PBE0 ionic forces computed with \texttt{SeA} ($\epsilon = 10^{-3.5}$) in $8{,}705$ \ce{(H2O)64} configurations.
    Also depicted are $P(\mathcal{E})$ for the following out-of-sample thermodynamic conditions: $T \in \{ 270~\text{K}, \, 330~\text{K}, \, 360~\text{K} \}$ and $p = 1$~Bar (dashed lines).
    In each case, $\mathcal{E}$ was evaluated over a DPMD trajectory of \ce{(H2O)64} at the corresponding thermodynamic conditions; all averages were evaluated over the ensemble of four DNN models and denoted by $\braket{\cdots}$.
    }
    \label{fig:app_model_dev}
\end{figure}

Doing so leads to a total of two active-learning iterations across a total of $8{,}705$ \ce{(H2O)64} configurations (i.e., yielding $95$ additional configurations not contained in the initial training set).
As shown in Fig.~\ref{fig:app_model_dev}, the final $P(\mathcal{E})$ at $T = 300$~K and $p = 1$~Bar (i.e., the thermodynamic conditions used during training) meets the convergence criterion outlined above and therefore depicts a converged DNN potential for ambient liquid water at the PBE0 level.
As an additional assessment of the precision in our \texttt{SeA}-trained DNN model, we also performed DPMD simulations (using the DNN potential trained at $T = 300$~K and $1$~Bar) in the $NpT$ ensemble at $270$~K, $330$~K, and $360$~K.
As expected, we find that $\mathcal{E}$ tends to increase with $T$ in Fig.~\ref{fig:app_model_dev}, as the additional thermal energy causes the system to explore larger sectors of configuration space which were sampled less thoroughly by the active-learning process at $300$~K.
However, the trained DNN model still remains quite precise with relatively smooth/continuous $P(\mathcal{E})$ distributions---the bulk of which are still below the desired $1.0$~mHa/Bohr threshold---even at such elevated temperatures.
While $P(\mathcal{E})$ can be further improved by performing additional active-learning steps, this is beyond the scope of this proof-of-principle high-throughput application, which illustrates the utility of \texttt{SeA} for training DNN potentials for large-scale finite-gap systems at the hybrid DFT level.
%
%
%
%

\subsection{Accuracy of the Model: Out-of-Sample Testing and a Beyond-One-Thousand-Atom Challenge \label{app:water512}}

\begin{figure}
    \centering
    \includegraphics[width=\linewidth]{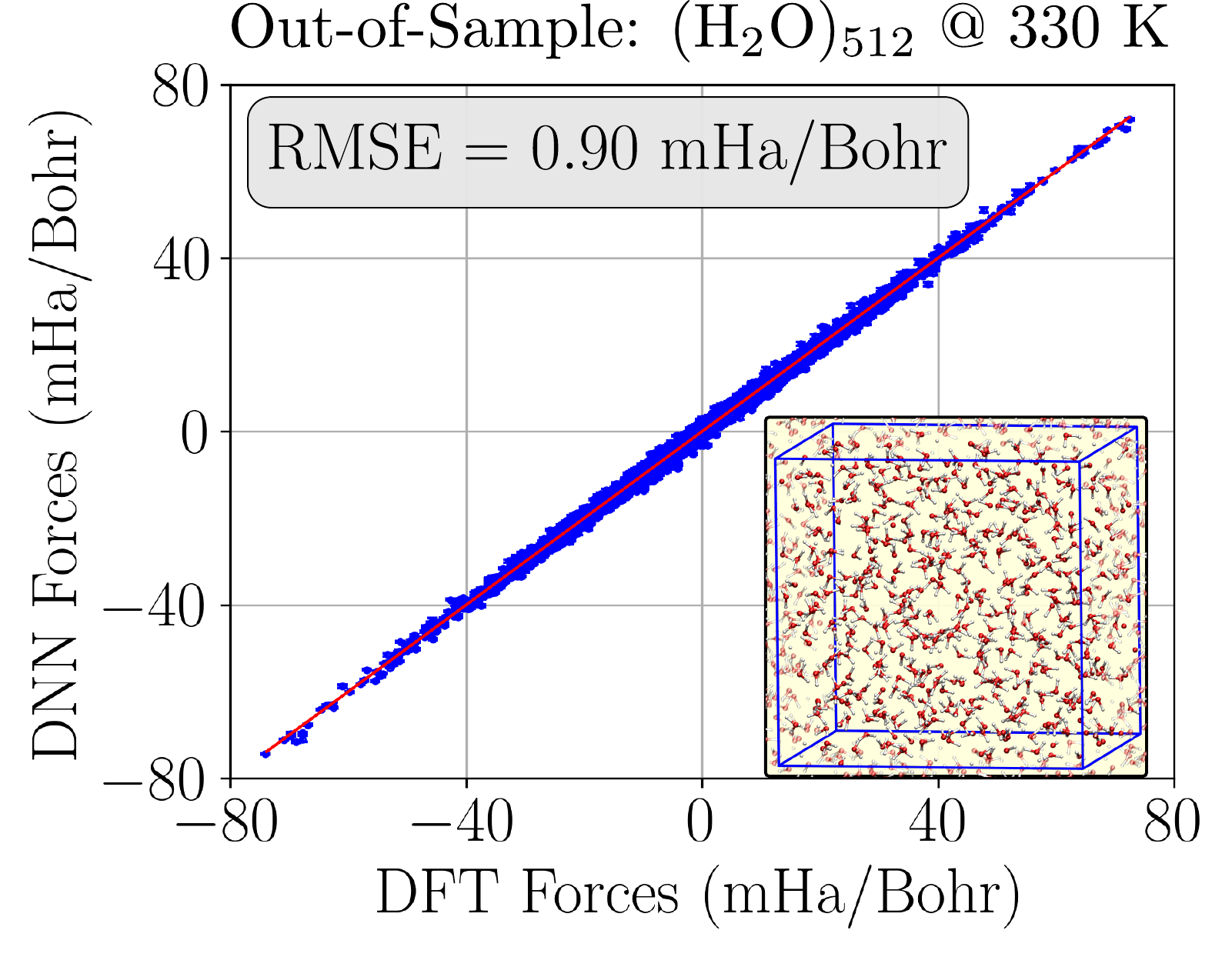}
    \caption{
    Correlation plot between DNN ionic force components (training set: $8{,}705$ \ce{(H2O)64} configurations actively learned at $T = 300$~K and $p = 1$~Bar) and ground-truth PBE0 ionic force components (computed using \texttt{SeA} ($\epsilon = 10^{-3.5}$)) for an out-of-sample test set containing two uncorrelated snapshots from a DPMD simulation of \ce{(H2O)512} at $T = 330$~K and $p = 1$~Bar.
    DNN ionic force components are plotted as blue dots (with error bars based on the mean and standard deviation among the four independently trained DNN models); the red line indicates perfect correlation with the ground-truth PBE0 ionic force components.
    }
    \label{fig:water512_validation}
\end{figure}

To assess the accuracy of this \texttt{SeA}-trained DNN model, we also compared its ionic force predictions against the ground truth (i.e., direct ionic force calculations at the PBE0 level using \texttt{SeA} with $\epsilon = 10^{-3.5}$).
As an initial assessment, we considered $16$ equispaced \ce{(H2O)64} snapshots from the $NpT$ DPMD simulation performed in Sec.~\ref{app:liq_water} at $T=300$~K and $1$~Bar (i.e., the training/active-learning thermodynamic conditions).
When compared against direct ionic force calculations with \texttt{SeA}, we found that the ionic forces furnished by the final DNN model(s) were highly accurate with a root-mean-square error (RMSE) of $0.75$~mHa/Bohr.
For a more stringent assessment, we prepared an out-of-sample test set, which included two equispaced snapshots from a $\approx 1$~ns DPMD simulation of \ce{(H2O)512} at $T = 330$~K and $1$~Bar (using the DNN potential trained on \ce{(H2O)64} at $T = 300$~K and $1$~Bar).
Using these uncorrelated snapshots, we again found that the ionic forces provided by the DNN model(s) were able to reproduce the ground-truth ionic forces with very high fidelity (Fig.~\ref{fig:water512_validation}).
As depicted in this figure, the RMSE for this test set (which is out-of-sample in both system size and $T$) was $0.90$~mHa/Bohr, which is only slightly larger than the RMSE found above for the \ce{(H2O)64} snapshots collected at the training/active-learning thermodynamic conditions.
We attribute this (rather small) RMSE increase to the additional structural disorder afforded by the extended system size (\ce{(H2O)64} $\rightarrow$ \ce{(H2O)512}) and elevated temperature ($300$~K $\rightarrow 330$~K) in this out-of-sample test set.
Here, we also note that both RMSE values are below typical force convergence thresholds ($\sim 1.0$~mHa/Bohr), which further highlights the high degree of accuracy achieved by this \texttt{SeA}-trained DNN model.
While direct AIMD simulations of liquid water at the PBE0 level have been reported (e.g., Ref.~\cite{distasio_jr._individual_2014}), there are a number of differences between those simulations and the DNN-based simulations performed herein (e.g., Car--Parrinello vs.\ Born--Oppenheimer, simulation lengths, $NpT$ vs.\ $NVT$, etc) which hinder a quantitative comparison of the resultant structural/equilibrium properties (e.g., radial distribution functions) as an additional assessment of the accuracy of our \texttt{SeA}-trained DNN model. In the same breath, the development of the high-throughput \texttt{SeA} framework in this work enables a more thorough investigation of such issues, which will be addressed in future work.

%
%
%
%
Besides validating the accuracy of our \texttt{SeA}-trained DNN model, direct calculations of the ionic forces in these two \ce{(H2O)512} snapshots (each containing $>1{,}500$ atoms) using \texttt{SeA} also highlight the capabilities of this approach when performing hybrid DFT calculations on large-scale condensed-phase systems.
For such challenging systems (which are currently beyond the reach of \texttt{PWSCF(ACE)}), \texttt{SeA} has enabled fully self-consistent calculations of their energies and ionic forces at the hybrid DFT level in $\approx 1.2$~hours using $25$ Cori-Haswell nodes (with a total of $100$~\mpi{} processes and $16$~\omp{} threads per \mpi{} process).
By performing such calculations with \texttt{SeA}, the overall cost is now dominated by the inner-loop \texttt{SCF\_Iteration} step instead of the outer-loop \texttt{ACE\_Construction} step---the typical bottleneck in \texttt{PWSCF(ACE)} for large-scale systems.
By harnessing three levels of computational savings, \texttt{SeA} has effectively removed the computational bottleneck that typically prohibits the routine use of hybrid DFT in high-throughput applications for systems like \ce{(H2O)64} (cf. Tables~\ref{tab:perf_breakdown_w064_ace_cons} and \ref{tab:perf_breakdown_w064_time_to_solution}) and beyond, e.g., \ce{(H2O)512}.
This observation again reiterates the need for scalable and efficient GGA-based KS-DFT algorithms (e.g., linear-scaling and/or real-space approaches) which could reduce the cost of the inner-loop \texttt{SCF\_Iteration} step in \texttt{SeA} and thereby enable hybrid DFT calculations for significantly larger systems.

\section{Conclusions \label{sec:conclusion}}

In this work, we have developed \texttt{SeA} (\texttt{SeA} = SCDM+\texttt{exx}+ACE)---a robust, accurate, and computationally efficient framework for performing high-throughput hybrid DFT calculations on large-scale finite-gap systems.
Implemented in the \texttt{PWSCF} module of \texttt{Quantum ESPRESSO} (\texttt{QE}), \texttt{SeA} combines and seamlessly integrates: (\textit{i}) the selected columns of the density matrix (SCDM) approach~\cite{damle_compressed_2015} (a direct/non-iterative orbital localization scheme that sidesteps the need for system-dependent optimization protocols), (\textit{ii}) a recently extended black-box version~\cite{paper3} of \texttt{exx}~\cite{paper1,paper2} (a linear-scaling real-space EXX algorithm that exploits the sparsity between localized orbitals when evaluating the action of the standard/full-rank EXX operator ($\hat{V}_{\rm xx}$)), and (\textit{iii}) the adaptively compressed exchange (ACE)~\cite{lin_adaptively_2016} formalism (an efficient low-rank $\hat{V}_{\rm xx}$ approximation that reduces the number of full-rank evaluations of the action during the iterative EXX-SCF procedure).
By construction, \texttt{SeA} is able to harness three distinct levels of computational savings during hybrid DFT calculations: two from SCDM+\texttt{exx} (\textit{pair selection} and \textit{domain truncation}: as this approach only considers spatially overlapping orbital pairs and evaluates their corresponding EXX interaction on orbital-pair-specific/system-size-independent real-space domains) and one from ACE (\textit{low-rank $\hat{V}_{\rm xx}$ approximation}: which reduces the number of SCDM+\texttt{exx} calls during the SCF solution to the KS-DFT equations).
To assess the accuracy and performance of \texttt{SeA}, we compared this approach against \texttt{PWSCF(ACE)} (the convolution-based ACE implementation in the \texttt{PWSCF} module of \texttt{QE}) across a diverse set of $200$ non-equilibrium \ce{(H2O)64} configurations (which include both intact and auto-ionized water molecules, and have system densities ranging from $\approx 0.4$~g/cm$^3$ to $\approx 1.7$~g/cm$^3$). 
In doing so, we found that \texttt{SeA} furnishes a one--two order-of-magnitude speedup ($\approx 11\times$$-$$116\times$) in the computational bottleneck \texttt{ACE\_Construction} step in \texttt{PWSCF(ACE)}, while reproducing the EXX energy and ionic forces with high fidelity (i.e., mean $E_{\rm xx}$ differences of $\approx 10^{-3}\%$ and RMSD in $\{\bm F_{I}\}$ of $\approx 0.2$~mHa/Bohr).
%
Hence, \texttt{SeA} effectively removes the computational bottleneck that typically prohibits the routine use of hybrid DFT in high-throughput applications involving systems with sizes similar to (and beyond that of) \ce{(H2O)64}, and enables single-point energy and ionic force evaluations for such systems with an order-of-magnitude speedup ($8\times$$-$$26\times$) in the overall time-to-solution compared to \texttt{PWSCF(ACE)} and a one--two order-of-magnitude speedup ($78\times$$-$$247\times$) in the overall time-to-solution compared to \texttt{PWSCF(FULL)}, the conventional (non-ACE) convolution-based EXX implementation.
With ionic force errors that are lower than typical convergence thresholds used during the structural relaxation of condensed-phase systems (e.g., $1.0$~mHa/Bohr) and order(s)-of-magnitude performance improvements, \texttt{SeA} paves the way towards the routine use of hybrid DFT in large-scale high-throughput applications such as materials screening and discovery, PES sampling, and quantum mechanical data generation for ML.

As a proof-of-principle high-throughput application, we used \texttt{SeA} to train a deep neural network (DNN) potential for ambient ($T = 300$~K, $p = 1$~Bar) liquid water at the PBE0 level, based on an actively learned data set containing $\approx 8{,}700$ \ce{(H2O)64} configurations.
The convergence/precision of this \texttt{SeA}-trained DNN model was demonstrated by small model deviations in the ionic forces for liquid water under ambient ($T=300$~K, $p=1$~Bar) and non-ambient/out-of-sample ($T \in  \{ 270~\text{K}, \, 330~\text{K}, \, 360~\text{K} \}$, $p=1$~Bar) thermodynamic conditions. 
By comparing the DNN ionic force predictions to the ground truth (direct PBE0 calculations using \texttt{SeA}) for a challenging out-of-sample test set (\ce{(H2O)512} at $T=330$~K, $p=1$~Bar), we also found that the \texttt{SeA}-trained DNN model was highly accurate with an RMSE of $0.90$~mHa/Bohr in the ionic force components.
This ground-truth comparison not only validated the accuracy of the DNN model, but also showcased the capabilities of \texttt{SeA} to perform hybrid DFT calculations for condensed-phase systems containing $>1{,}500$ atoms.

\section{Future Outlook \label{sec:future_outlook}}

While \texttt{SeA} in its current form already provides a robust, accurate, and computationally efficient framework for performing high-throughput condensed-phase hybrid DFT calculations on large-scale finite-gap systems, there still remains room for further improvement.
%
In particular, we are actively working on the following research thrusts to enhance the features and capabilities of \texttt{SeA}: (\textit{i}) support for general Monkhorst-Pack (i.e., beyond $\Gamma$-point) Brillouin zone sampling, (\textit{ii}) an extension to range-separated hybrid (RSH) functionals,~\cite{heyd_hybrid_2003,gerber_hybrid_2005,vydrov_assessment_2006,janesko_screened_2009,baer_tuned_2010,kronik_excitation_2012,karolewski_using_2013} and (\textit{iii}) the development of a consistent \texttt{SeA} algorithm for implementation in real-space codes.
We are also actively working to further improve the overall efficiency of \texttt{SeA} (by enabling GPU support and reducing the prefactor/scaling in the \texttt{SCDM} orbital localization scheme) and finalizing a public release of \texttt{SeA} in the \texttt{PWSCF} module of \texttt{QE}.
We have also completely overhauled the \texttt{exx} codebase with a comprehensive three-pronged algorithmic strategy designed to increase computational efficiency, decrease communication overhead, and minimize processor idling; to maximize the impact of these developments, we are also working on the public release of this EXX engine as a standalone free software library (\texttt{exxl}).

Here, we would also emphasize that the applicability of \texttt{SeA} extends beyond high-throughput EXX-SCF calculations (i.e., single-point energy and ionic force evaluations) at the hybrid DFT level.
For instance, \texttt{SeA} (in its current form) can be used to accelerate Born--Oppenheimer AIMD (BOMD) simulations (to a significantly larger degree than the original implementation of \texttt{exx}~\cite{paper1,paper2} in the \texttt{CP} module of \texttt{QE} due to the additional computational savings from the ACE operator), and also has the potential to increase the length- and time-scales accessible by AIMD (both CPMD and BOMD) when used in conjunction with the multiple time scale approach~\cite{tuckerman_reversible_1992,tuckerman_integrating_1994} based on the ACE operator.~\cite{mandal_speeding-up_2019,mandal_achieving_2021,mandal_improving_2022}
Due to the continuous time evolution of the trajectory during AIMD simulations, \texttt{SeA} is not restricted to SCDM orbitals and could also be used in conjunction with other localization schemes (e.g., MLWFs~\cite{marzari_maximally_1997,marzari_maximally_2012}) by keeping track of (and continuously refining~\cite{sharma_ab_2003}) the unitary operator connecting the local and canonical representations of the occupied space.
When performing hybrid DFT based CPMD, a gauge-invariant sampling can be achieved using the field-theoretic approach proposed by Tuckerman and co-workers~\cite{thomas_field_2004} or direct propagation in the canonical orbital representation while evaluating all EXX-related quantities in the localized orbital representation (i.e., akin to what was done for \texttt{SeA} in this work).
Although not explicitly described in this work, the current version of \texttt{SeA} computes the EXX contribution to the cell forces/stress tensor (i.e., $\bm{\sigma}_{\rm xx}$) in real space via \texttt{exx},~\cite{paper2} and can therefore be used to perform constant-pressure calculations at the hybrid DFT level (e.g., variable cell relaxations and $NpH$/$NpT$ simulations) as well.
Since the real-space evaluation of $\bm{\sigma}_{\rm xx}$ is subject to small (but non-negligible) aliasing errors when performed in a planewave code, we are also working on an extension to \texttt{SeA} that will compute this quantity directly from $\hat{V}_{\rm xx}^{\rm ACE}$ (in analogy to the evaluation of $E_{\rm xx}$ currently done in \texttt{SeA} via Eq.~\eqref{eq:exx_ace}).

\begin{acknowledgements}
All authors thank Anil Damle for helpful scientific discussions.
This work was supported as part of the Center for Alkaline Based Energy Solutions (CABES), an Energy Frontier Research Center funded by the U.S.\ Department of Energy, Office of Science, Basic Energy Sciences at Cornell University under Award No.\ DE-SC0019445.
In addition, RAD also gratefully acknowledges financial support from an Alfred P.\ Sloan Research Fellowship.
MCA acknowledges financial support from Lawrence Livermore National Lab (LLNL). The work at LLNL was performed under the auspices of the U.S.\ Department of Energy under Contract No.\ DE-AC52-07NA27344.
This research used resources of the National Energy Research Scientific Computing Center, which is supported by the Office of Science of the U.S.\ Department of Energy under Contract No.\ DE-AC02-05CH11231.
\end{acknowledgements}

\appendix

\section{Alternative ACE Decomposition Scheme for Approximate EXX Calculations 
\label{app:es_comp_details_decomp}}

\begin{small}
During the \texttt{ACE\_Construction} step, an approximate evaluation of $\{D_{\rm xx}^{i}\}$ can lead to a non-symmetric and/or non-positive-semi-definite $\bm M$ matrix (with elements $M_{ij} \equiv \left< \phi_i \left| D_{\rm xx}^{j} \right. \right> = - \left< \phi_i \right| \hat{V}_{\rm xx} \left| \phi_{j} \right>$), making this matrix unsuitable for Cholesky decomposition in the subsequent \texttt{decomp} function (see Eq.~\eqref{eq:Vace} and surrounding discussion).
Here, we remind the reader that the use of $\left| D_{\rm xx}^{i} \right>$ instead of $\hat{V}_{\rm xx}\left| \phi_{i} \right>$ leads to a sign difference in $M_{ij}$ when compared to Lin's original formulation.~\cite{lin_adaptively_2016}
To fulfill the symmetry requirement, we symmetrize $\bm M$ via $\bm{\overline{M}} \equiv \frac{1}{2}\left( \bm M + \bm M^T \right)$ to remove the (albeit small) skew-symmetric contributions arising from the approximate evaluation of $\{D_{\rm xx}^{i}\}$; this approach has also been used in the L-ACE approach.~\cite{carnimeo_fast_2018}
For the $\approx 8{,}700$ aqueous configurations (\ce{(H2O)64} and \ce{(H2O)512}) considered in this work, symmetrization of $\bm{M}$ was sufficient to allow for Cholesky decomposition to proceed without issue (i.e., $\bm{\overline{M}}$ remained positive-semi-definite in all of these cases).
However, Cholesky decomposition will fail if the approximate evaluation of $\{D_{\rm xx}^{i}\}$ violates the positive-semi-definiteness requirement on $\bm{\overline{M}}$; in \texttt{SeA}, this issue can arise when using loose $\epsilon$ settings and/or treating systems with more substantial finite-size effects (i.e., systems with highly delocalized orbitals relative to the size of the unit cell).

To improve the robustness of \texttt{SeA}, we therefore implemented an alternative/generalized low-rank decomposition routine based on the following eigen-decomposition of the real symmetric $\bm{\overline{M}}$ matrix:
\begin{align}
    \bm{\overline{M}} = \bm Q \bm \Lambda \bm Q^T = \left( \bm Q \bm \Lambda ^{1/2} \right) \left( \bm \Lambda^{1/2}\bm Q^T \right) \equiv \bm Z \bm Z^T ,
    \label{eq:mzzt}
\end{align}
in which $\bm Q$ is an $N_{\rm occ}\times N_{\rm occ}$ orthogonal eigenvector matrix (not to be confused with the orthogonal matrix in the SCDM QR factorization in Eq.~\eqref{eq:phi_qr}), $\bm \Lambda$ is the corresponding $N_{\rm occ}\times N_{\rm occ}$ diagonal eigenvalue matrix, and $\bm Z$ is an $N_{\rm occ}\times N_{\rm occ}$ matrix alternative to the Cholesky factor ($\bm L$).
When constructed from the \textit{exact} $\{D_{\rm xx}^{i}\}$, the eigenvalues in $\bm \Lambda$ would be real and non-negative, and hence $\bm \Lambda^{1/2}$ would be real.
When constructed from an approximate $\{D_{\rm xx}^{i}\}$, $\bm{\overline{M}}$ is no longer guaranteed to be positive-semi-definite; in this case, the eigenvalues in $\bm \Lambda$ will be real but not necessarily non-negative, hence $\bm \Lambda^{1/2}$ (and $\bm Z$) can be complex.
In this alternative scheme, the analogous ACE operator can be constructed as follows:
\begin{align}
  \hat{V}_{\rm xx}^{\rm ACE} &= -\sum_{k} \left| \zeta_{k} \right>\left< \zeta_{k} \right| ,
 \label{eq:Vace_alter} 
\end{align}
in which
\begin{align}
  \left| \zeta_{k} \right> &\equiv -\sum_i \left| D_{\rm xx}^{i} \right> \left( {\bm Z}^{-T} \right)_{ik} .
\end{align}
Eq.~\eqref{eq:Vace_alter} is the analog of Eq.~\eqref{eq:Vace} with $\left| \xi_{k} \right>$ replaced by $\left| \zeta_{k} \right>$; since both of these quantities are complex-valued (as $\{D_{\rm xx}^{i}(\bm G)\}$ is complex), no downstream changes in the code are needed to implement this alternative decomposition scheme.
In this approach, ${\bm Z}^{-T}$ can be conveniently computed as:
\begin{align}
  {\bm Z}^{-T} &= \left( \bm Q \bm \Lambda^{1/2} \right)^{-T} = \bm Q \bm \Lambda^{-1/2} 
\end{align}
since $\bm Q$ is orthogonal ($\bm{Q}^T = \bm Q^{-1}$) and $\bm \Lambda$ is diagonal.
In practice, we have found that this alternative decomposition scheme is quite robust across a broad range of systems; as such, it would be interesting to see if this approach also increases the robustness of other approximate ACE-based methods.~\cite{carnimeo_fast_2018,mandal_achieving_2021}
\end{small}

\section{Outer-Loop Convergence in \texttt{SeA} via Successive Idempotency \label{app:es_comp_details_conv}}

\begin{small}
SCF convergence in \texttt{SeA} is measured at the $n$-th outer-loop step (see Algorithm~\ref{alg:ace}) via successive idempotency, i.e., the change in idempotency between density matrices at the current ($\hat{P}^{(n)}$) and previous ($\hat{P}^{(n-1)}$) steps:
\begin{align}
  \Delta I^{(n)} &\equiv 1 - \frac{\mathrm{Tr} \, \left[ \hat{P}^{(n)} \hat{P}^{(n-1)} \right]}{\mathrm{Tr} \, \left[ \hat{P}^{(n)} \hat{P}^{(n)} \right]} \nonumber \\
  &= 1 - \frac{\sum_{ij}\left.\left|\left< \phi_{i}^{(n)}\right| \phi_{j}^{(n-1)}\right>\right|^2}{N_{\rm occ}} ,
\end{align}
in which $\left\{ \left| \phi_{i}^{(n)} \right> \right\}$ is the set of proto-KS orbitals generated at the end of the $n$-th outer-loop step (i.e., the self-consistent orbitals obtained using the fixed $\hat{V}_{\rm xx}^{\rm ACE}$ at this step).
For the $\approx 8{,}700$ aqueous configurations (\ce{(H2O)64} and \ce{(H2O)512}) considered in this work, we used $10^{-7}$ as the outer-loop convergence criterion, i.e., the EXX-SCF procedure in \texttt{SeA} was considered converged when $\Delta I^{(n)} \leq 10^{-7}$. 
With the use of this setting, $E_{\rm xx}$ converges to within $\approx 10^{-2}\%$$-$$10^{-3}\%$ of \texttt{PWSCF(ACE)} (i.e., the \textit{a priori} estimated level of accuracy in \texttt{SeA} for $10^{-3.0} \geq\epsilon\geq 10^{-4.0}$, see Fig.~\ref{fig:sea_acc_prf} and Table~\ref{tab:acc_scalar_vector}) and no appreciable oscillatory behavior was observed in the total energy during the EXX-SCF procedure. 
We empirically found that this convergence criterion requires essentially the same number of outer-loop steps when compared to the default convergence criteria in \texttt{PWSCF(ACE)} for these aqueous systems.
While further tightening of this convergence criterion has a negligible effect on $E_{\rm xx}$ (i.e., $\approx 10^{-4}\%$, which is smaller than the \textit{a priori} estimated accuracy of \texttt{SeA}), the use of a significantly tighter convergence criterion (e.g., $\approx 10^{-12}$) will ultimately lead to oscillations in $E_{\rm xx}$ (and the total energy) with a magnitude of $\approx 10^{-4}\%$.
\end{small}

\section{Computational Details (Electronic Structure Calculations) \label{app:es_comp_details}}

\begin{small} 
For all hybrid DFT calculations (with \texttt{SeA} or \texttt{PWSCF(ACE)}), we used an in-house implementation built upon the \texttt{PWSCF} module of \texttt{Quantum ESPRESSO} (git version: qe-7.0, in which the \texttt{vexx} function in \texttt{PWSCF(ACE)} can effectively exploit both \texttt{MPI} and \texttt{OpenMP} parallelization).
We modeled the interactions between the valence electrons and ions (nuclei and their frozen-core electrons) using the Hamann--Schl{\" u}ter--Chiang--Vanderbilt (HSCV) type norm-conserving pseudopotentials~\cite{hamann_norm-conserving_1979,vanderbilt_optimally_1985} distributed with the \texttt{Qbox} package.~\cite{gygi_architecture_2008}
Pseudo-wavefunctions for the valence electrons were represented using a planewave basis truncated at a maximum kinetic energy of $85$~Ry.
Unless otherwise specified, default settings (e.g., Davidson diagonalization, convergence thresholds, and density mixing parameters) were used throughout.
For all \texttt{PWSCF(ACE)} calculations, the integrable divergence at $\bm G = 0$ during \texttt{vexx} evaluations was treated using the Gygi--Baldereschi approach~\cite{gygi_self-consistent_1986} in conjunction with the finite-size correction proposed by Nguyen and de~Gironcoli (i.e., \texttt{x\_gamma\_extrapolation}).~\cite{nguyen_efficient_2009}
For all \texttt{SeA} calculations, we solved the near-field PE using an $\mathcal{O}(h^{12})$ discrete Laplacian operator and a final residual of $10^{-3}$~Bohr$^{-3}$ in the conjugate gradient (CG) solver; each far-field ME was computed with a maximum angular momentum of $l_{\rm max} = 6$.
\end{small}

\section{Computational Details (DNN Training and DPMD Simulations) \label{app:train_comp_detail}}

\begin{small} 
The \texttt{DeepMD-kit}~\cite{wang_deepmd-kit:_2018} (git version: v1.3.3) interfaced with the \texttt{TensorFlow} library~\cite{tensorflow2015-whitepaper} (git version: v2.4.0) was used to train the DNN model for liquid water based on the total energy $E$ and ionic forces $\{\bm{F}_I\}$ from each selected configuration in the training set.
We used DeepPot-SE~\cite{zhang_end--end_2018} to smoothly map the local chemical environment around each atom (i.e., the relative atomic coordinates within a radius of $6$~\AA{}) into input for a DNN model containing two coupled three-layer DNNs (i.e., an embedding network with $(25,50,100)$ neurons and a fitting network with $(60,60,60)$ neurons).
During the training/active-learning process, we followed Refs.~\onlinecite{zhang_deep_2018,zhang_end--end_2018} and minimized the following loss function $\mathcal{L}$:
\begin{align}
  \mathcal L(\lambda, \eta) = \frac{\lambda}{N^2} \vert \Delta E \vert^2 + \frac{\eta}{3N} \sum_I \vert \Delta\bm{F}_I\vert^2 ,
  \label{eqn:loss}
\end{align}
in which $\Delta E$ and $\Delta \bm{F}_I$ are the differences between the DPMD model prediction and the training data for $E$ and $\bm{F}_I$, respectively, and $N$ is the number of atoms in a given configuration.
The Adam stochastic gradient descent method~\cite{Kingma2015adam} was performed for $0.5$~M steps to train the DPMD model parameters ($\lambda$ and $\eta$) with an initial learning rate of $5.0 \times 10^{-3}$ that exponentially decayed with respect to the number of training steps to $1.0 \times 10^{-8}$.
As a function of the learning rate, we linearly varied $\lambda$ (from $10^{-2}$ to $1$) and $\eta$ (from $10^{3}$ to $1$) to achieve an efficient and well-balanced training procedure.

All DPMD simulations were performed using \texttt{LAMMPS}~\cite{plimpton_fast_1995} (git version: stable\_29Oct2020) with Nos\'e-Hoover thermostat and barostat chains~\cite{tuckerman_liouville-operator_2006} (with chain lengths of $3$ and characteristic timescales of $0.1$~ps and $0.5$~ps, respectively).
All hydrogen atoms were replaced with deuterium to allow for a $0.5$~fs integration time step.
For each temperature setting, we performed a $100$~ps equilibration run followed by a $1$~ns production calculation to sample the $NpT$ ensemble using a cubic periodic cell with $64$ water molecules.
Snapshots were taken every $100$ DPMD steps to form a discrete trajectory for analysis.
\end{small}


\begin{thebibliography}{102}%
\makeatletter
\providecommand \@ifxundefined [1]{%
 \@ifx{#1\undefined}
}%
\providecommand \@ifnum [1]{%
 \ifnum #1\expandafter \@firstoftwo
 \else \expandafter \@secondoftwo
 \fi
}%
\providecommand \@ifx [1]{%
 \ifx #1\expandafter \@firstoftwo
 \else \expandafter \@secondoftwo
 \fi
}%
\providecommand \natexlab [1]{#1}%
\providecommand \enquote  [1]{``#1''}%
\providecommand \bibnamefont  [1]{#1}%
\providecommand \bibfnamefont [1]{#1}%
\providecommand \citenamefont [1]{#1}%
\providecommand \href@noop [0]{\@secondoftwo}%
\providecommand \href [0]{\begingroup \@sanitize@url \@href}%
\providecommand \@href[1]{\@@startlink{#1}\@@href}%
\providecommand \@@href[1]{\endgroup#1\@@endlink}%
\providecommand \@sanitize@url [0]{\catcode `\\12\catcode `\$12\catcode
  `\&12\catcode `\#12\catcode `\^12\catcode `\_12\catcode `\%12\relax}%
\providecommand \@@startlink[1]{}%
\providecommand \@@endlink[0]{}%
\providecommand \url  [0]{\begingroup\@sanitize@url \@url }%
\providecommand \@url [1]{\endgroup\@href {#1}{\urlprefix }}%
\providecommand \urlprefix  [0]{URL }%
\providecommand \Eprint [0]{\href }%
\providecommand \doibase [0]{http://dx.doi.org/}%
\providecommand \selectlanguage [0]{\@gobble}%
\providecommand \bibinfo  [0]{\@secondoftwo}%
\providecommand \bibfield  [0]{\@secondoftwo}%
\providecommand \translation [1]{[#1]}%
\providecommand \BibitemOpen [0]{}%
\providecommand \bibitemStop [0]{}%
\providecommand \bibitemNoStop [0]{.\EOS\space}%
\providecommand \EOS [0]{\spacefactor3000\relax}%
\providecommand \BibitemShut  [1]{\csname bibitem#1\endcsname}%
\let\auto@bib@innerbib\@empty
\bibitem [{\citenamefont {Curtarolo}\ \emph {et~al.}(2013)\citenamefont
  {Curtarolo}, \citenamefont {Hart}, \citenamefont {Nardelli}, \citenamefont
  {Mingo}, \citenamefont {Sanvito},\ and\ \citenamefont
  {Levy}}]{curtarolo_high-throughput_2013}%
  \BibitemOpen
  \bibfield  {author} {\bibinfo {author} {\bibfnamefont {S.}~\bibnamefont
  {Curtarolo}}, \bibinfo {author} {\bibfnamefont {G.~L.~W.}\ \bibnamefont
  {Hart}}, \bibinfo {author} {\bibfnamefont {M.~B.}\ \bibnamefont {Nardelli}},
  \bibinfo {author} {\bibfnamefont {N.}~\bibnamefont {Mingo}}, \bibinfo
  {author} {\bibfnamefont {S.}~\bibnamefont {Sanvito}}, \ and\ \bibinfo
  {author} {\bibfnamefont {O.}~\bibnamefont {Levy}},\ }\bibfield  {title}
  {\enquote {\bibinfo {title} {The high-throughput highway to computational
  materials design},}\ }\href@noop {} {\bibfield  {journal} {\bibinfo
  {journal} {Nat. Mater.}\ }\textbf {\bibinfo {volume} {12}},\ \bibinfo {pages}
  {191--201} (\bibinfo {year} {2013})}\BibitemShut {NoStop}%
\bibitem [{\citenamefont {von Lilienfeld}(2014)}]{von_lilienfeld_towards_2014}%
  \BibitemOpen
  \bibfield  {author} {\bibinfo {author} {\bibfnamefont {O.~A.}\ \bibnamefont
  {von Lilienfeld}},\ }\bibfield  {title} {\enquote {\bibinfo {title} {Towards
  the computational design of compounds from first principles},}\ }in\
  \href@noop {} {\emph {\bibinfo {booktitle} {Many-{Electron} {Approaches} in
  {Physics}, {Chemistry} and {Mathematics}: {A} {Multidisciplinary} {View}}}},\
  \bibinfo {series and number} {Mathematical {Physics} {Studies}},\ \bibinfo
  {editor} {edited by\ \bibinfo {editor} {\bibfnamefont {V.}~\bibnamefont
  {Bach}}\ and\ \bibinfo {editor} {\bibfnamefont {L.}~\bibnamefont
  {Delle~Site}}}\ (\bibinfo  {publisher} {Springer International Publishing},\
  \bibinfo {address} {Cham},\ \bibinfo {year} {2014})\ pp.\ \bibinfo {pages}
  {169--189}\BibitemShut {NoStop}%
\bibitem [{\citenamefont {Unke}\ \emph {et~al.}(2021)\citenamefont {Unke},
  \citenamefont {Chmiela}, \citenamefont {Sauceda}, \citenamefont {Gastegger},
  \citenamefont {Poltavsky}, \citenamefont {Sch{\"u}tt}, \citenamefont
  {Tkatchenko},\ and\ \citenamefont {M{\"u}ller}}]{unke_machine_2021}%
  \BibitemOpen
  \bibfield  {author} {\bibinfo {author} {\bibfnamefont {O.~T.}\ \bibnamefont
  {Unke}}, \bibinfo {author} {\bibfnamefont {S.}~\bibnamefont {Chmiela}},
  \bibinfo {author} {\bibfnamefont {H.~E.}\ \bibnamefont {Sauceda}}, \bibinfo
  {author} {\bibfnamefont {M.}~\bibnamefont {Gastegger}}, \bibinfo {author}
  {\bibfnamefont {I.}~\bibnamefont {Poltavsky}}, \bibinfo {author}
  {\bibfnamefont {K.~T.}\ \bibnamefont {Sch{\"u}tt}}, \bibinfo {author}
  {\bibfnamefont {A.}~\bibnamefont {Tkatchenko}}, \ and\ \bibinfo {author}
  {\bibfnamefont {K.-R.}\ \bibnamefont {M{\"u}ller}},\ }\bibfield  {title}
  {\enquote {\bibinfo {title} {Machine learning force fields},}\ }\href@noop {}
  {\bibfield  {journal} {\bibinfo  {journal} {Chem. Rev.}\ }\textbf {\bibinfo
  {volume} {121}},\ \bibinfo {pages} {10142--10186} (\bibinfo {year}
  {2021})}\BibitemShut {NoStop}%
\bibitem [{\citenamefont {Meuwly}(2021)}]{meuwly_machine_2021}%
  \BibitemOpen
  \bibfield  {author} {\bibinfo {author} {\bibfnamefont {M.}~\bibnamefont
  {Meuwly}},\ }\bibfield  {title} {\enquote {\bibinfo {title} {Machine learning
  for chemical reactions},}\ }\href@noop {} {\bibfield  {journal} {\bibinfo
  {journal} {Chem. Rev.}\ }\textbf {\bibinfo {volume} {121}},\ \bibinfo {pages}
  {10218--10239} (\bibinfo {year} {2021})}\BibitemShut {NoStop}%
\bibitem [{\citenamefont {Huang}\ and\ \citenamefont {von
  Lilienfeld}(2021)}]{huang_ab_2021}%
  \BibitemOpen
  \bibfield  {author} {\bibinfo {author} {\bibfnamefont {B.}~\bibnamefont
  {Huang}}\ and\ \bibinfo {author} {\bibfnamefont {O.~A.}\ \bibnamefont {von
  Lilienfeld}},\ }\bibfield  {title} {\enquote {\bibinfo {title} {\textit{Ab
  initio} machine learning in chemical compound space},}\ }\href@noop {}
  {\bibfield  {journal} {\bibinfo  {journal} {Chem. Rev.}\ }\textbf {\bibinfo
  {volume} {121}},\ \bibinfo {pages} {10001--10036} (\bibinfo {year}
  {2021})}\BibitemShut {NoStop}%
\bibitem [{\citenamefont {Hohenberg}\ and\ \citenamefont
  {Kohn}(1964)}]{hohenberg_inhomogeneous_1964}%
  \BibitemOpen
  \bibfield  {author} {\bibinfo {author} {\bibfnamefont {P.}~\bibnamefont
  {Hohenberg}}\ and\ \bibinfo {author} {\bibfnamefont {W.}~\bibnamefont
  {Kohn}},\ }\bibfield  {title} {\enquote {\bibinfo {title} {Inhomogeneous
  electron gas},}\ }\href@noop {} {\bibfield  {journal} {\bibinfo  {journal}
  {Phys. Rev.}\ }\textbf {\bibinfo {volume} {136}},\ \bibinfo {pages} {B864}
  (\bibinfo {year} {1964})}\BibitemShut {NoStop}%
\bibitem [{\citenamefont {Kohn}\ and\ \citenamefont
  {Sham}(1965)}]{kohn_self-consistent_1965}%
  \BibitemOpen
  \bibfield  {author} {\bibinfo {author} {\bibfnamefont {W.}~\bibnamefont
  {Kohn}}\ and\ \bibinfo {author} {\bibfnamefont {L.~J.}\ \bibnamefont
  {Sham}},\ }\bibfield  {title} {\enquote {\bibinfo {title} {Self-consistent
  equations including exchange and correlation effects},}\ }\href@noop {}
  {\bibfield  {journal} {\bibinfo  {journal} {Phys. Rev.}\ }\textbf {\bibinfo
  {volume} {140}},\ \bibinfo {pages} {A1133} (\bibinfo {year}
  {1965})}\BibitemShut {NoStop}%
\bibitem [{\citenamefont {Parr}\ and\ \citenamefont
  {Yang}(1989)}]{parr_density-functional_1989}%
  \BibitemOpen
  \bibfield  {author} {\bibinfo {author} {\bibfnamefont {R.~G.}\ \bibnamefont
  {Parr}}\ and\ \bibinfo {author} {\bibfnamefont {W.}~\bibnamefont {Yang}},\
  }\href@noop {} {\emph {\bibinfo {title} {Density-Functional Theory of Atoms
  and Molecules}}}\ (\bibinfo  {publisher} {Oxford University Press},\ \bibinfo
  {address} {New York},\ \bibinfo {year} {1989})\BibitemShut {NoStop}%
\bibitem [{fio(2003)}]{fiolhais_primer_2003}%
  \BibitemOpen
  \bibfield  {title} {\enquote {\bibinfo {title} {A primer in density
  functional theory},}\ }in\ \href@noop {} {\emph {\bibinfo {booktitle}
  {{Lecture Notes in Physics}}}},\ Vol.\ \bibinfo {volume} {620},\ \bibinfo
  {editor} {edited by\ \bibinfo {editor} {\bibfnamefont {C.}~\bibnamefont
  {Fiolhais}}, \bibinfo {editor} {\bibfnamefont {F.}~\bibnamefont {Nogueira}},
  \ and\ \bibinfo {editor} {\bibfnamefont {M.}~\bibnamefont {Marques}}}\
  (\bibinfo  {publisher} {Springer},\ \bibinfo {address} {New York},\ \bibinfo
  {year} {2003})\BibitemShut {NoStop}%
\bibitem [{\citenamefont {Becke}(2014)}]{becke_perspective:_2014}%
  \BibitemOpen
  \bibfield  {author} {\bibinfo {author} {\bibfnamefont {A.~D.}\ \bibnamefont
  {Becke}},\ }\bibfield  {title} {\enquote {\bibinfo {title} {Perspective:
  Fifty years of density-functional theory in chemical physics},}\ }\href@noop
  {} {\bibfield  {journal} {\bibinfo  {journal} {J. Chem. Phys.}\ }\textbf
  {\bibinfo {volume} {140}},\ \bibinfo {pages} {18A301} (\bibinfo {year}
  {2014})}\BibitemShut {NoStop}%
\bibitem [{\citenamefont {Mardirossian}\ and\ \citenamefont
  {Head-Gordon}(2017)}]{mardirossian_thirty_2017}%
  \BibitemOpen
  \bibfield  {author} {\bibinfo {author} {\bibfnamefont {N.}~\bibnamefont
  {Mardirossian}}\ and\ \bibinfo {author} {\bibfnamefont {M.}~\bibnamefont
  {Head-Gordon}},\ }\bibfield  {title} {\enquote {\bibinfo {title} {Thirty
  years of density functional theory in computational chemistry: An overview
  and extensive assessment of $200$ density functionals},}\ }\href@noop {}
  {\bibfield  {journal} {\bibinfo  {journal} {Mol. Phys.}\ }\textbf {\bibinfo
  {volume} {115}},\ \bibinfo {pages} {2315--2372} (\bibinfo {year}
  {2017})}\BibitemShut {NoStop}%
\bibitem [{\citenamefont {Medvedev}\ \emph
  {et~al.}(2017{\natexlab{a}})\citenamefont {Medvedev}, \citenamefont
  {Bushmarinov}, \citenamefont {Sun}, \citenamefont {Perdew},\ and\
  \citenamefont {Lyssenko}}]{medvedev_density_2017}%
  \BibitemOpen
  \bibfield  {author} {\bibinfo {author} {\bibfnamefont {M.~G.}\ \bibnamefont
  {Medvedev}}, \bibinfo {author} {\bibfnamefont {I.~S.}\ \bibnamefont
  {Bushmarinov}}, \bibinfo {author} {\bibfnamefont {J.}~\bibnamefont {Sun}},
  \bibinfo {author} {\bibfnamefont {J.~P.}\ \bibnamefont {Perdew}}, \ and\
  \bibinfo {author} {\bibfnamefont {K.~A.}\ \bibnamefont {Lyssenko}},\
  }\bibfield  {title} {\enquote {\bibinfo {title} {Density functional theory is
  straying from the path toward the exact functional},}\ }\href@noop {}
  {\bibfield  {journal} {\bibinfo  {journal} {Science}\ }\textbf {\bibinfo
  {volume} {355}},\ \bibinfo {pages} {49--52} (\bibinfo {year}
  {2017}{\natexlab{a}})}\BibitemShut {NoStop}%
\bibitem [{\citenamefont {Kepp}(2017)}]{kepp_comment_2017}%
  \BibitemOpen
  \bibfield  {author} {\bibinfo {author} {\bibfnamefont {K.~P.}\ \bibnamefont
  {Kepp}},\ }\bibfield  {title} {\enquote {\bibinfo {title} {\textit{Comment
  on:} {Density} functional theory is straying from the path toward the exact
  functional},}\ }\href@noop {} {\bibfield  {journal} {\bibinfo  {journal}
  {Science}\ }\textbf {\bibinfo {volume} {356}},\ \bibinfo {pages} {496--496}
  (\bibinfo {year} {2017})}\BibitemShut {NoStop}%
\bibitem [{\citenamefont
  {Hammes-Schiffer}(2017)}]{hammes-schiffer_conundrum_2017}%
  \BibitemOpen
  \bibfield  {author} {\bibinfo {author} {\bibfnamefont {S.}~\bibnamefont
  {Hammes-Schiffer}},\ }\bibfield  {title} {\enquote {\bibinfo {title} {A
  conundrum for density functional theory},}\ }\href@noop {} {\bibfield
  {journal} {\bibinfo  {journal} {Science}\ }\textbf {\bibinfo {volume}
  {355}},\ \bibinfo {pages} {28--29} (\bibinfo {year} {2017})}\BibitemShut
  {NoStop}%
\bibitem [{\citenamefont {Medvedev}\ \emph
  {et~al.}(2017{\natexlab{b}})\citenamefont {Medvedev}, \citenamefont
  {Bushmarinov}, \citenamefont {Sun}, \citenamefont {Perdew},\ and\
  \citenamefont {Lyssenko}}]{medvedev_response_2017}%
  \BibitemOpen
  \bibfield  {author} {\bibinfo {author} {\bibfnamefont {M.~G.}\ \bibnamefont
  {Medvedev}}, \bibinfo {author} {\bibfnamefont {I.~S.}\ \bibnamefont
  {Bushmarinov}}, \bibinfo {author} {\bibfnamefont {J.}~\bibnamefont {Sun}},
  \bibinfo {author} {\bibfnamefont {J.~P.}\ \bibnamefont {Perdew}}, \ and\
  \bibinfo {author} {\bibfnamefont {K.~A.}\ \bibnamefont {Lyssenko}},\
  }\bibfield  {title} {\enquote {\bibinfo {title} {\textit{Response to Comment
  on:} {Density} functional theory is straying from the path toward the exact
  functional},}\ }\href@noop {} {\bibfield  {journal} {\bibinfo  {journal}
  {Science}\ }\textbf {\bibinfo {volume} {356}},\ \bibinfo {pages} {496--496}
  (\bibinfo {year} {2017}{\natexlab{b}})}\BibitemShut {NoStop}%
\bibitem [{\citenamefont {Lehtola}\ \emph {et~al.}(2018)\citenamefont
  {Lehtola}, \citenamefont {Steigemann}, \citenamefont {Oliveira},\ and\
  \citenamefont {Marques}}]{lehtola_recent_2018}%
  \BibitemOpen
  \bibfield  {author} {\bibinfo {author} {\bibfnamefont {S.}~\bibnamefont
  {Lehtola}}, \bibinfo {author} {\bibfnamefont {C.}~\bibnamefont {Steigemann}},
  \bibinfo {author} {\bibfnamefont {M.~J.~T.}\ \bibnamefont {Oliveira}}, \ and\
  \bibinfo {author} {\bibfnamefont {M.~A.~L.}\ \bibnamefont {Marques}},\
  }\bibfield  {title} {\enquote {\bibinfo {title} {Recent developments in
  \texttt{libxc}---a comprehensive library of functionals for density
  functional theory},}\ }\href@noop {} {\bibfield  {journal} {\bibinfo
  {journal} {SoftwareX}\ }\textbf {\bibinfo {volume} {7}},\ \bibinfo {pages}
  {1--5} (\bibinfo {year} {2018})}\BibitemShut {NoStop}%
\bibitem [{\citenamefont {Perdew}\ and\ \citenamefont
  {Zunger}(1981)}]{perdew_self-interaction_1981}%
  \BibitemOpen
  \bibfield  {author} {\bibinfo {author} {\bibfnamefont {J.~P.}\ \bibnamefont
  {Perdew}}\ and\ \bibinfo {author} {\bibfnamefont {A.}~\bibnamefont
  {Zunger}},\ }\bibfield  {title} {\enquote {\bibinfo {title} {Self-interaction
  correction to density-functional approximations for many-electron systems},}\
  }\href {\doibase 10.1103/PhysRevB.23.5048} {\bibfield  {journal} {\bibinfo
  {journal} {Phys. Rev. B}\ }\textbf {\bibinfo {volume} {23}},\ \bibinfo
  {pages} {5048--5079} (\bibinfo {year} {1981})}\BibitemShut {NoStop}%
\bibitem [{\citenamefont {Cohen}\ \emph {et~al.}(2008)\citenamefont {Cohen},
  \citenamefont {Mori-S{\'a}nchez},\ and\ \citenamefont
  {Yang}}]{cohen_insights_2008}%
  \BibitemOpen
  \bibfield  {author} {\bibinfo {author} {\bibfnamefont {A.~J.}\ \bibnamefont
  {Cohen}}, \bibinfo {author} {\bibfnamefont {P.}~\bibnamefont
  {Mori-S{\'a}nchez}}, \ and\ \bibinfo {author} {\bibfnamefont
  {W.}~\bibnamefont {Yang}},\ }\bibfield  {title} {\enquote {\bibinfo {title}
  {Insights into current limitations of density functional theory},}\
  }\href@noop {} {\bibfield  {journal} {\bibinfo  {journal} {Science}\ }\textbf
  {\bibinfo {volume} {321}},\ \bibinfo {pages} {792--794} (\bibinfo {year}
  {2008})}\BibitemShut {NoStop}%
\bibitem [{\citenamefont {Becke}(1993)}]{becke_densityfunctional_1993}%
  \BibitemOpen
  \bibfield  {author} {\bibinfo {author} {\bibfnamefont {A.~D.}\ \bibnamefont
  {Becke}},\ }\bibfield  {title} {\enquote {\bibinfo {title}
  {Density-functional thermochemistry. {III.} {The} role of exact exchange},}\
  }\href@noop {} {\bibfield  {journal} {\bibinfo  {journal} {J. Chem. Phys.}\
  }\textbf {\bibinfo {volume} {98}},\ \bibinfo {pages} {5648--5652} (\bibinfo
  {year} {1993})}\BibitemShut {NoStop}%
\bibitem [{\citenamefont {Gygi}\ and\ \citenamefont
  {Baldereschi}(1986)}]{gygi_self-consistent_1986}%
  \BibitemOpen
  \bibfield  {author} {\bibinfo {author} {\bibfnamefont {F.}~\bibnamefont
  {Gygi}}\ and\ \bibinfo {author} {\bibfnamefont {A.}~\bibnamefont
  {Baldereschi}},\ }\bibfield  {title} {\enquote {\bibinfo {title}
  {Self-consistent {Hartree--Fock} and screened-exchange calculations in
  solids: {Application} to silicon},}\ }\href@noop {} {\bibfield  {journal}
  {\bibinfo  {journal} {Phys. Rev. B}\ }\textbf {\bibinfo {volume} {34}},\
  \bibinfo {pages} {4405--4408} (\bibinfo {year} {1986})}\BibitemShut {NoStop}%
\bibitem [{\citenamefont {Chawla}\ and\ \citenamefont
  {Voth}(1998)}]{chawla_exact_1998}%
  \BibitemOpen
  \bibfield  {author} {\bibinfo {author} {\bibfnamefont {S.}~\bibnamefont
  {Chawla}}\ and\ \bibinfo {author} {\bibfnamefont {G.~A.}\ \bibnamefont
  {Voth}},\ }\bibfield  {title} {\enquote {\bibinfo {title} {Exact exchange in
  ab initio molecular dynamics: {An} efficient plane-wave based algorithm},}\
  }\href@noop {} {\bibfield  {journal} {\bibinfo  {journal} {J. Chem. Phys.}\
  }\textbf {\bibinfo {volume} {108}},\ \bibinfo {pages} {4697--4700} (\bibinfo
  {year} {1998})}\BibitemShut {NoStop}%
\bibitem [{\citenamefont {Izmaylov}\ \emph {et~al.}(2006)\citenamefont
  {Izmaylov}, \citenamefont {Scuseria},\ and\ \citenamefont
  {Frisch}}]{izmaylov_efficient_2006}%
  \BibitemOpen
  \bibfield  {author} {\bibinfo {author} {\bibfnamefont {A.~F.}\ \bibnamefont
  {Izmaylov}}, \bibinfo {author} {\bibfnamefont {G.~E.}\ \bibnamefont
  {Scuseria}}, \ and\ \bibinfo {author} {\bibfnamefont {M.~J.}\ \bibnamefont
  {Frisch}},\ }\bibfield  {title} {\enquote {\bibinfo {title} {Efficient
  evaluation of short-range {Hartree--Fock} exchange in large molecules and
  periodic systems},}\ }\href@noop {} {\bibfield  {journal} {\bibinfo
  {journal} {J. Chem. Phys.}\ }\textbf {\bibinfo {volume} {125}},\ \bibinfo
  {pages} {104103} (\bibinfo {year} {2006})}\BibitemShut {NoStop}%
\bibitem [{\citenamefont {Sorouri}\ \emph {et~al.}(2006)\citenamefont
  {Sorouri}, \citenamefont {Foulkes},\ and\ \citenamefont
  {Hine}}]{sorouri_accurate_2006}%
  \BibitemOpen
  \bibfield  {author} {\bibinfo {author} {\bibfnamefont {A.}~\bibnamefont
  {Sorouri}}, \bibinfo {author} {\bibfnamefont {W.~M.~C.}\ \bibnamefont
  {Foulkes}}, \ and\ \bibinfo {author} {\bibfnamefont {N.~D.~M.}\ \bibnamefont
  {Hine}},\ }\bibfield  {title} {\enquote {\bibinfo {title} {Accurate and
  efficient method for the treatment of exchange in a plane-wave basis},}\
  }\href@noop {} {\bibfield  {journal} {\bibinfo  {journal} {J. Chem. Phys.}\
  }\textbf {\bibinfo {volume} {124}},\ \bibinfo {pages} {064105} (\bibinfo
  {year} {2006})}\BibitemShut {NoStop}%
\bibitem [{\citenamefont {Guidon}\ \emph {et~al.}(2008)\citenamefont {Guidon},
  \citenamefont {Schiffmann}, \citenamefont {Hutter},\ and\ \citenamefont
  {VandeVondele}}]{guidon_ab_2008}%
  \BibitemOpen
  \bibfield  {author} {\bibinfo {author} {\bibfnamefont {M.}~\bibnamefont
  {Guidon}}, \bibinfo {author} {\bibfnamefont {F.}~\bibnamefont {Schiffmann}},
  \bibinfo {author} {\bibfnamefont {J.}~\bibnamefont {Hutter}}, \ and\ \bibinfo
  {author} {\bibfnamefont {J.}~\bibnamefont {VandeVondele}},\ }\bibfield
  {title} {\enquote {\bibinfo {title} {\textit{Ab initio} molecular dynamics
  using hybrid density functionals},}\ }\href@noop {} {\bibfield  {journal}
  {\bibinfo  {journal} {J. Chem. Phys.}\ }\textbf {\bibinfo {volume} {128}},\
  \bibinfo {pages} {214104} (\bibinfo {year} {2008})}\BibitemShut {NoStop}%
\bibitem [{\citenamefont {Guidon}\ \emph {et~al.}(2009)\citenamefont {Guidon},
  \citenamefont {Hutter},\ and\ \citenamefont
  {VandeVondele}}]{guidon_robust_2009}%
  \BibitemOpen
  \bibfield  {author} {\bibinfo {author} {\bibfnamefont {M.}~\bibnamefont
  {Guidon}}, \bibinfo {author} {\bibfnamefont {J.}~\bibnamefont {Hutter}}, \
  and\ \bibinfo {author} {\bibfnamefont {J.}~\bibnamefont {VandeVondele}},\
  }\bibfield  {title} {\enquote {\bibinfo {title} {Robust periodic
  {Hartree--Fock} exchange for large-scale simulations using gaussian basis
  sets},}\ }\href@noop {} {\bibfield  {journal} {\bibinfo  {journal} {J. Chem.
  Theory Comput.}\ }\textbf {\bibinfo {volume} {5}},\ \bibinfo {pages}
  {3010--3021} (\bibinfo {year} {2009})}\BibitemShut {NoStop}%
\bibitem [{\citenamefont {Wu}\ \emph {et~al.}(2009{\natexlab{a}})\citenamefont
  {Wu}, \citenamefont {Selloni},\ and\ \citenamefont {Car}}]{wu_order-n_2009}%
  \BibitemOpen
  \bibfield  {author} {\bibinfo {author} {\bibfnamefont {X.}~\bibnamefont
  {Wu}}, \bibinfo {author} {\bibfnamefont {A.}~\bibnamefont {Selloni}}, \ and\
  \bibinfo {author} {\bibfnamefont {R.}~\bibnamefont {Car}},\ }\bibfield
  {title} {\enquote {\bibinfo {title} {Order-$N$ implementation of exact
  exchange in extended insulating systems},}\ }\href@noop {} {\bibfield
  {journal} {\bibinfo  {journal} {Phys. Rev. B}\ }\textbf {\bibinfo {volume}
  {79}},\ \bibinfo {pages} {085102} (\bibinfo {year}
  {2009}{\natexlab{a}})}\BibitemShut {NoStop}%
\bibitem [{\citenamefont {Gygi}(2009)}]{gygi_compact_2009}%
  \BibitemOpen
  \bibfield  {author} {\bibinfo {author} {\bibfnamefont {F.}~\bibnamefont
  {Gygi}},\ }\bibfield  {title} {\enquote {\bibinfo {title} {Compact
  representations of {Kohn--Sham} invariant subspaces},}\ }\href@noop {}
  {\bibfield  {journal} {\bibinfo  {journal} {Phys. Rev. Lett.}\ }\textbf
  {\bibinfo {volume} {102}},\ \bibinfo {pages} {166406} (\bibinfo {year}
  {2009})}\BibitemShut {NoStop}%
\bibitem [{\citenamefont {Guidon}\ \emph {et~al.}(2010)\citenamefont {Guidon},
  \citenamefont {Hutter},\ and\ \citenamefont
  {VandeVondele}}]{guidon_auxiliary_2010}%
  \BibitemOpen
  \bibfield  {author} {\bibinfo {author} {\bibfnamefont {M.}~\bibnamefont
  {Guidon}}, \bibinfo {author} {\bibfnamefont {J.}~\bibnamefont {Hutter}}, \
  and\ \bibinfo {author} {\bibfnamefont {J.}~\bibnamefont {VandeVondele}},\
  }\bibfield  {title} {\enquote {\bibinfo {title} {Auxiliary density matrix
  methods for {Hartree--Fock} exchange calculations},}\ }\href@noop {}
  {\bibfield  {journal} {\bibinfo  {journal} {J. Chem. Theory Comput.}\
  }\textbf {\bibinfo {volume} {6}},\ \bibinfo {pages} {2348--2364} (\bibinfo
  {year} {2010})}\BibitemShut {NoStop}%
\bibitem [{\citenamefont {Duchemin}\ and\ \citenamefont
  {Gygi}(2010)}]{duchemin_scalable_2010}%
  \BibitemOpen
  \bibfield  {author} {\bibinfo {author} {\bibfnamefont {I.}~\bibnamefont
  {Duchemin}}\ and\ \bibinfo {author} {\bibfnamefont {F.}~\bibnamefont
  {Gygi}},\ }\bibfield  {title} {\enquote {\bibinfo {title} {A scalable and
  accurate algorithm for the computation of {Hartree--Fock} exchange},}\
  }\href@noop {} {\bibfield  {journal} {\bibinfo  {journal} {Comput. Phys.
  Commun.}\ }\textbf {\bibinfo {volume} {181}},\ \bibinfo {pages} {855--860}
  (\bibinfo {year} {2010})}\BibitemShut {NoStop}%
\bibitem [{\citenamefont {Bylaska}\ \emph {et~al.}(2011)\citenamefont
  {Bylaska}, \citenamefont {Tsemekhman}, \citenamefont {Baden}, \citenamefont
  {Weare},\ and\ \citenamefont {Jonsson}}]{bylaska_parallel_2011}%
  \BibitemOpen
  \bibfield  {author} {\bibinfo {author} {\bibfnamefont {E.~J.}\ \bibnamefont
  {Bylaska}}, \bibinfo {author} {\bibfnamefont {K.}~\bibnamefont {Tsemekhman}},
  \bibinfo {author} {\bibfnamefont {S.~B.}\ \bibnamefont {Baden}}, \bibinfo
  {author} {\bibfnamefont {J.~H.}\ \bibnamefont {Weare}}, \ and\ \bibinfo
  {author} {\bibfnamefont {H.}~\bibnamefont {Jonsson}},\ }\bibfield  {title}
  {\enquote {\bibinfo {title} {Parallel implementation of {$\Gamma$}-point
  pseudopotential plane-wave {DFT} with exact exchange},}\ }\href@noop {}
  {\bibfield  {journal} {\bibinfo  {journal} {J. Comput. Chem.}\ }\textbf
  {\bibinfo {volume} {32}},\ \bibinfo {pages} {54--69} (\bibinfo {year}
  {2011})}\BibitemShut {NoStop}%
\bibitem [{\citenamefont {Varini}\ \emph {et~al.}(2013)\citenamefont {Varini},
  \citenamefont {Ceresoli}, \citenamefont {Martin-Samos}, \citenamefont
  {Girotto},\ and\ \citenamefont {Cavazzoni}}]{varini_enhancement_2013}%
  \BibitemOpen
  \bibfield  {author} {\bibinfo {author} {\bibfnamefont {N.}~\bibnamefont
  {Varini}}, \bibinfo {author} {\bibfnamefont {D.}~\bibnamefont {Ceresoli}},
  \bibinfo {author} {\bibfnamefont {L.}~\bibnamefont {Martin-Samos}}, \bibinfo
  {author} {\bibfnamefont {I.}~\bibnamefont {Girotto}}, \ and\ \bibinfo
  {author} {\bibfnamefont {C.}~\bibnamefont {Cavazzoni}},\ }\bibfield  {title}
  {\enquote {\bibinfo {title} {Enhancement of {DFT}-calculations at petascale:
  Nuclear magnetic resonance, hybrid density functional theory and
  {Car--Parrinello} calculations},}\ }\href@noop {} {\bibfield  {journal}
  {\bibinfo  {journal} {Comput. Phys. Commun.}\ }\textbf {\bibinfo {volume}
  {184}},\ \bibinfo {pages} {1827--1833} (\bibinfo {year} {2013})}\BibitemShut
  {NoStop}%
\bibitem [{\citenamefont {Gygi}\ and\ \citenamefont
  {Duchemin}(2013)}]{gygi_efficient_2013}%
  \BibitemOpen
  \bibfield  {author} {\bibinfo {author} {\bibfnamefont {F.}~\bibnamefont
  {Gygi}}\ and\ \bibinfo {author} {\bibfnamefont {I.}~\bibnamefont
  {Duchemin}},\ }\bibfield  {title} {\enquote {\bibinfo {title} {Efficient
  computation of {Hartree}-{Fock} exchange using recursive subspace
  bisection},}\ }\href@noop {} {\bibfield  {journal} {\bibinfo  {journal} {J.
  Chem. Theory Comput.}\ }\textbf {\bibinfo {volume} {9}},\ \bibinfo {pages}
  {582--587} (\bibinfo {year} {2013})}\BibitemShut {NoStop}%
\bibitem [{\citenamefont {{DiStasio Jr.}}\ \emph {et~al.}(2014)\citenamefont
  {{DiStasio Jr.}}, \citenamefont {Santra}, \citenamefont {Li}, \citenamefont
  {Wu},\ and\ \citenamefont {Car}}]{distasio_jr._individual_2014}%
  \BibitemOpen
  \bibfield  {author} {\bibinfo {author} {\bibfnamefont {R.~A.}\ \bibnamefont
  {{DiStasio Jr.}}}, \bibinfo {author} {\bibfnamefont {B.}~\bibnamefont
  {Santra}}, \bibinfo {author} {\bibfnamefont {Z.}~\bibnamefont {Li}}, \bibinfo
  {author} {\bibfnamefont {X.}~\bibnamefont {Wu}}, \ and\ \bibinfo {author}
  {\bibfnamefont {R.}~\bibnamefont {Car}},\ }\bibfield  {title} {\enquote
  {\bibinfo {title} {The individual and collective effects of exact exchange
  and dispersion interactions on the \textit{ab initio} structure of liquid
  water},}\ }\href@noop {} {\bibfield  {journal} {\bibinfo  {journal} {J. Chem.
  Phys.}\ }\textbf {\bibinfo {volume} {141}},\ \bibinfo {pages} {084502}
  (\bibinfo {year} {2014})}\BibitemShut {NoStop}%
\bibitem [{\citenamefont {Dawson}\ and\ \citenamefont
  {Gygi}(2015)}]{dawson_performance_2015}%
  \BibitemOpen
  \bibfield  {author} {\bibinfo {author} {\bibfnamefont {W.}~\bibnamefont
  {Dawson}}\ and\ \bibinfo {author} {\bibfnamefont {F.}~\bibnamefont {Gygi}},\
  }\bibfield  {title} {\enquote {\bibinfo {title} {Performance and accuracy of
  recursive subspace bisection for hybrid {DFT} calculations in inhomogeneous
  systems},}\ }\href@noop {} {\bibfield  {journal} {\bibinfo  {journal} {J.
  Chem. Theory Comput.}\ }\textbf {\bibinfo {volume} {11}},\ \bibinfo {pages}
  {4655--4663} (\bibinfo {year} {2015})}\BibitemShut {NoStop}%
\bibitem [{\citenamefont {Damle}\ \emph {et~al.}(2015)\citenamefont {Damle},
  \citenamefont {Lin},\ and\ \citenamefont {Ying}}]{damle_compressed_2015}%
  \BibitemOpen
  \bibfield  {author} {\bibinfo {author} {\bibfnamefont {A.}~\bibnamefont
  {Damle}}, \bibinfo {author} {\bibfnamefont {L.}~\bibnamefont {Lin}}, \ and\
  \bibinfo {author} {\bibfnamefont {L.}~\bibnamefont {Ying}},\ }\bibfield
  {title} {\enquote {\bibinfo {title} {Compressed representation of {Kohn}--{Sham}
  orbitals via selected columns of the density matrix},}\ }\href@noop {}
  {\bibfield  {journal} {\bibinfo  {journal} {J. Chem. Theory Comput.}\
  }\textbf {\bibinfo {volume} {11}},\ \bibinfo {pages} {1463--1469} (\bibinfo
  {year} {2015})}\BibitemShut {NoStop}%
\bibitem [{\citenamefont {Lin}(2016)}]{lin_adaptively_2016}%
  \BibitemOpen
  \bibfield  {author} {\bibinfo {author} {\bibfnamefont {L.}~\bibnamefont
  {Lin}},\ }\bibfield  {title} {\enquote {\bibinfo {title} {Adaptively
  compressed exchange operator},}\ }\href@noop {} {\bibfield  {journal}
  {\bibinfo  {journal} {J. Chem. Theory Comput.}\ }\textbf {\bibinfo {volume}
  {12}},\ \bibinfo {pages} {2242--2249} (\bibinfo {year} {2016})}\BibitemShut
  {NoStop}%
\bibitem [{\citenamefont {Boffi}\ \emph {et~al.}(2016)\citenamefont {Boffi},
  \citenamefont {Jain},\ and\ \citenamefont {Natan}}]{boffi_efficient_2016}%
  \BibitemOpen
  \bibfield  {author} {\bibinfo {author} {\bibfnamefont {N.~M.}\ \bibnamefont
  {Boffi}}, \bibinfo {author} {\bibfnamefont {M.}~\bibnamefont {Jain}}, \ and\
  \bibinfo {author} {\bibfnamefont {A.}~\bibnamefont {Natan}},\ }\bibfield
  {title} {\enquote {\bibinfo {title} {Efficient computation of the
  {Hartree--Fock} exchange in real-space with projection operators},}\
  }\href@noop {} {\bibfield  {journal} {\bibinfo  {journal} {J. Chem. Theory
  Comput.}\ }\textbf {\bibinfo {volume} {12}},\ \bibinfo {pages} {3614--3622}
  (\bibinfo {year} {2016})}\BibitemShut {NoStop}%
\bibitem [{\citenamefont {Barnes}\ \emph {et~al.}(2017)\citenamefont {Barnes},
  \citenamefont {Kurth}, \citenamefont {Carrier}, \citenamefont {Wichmann},
  \citenamefont {Prendergast}, \citenamefont {Kent},\ and\ \citenamefont
  {Deslippe}}]{barnes_improved_2017}%
  \BibitemOpen
  \bibfield  {author} {\bibinfo {author} {\bibfnamefont {T.~A.}\ \bibnamefont
  {Barnes}}, \bibinfo {author} {\bibfnamefont {T.}~\bibnamefont {Kurth}},
  \bibinfo {author} {\bibfnamefont {P.}~\bibnamefont {Carrier}}, \bibinfo
  {author} {\bibfnamefont {N.}~\bibnamefont {Wichmann}}, \bibinfo {author}
  {\bibfnamefont {D.}~\bibnamefont {Prendergast}}, \bibinfo {author}
  {\bibfnamefont {P.~R.~C.}\ \bibnamefont {Kent}}, \ and\ \bibinfo {author}
  {\bibfnamefont {J.}~\bibnamefont {Deslippe}},\ }\bibfield  {title} {\enquote
  {\bibinfo {title} {Improved treatment of exact exchange in \texttt{Quantum
  ESPRESSO}},}\ }\href@noop {} {\bibfield  {journal} {\bibinfo  {journal}
  {Comput. Phys. Commun.}\ }\textbf {\bibinfo {volume} {214}},\ \bibinfo
  {pages} {52--58} (\bibinfo {year} {2017})}\BibitemShut {NoStop}%
\bibitem [{\citenamefont {Hu}\ \emph {et~al.}(2017{\natexlab{a}})\citenamefont
  {Hu}, \citenamefont {Lin},\ and\ \citenamefont
  {Yang}}]{hu_interpolative_2017}%
  \BibitemOpen
  \bibfield  {author} {\bibinfo {author} {\bibfnamefont {W.}~\bibnamefont
  {Hu}}, \bibinfo {author} {\bibfnamefont {L.}~\bibnamefont {Lin}}, \ and\
  \bibinfo {author} {\bibfnamefont {C.}~\bibnamefont {Yang}},\ }\bibfield
  {title} {\enquote {\bibinfo {title} {Interpolative separable density fitting
  decomposition for accelerating hybrid density functional calculations with
  applications to defects in silicon},}\ }\href@noop {} {\bibfield  {journal}
  {\bibinfo  {journal} {J. Chem. Theory Comput.}\ }\textbf {\bibinfo {volume}
  {13}},\ \bibinfo {pages} {5420--5431} (\bibinfo {year}
  {2017}{\natexlab{a}})}\BibitemShut {NoStop}%
\bibitem [{\citenamefont {Hu}\ \emph {et~al.}(2017{\natexlab{b}})\citenamefont
  {Hu}, \citenamefont {Lin},\ and\ \citenamefont {Yang}}]{hu_projected_2017}%
  \BibitemOpen
  \bibfield  {author} {\bibinfo {author} {\bibfnamefont {W.}~\bibnamefont
  {Hu}}, \bibinfo {author} {\bibfnamefont {L.}~\bibnamefont {Lin}}, \ and\
  \bibinfo {author} {\bibfnamefont {C.}~\bibnamefont {Yang}},\ }\bibfield
  {title} {\enquote {\bibinfo {title} {Projected commutator {DIIS} method for
  accelerating hybrid functional electronic structure calculations},}\
  }\href@noop {} {\bibfield  {journal} {\bibinfo  {journal} {J. Chem. Theory
  Comput.}\ }\textbf {\bibinfo {volume} {13}},\ \bibinfo {pages} {5458--5467}
  (\bibinfo {year} {2017}{\natexlab{b}})}\BibitemShut {NoStop}%
\bibitem [{\citenamefont {Mountjoy}\ \emph {et~al.}(2017)\citenamefont
  {Mountjoy}, \citenamefont {Todd},\ and\ \citenamefont
  {Mosey}}]{mountjoy_exact_2017}%
  \BibitemOpen
  \bibfield  {author} {\bibinfo {author} {\bibfnamefont {J.}~\bibnamefont
  {Mountjoy}}, \bibinfo {author} {\bibfnamefont {M.}~\bibnamefont {Todd}}, \
  and\ \bibinfo {author} {\bibfnamefont {N.~J.}\ \bibnamefont {Mosey}},\
  }\bibfield  {title} {\enquote {\bibinfo {title} {Exact exchange with
  non-orthogonal generalized {Wannier} functions},}\ }\href@noop {} {\bibfield
  {journal} {\bibinfo  {journal} {J. Chem. Phys.}\ }\textbf {\bibinfo {volume}
  {146}},\ \bibinfo {pages} {104108} (\bibinfo {year} {2017})}\BibitemShut
  {NoStop}%
\bibitem [{\citenamefont {Dong}\ \emph {et~al.}(2018)\citenamefont {Dong},
  \citenamefont {Hu},\ and\ \citenamefont {Lin}}]{dong_interpolative_2018}%
  \BibitemOpen
  \bibfield  {author} {\bibinfo {author} {\bibfnamefont {K.}~\bibnamefont
  {Dong}}, \bibinfo {author} {\bibfnamefont {W.}~\bibnamefont {Hu}}, \ and\
  \bibinfo {author} {\bibfnamefont {L.}~\bibnamefont {Lin}},\ }\bibfield
  {title} {\enquote {\bibinfo {title} {Interpolative separable density fitting
  through centroidal voronoi tessellation with applications to hybrid
  functional electronic structure calculations},}\ }\href@noop {} {\bibfield
  {journal} {\bibinfo  {journal} {J. Chem. Theory Comput.}\ }\textbf {\bibinfo
  {volume} {14}},\ \bibinfo {pages} {1311--1320} (\bibinfo {year}
  {2018})}\BibitemShut {NoStop}%
\bibitem [{\citenamefont {Carnimeo}\ \emph {et~al.}(2019)\citenamefont
  {Carnimeo}, \citenamefont {Baroni},\ and\ \citenamefont
  {Giannozzi}}]{carnimeo_fast_2018}%
  \BibitemOpen
  \bibfield  {author} {\bibinfo {author} {\bibfnamefont {I.}~\bibnamefont
  {Carnimeo}}, \bibinfo {author} {\bibfnamefont {S.}~\bibnamefont {Baroni}}, \
  and\ \bibinfo {author} {\bibfnamefont {P.}~\bibnamefont {Giannozzi}},\
  }\bibfield  {title} {\enquote {\bibinfo {title} {Fast hybrid
  density-functional computations using plane-wave basis sets},}\ }\href@noop
  {} {\bibfield  {journal} {\bibinfo  {journal} {Electron. Struct.}\ }\textbf
  {\bibinfo {volume} {1}},\ \bibinfo {pages} {015009} (\bibinfo {year}
  {2019})}\BibitemShut {NoStop}%
\bibitem [{\citenamefont {Mandal}\ \emph {et~al.}(2018)\citenamefont {Mandal},
  \citenamefont {Debnath}, \citenamefont {Meyer},\ and\ \citenamefont
  {Nair}}]{mandal_enhanced_2018}%
  \BibitemOpen
  \bibfield  {author} {\bibinfo {author} {\bibfnamefont {S.}~\bibnamefont
  {Mandal}}, \bibinfo {author} {\bibfnamefont {J.}~\bibnamefont {Debnath}},
  \bibinfo {author} {\bibfnamefont {B.}~\bibnamefont {Meyer}}, \ and\ \bibinfo
  {author} {\bibfnamefont {N.~N.}\ \bibnamefont {Nair}},\ }\bibfield  {title}
  {\enquote {\bibinfo {title} {Enhanced sampling and free energy calculations
  with hybrid functionals and plane waves for chemical reactions},}\ }\href
  {\doibase 10.1063/1.5049700} {\bibfield  {journal} {\bibinfo  {journal} {J.
  Chem. Phys.}\ }\textbf {\bibinfo {volume} {149}},\ \bibinfo {pages} {144113}
  (\bibinfo {year} {2018})}\BibitemShut {NoStop}%
\bibitem [{\citenamefont {Mandal}\ and\ \citenamefont
  {Nair}(2019)}]{mandal_speeding-up_2019}%
  \BibitemOpen
  \bibfield  {author} {\bibinfo {author} {\bibfnamefont {S.}~\bibnamefont
  {Mandal}}\ and\ \bibinfo {author} {\bibfnamefont {N.~N.}\ \bibnamefont
  {Nair}},\ }\bibfield  {title} {\enquote {\bibinfo {title} {Speeding-up ab
  initio molecular dynamics with hybrid functionals using adaptively compressed
  exchange operator based multiple timestepping},}\ }\href@noop {} {\bibfield
  {journal} {\bibinfo  {journal} {J. Chem. Phys.}\ }\textbf {\bibinfo {volume}
  {151}},\ \bibinfo {pages} {151102} (\bibinfo {year} {2019})}\BibitemShut
  {NoStop}%
\bibitem [{\citenamefont {Mandal}\ and\ \citenamefont
  {Nair}(2020)}]{mandal_efficient_2020}%
  \BibitemOpen
  \bibfield  {author} {\bibinfo {author} {\bibfnamefont {S.}~\bibnamefont
  {Mandal}}\ and\ \bibinfo {author} {\bibfnamefont {N.~N.}\ \bibnamefont
  {Nair}},\ }\bibfield  {title} {\enquote {\bibinfo {title} {Efficient
  computation of free energy surfaces of chemical reactions using ab initio
  molecular dynamics with hybrid functionals and plane waves},}\ }\href@noop {}
  {\bibfield  {journal} {\bibinfo  {journal} {J. Comput. Chem.}\ }\textbf
  {\bibinfo {volume} {41}},\ \bibinfo {pages} {1790--1797} (\bibinfo {year}
  {2020})}\BibitemShut {NoStop}%
\bibitem [{\citenamefont {Ko}\ \emph {et~al.}(2020)\citenamefont {Ko},
  \citenamefont {Jia}, \citenamefont {Santra}, \citenamefont {Wu},
  \citenamefont {Car},\ and\ \citenamefont {{DiStasio Jr.}}}]{paper1}%
  \BibitemOpen
  \bibfield  {author} {\bibinfo {author} {\bibfnamefont {H.-Y.}\ \bibnamefont
  {Ko}}, \bibinfo {author} {\bibfnamefont {J.}~\bibnamefont {Jia}}, \bibinfo
  {author} {\bibfnamefont {B.}~\bibnamefont {Santra}}, \bibinfo {author}
  {\bibfnamefont {X.}~\bibnamefont {Wu}}, \bibinfo {author} {\bibfnamefont
  {R.}~\bibnamefont {Car}}, \ and\ \bibinfo {author} {\bibfnamefont {R.~A.}\
  \bibnamefont {{DiStasio Jr.}}},\ }\bibfield  {title} {\enquote {\bibinfo
  {title} {Enabling large-scale condensed-phase hybrid density functional
  theory based \textit{ab initio} molecular dynamics. {1.} {Theory}, algorithm,
  and performance},}\ }\href@noop {} {\bibfield  {journal} {\bibinfo  {journal}
  {J. Chem. Theory Comput.}\ }\textbf {\bibinfo {volume} {16}},\ \bibinfo
  {pages} {3757--3785} (\bibinfo {year} {2020})}\BibitemShut {NoStop}%
\bibitem [{\citenamefont {Mandal}\ \emph {et~al.}(2021)\citenamefont {Mandal},
  \citenamefont {Thakkur},\ and\ \citenamefont {Nair}}]{mandal_achieving_2021}%
  \BibitemOpen
  \bibfield  {author} {\bibinfo {author} {\bibfnamefont {S.}~\bibnamefont
  {Mandal}}, \bibinfo {author} {\bibfnamefont {V.}~\bibnamefont {Thakkur}}, \
  and\ \bibinfo {author} {\bibfnamefont {N.~N.}\ \bibnamefont {Nair}},\
  }\bibfield  {title} {\enquote {\bibinfo {title} {Achieving an order of
  magnitude speedup in hybrid-functional- and plane-wave-based \textit{ab
  initio} molecular dynamics: Applications to proton-transfer reactions in
  enzymes and in solution},}\ }\href@noop {} {\bibfield  {journal} {\bibinfo
  {journal} {J. Chem. Theory Comput.}\ }\textbf {\bibinfo {volume} {17}},\
  \bibinfo {pages} {2244--2255} (\bibinfo {year} {2021})}\BibitemShut {NoStop}%
\bibitem [{\citenamefont {Ko}\ \emph {et~al.}(2021)\citenamefont {Ko},
  \citenamefont {Santra},\ and\ \citenamefont {{DiStasio Jr.}}}]{paper2}%
  \BibitemOpen
  \bibfield  {author} {\bibinfo {author} {\bibfnamefont {H.-Y.}\ \bibnamefont
  {Ko}}, \bibinfo {author} {\bibfnamefont {B.}~\bibnamefont {Santra}}, \ and\
  \bibinfo {author} {\bibfnamefont {R.~A.}\ \bibnamefont {{DiStasio Jr.}}},\
  }\bibfield  {title} {\enquote {\bibinfo {title} {Enabling large-scale
  condensed-phase hybrid density functional theory-based \textit{ab initio}
  molecular dynamics. {2.} {Extensions} to the isobaric--isoenthalpic and
  isobaric--isothermal ensembles},}\ }\href@noop {} {\bibfield  {journal}
  {\bibinfo  {journal} {J. Chem. Theory Comput.}\ }\textbf {\bibinfo {volume}
  {17}},\ \bibinfo {pages} {7789--7813} (\bibinfo {year} {2021})}\BibitemShut
  {NoStop}%
\bibitem [{\citenamefont {Marzari}\ and\ \citenamefont
  {Vanderbilt}(1997)}]{marzari_maximally_1997}%
  \BibitemOpen
  \bibfield  {author} {\bibinfo {author} {\bibfnamefont {N.}~\bibnamefont
  {Marzari}}\ and\ \bibinfo {author} {\bibfnamefont {D.}~\bibnamefont
  {Vanderbilt}},\ }\bibfield  {title} {\enquote {\bibinfo {title} {Maximally
  localized generalized {Wannier} functions for composite energy bands},}\
  }\href@noop {} {\bibfield  {journal} {\bibinfo  {journal} {Phys. Rev. B}\
  }\textbf {\bibinfo {volume} {56}},\ \bibinfo {pages} {12847--12865} (\bibinfo
  {year} {1997})}\BibitemShut {NoStop}%
\bibitem [{\citenamefont {Marzari}\ \emph {et~al.}(2012)\citenamefont
  {Marzari}, \citenamefont {Mostofi}, \citenamefont {Yates}, \citenamefont
  {Souza},\ and\ \citenamefont {Vanderbilt}}]{marzari_maximally_2012}%
  \BibitemOpen
  \bibfield  {author} {\bibinfo {author} {\bibfnamefont {N.}~\bibnamefont
  {Marzari}}, \bibinfo {author} {\bibfnamefont {A.~A.}\ \bibnamefont
  {Mostofi}}, \bibinfo {author} {\bibfnamefont {J.~R.}\ \bibnamefont {Yates}},
  \bibinfo {author} {\bibfnamefont {I.}~\bibnamefont {Souza}}, \ and\ \bibinfo
  {author} {\bibfnamefont {D.}~\bibnamefont {Vanderbilt}},\ }\bibfield  {title}
  {\enquote {\bibinfo {title} {Maximally localized {Wannier} functions:
  {Theory} and applications},}\ }\href@noop {} {\bibfield  {journal} {\bibinfo
  {journal} {Rev. Mod. Phys.}\ }\textbf {\bibinfo {volume} {84}},\ \bibinfo
  {pages} {1419--1475} (\bibinfo {year} {2012})}\BibitemShut {NoStop}%
\bibitem [{\citenamefont {Giannozzi}\ \emph {et~al.}(2017)\citenamefont
  {Giannozzi}, \citenamefont {Andreussi}, \citenamefont {Brumme}, \citenamefont
  {Bunau}, \citenamefont {Nardelli}, \citenamefont {Calandra}, \citenamefont
  {Car}, \citenamefont {Cavazzoni}, \citenamefont {Ceresoli}, \citenamefont
  {Cococcioni}, \citenamefont {Colonna}, \citenamefont {Carnimeo},
  \citenamefont {Corso}, \citenamefont {{de Gironcoli}}, \citenamefont
  {Delugas}, \citenamefont {{DiStasio Jr.}}, \citenamefont {Ferretti},
  \citenamefont {Floris}, \citenamefont {Fratesi}, \citenamefont {Fugallo},
  \citenamefont {Gebauer}, \citenamefont {Gerstmann}, \citenamefont {Giustino},
  \citenamefont {Gorni}, \citenamefont {Jia}, \citenamefont {Kawamura},
  \citenamefont {Ko}, \citenamefont {Kokalj}, \citenamefont {K{\"u}{\c
  c}{\"u}kbenli}, \citenamefont {Lazzeri}, \citenamefont {Marsili},
  \citenamefont {Marzari}, \citenamefont {Mauri}, \citenamefont {Nguyen},
  \citenamefont {Nguyen}, \citenamefont {{Otero-de-la-Roza}}, \citenamefont
  {Paulatto}, \citenamefont {Ponc{\'e}}, \citenamefont {Rocca}, \citenamefont
  {Sabatini}, \citenamefont {Santra}, \citenamefont {Schlipf}, \citenamefont
  {Seitsonen}, \citenamefont {Smogunov}, \citenamefont {Timrov}, \citenamefont
  {Thonhauser}, \citenamefont {Umari}, \citenamefont {Vast}, \citenamefont
  {Wu},\ and\ \citenamefont {Baroni}}]{giannozzi_advanced_2017}%
  \BibitemOpen
  \bibfield  {author} {\bibinfo {author} {\bibfnamefont {P.}~\bibnamefont
  {Giannozzi}}, \bibinfo {author} {\bibfnamefont {O.}~\bibnamefont
  {Andreussi}}, \bibinfo {author} {\bibfnamefont {T.}~\bibnamefont {Brumme}},
  \bibinfo {author} {\bibfnamefont {O.}~\bibnamefont {Bunau}}, \bibinfo
  {author} {\bibfnamefont {M.~B.}\ \bibnamefont {Nardelli}}, \bibinfo {author}
  {\bibfnamefont {M.}~\bibnamefont {Calandra}}, \bibinfo {author}
  {\bibfnamefont {R.}~\bibnamefont {Car}}, \bibinfo {author} {\bibfnamefont
  {C.}~\bibnamefont {Cavazzoni}}, \bibinfo {author} {\bibfnamefont
  {D.}~\bibnamefont {Ceresoli}}, \bibinfo {author} {\bibfnamefont
  {M.}~\bibnamefont {Cococcioni}}, \bibinfo {author} {\bibfnamefont
  {N.}~\bibnamefont {Colonna}}, \bibinfo {author} {\bibfnamefont
  {I.}~\bibnamefont {Carnimeo}}, \bibinfo {author} {\bibfnamefont {A.~D.}\
  \bibnamefont {Corso}}, \bibinfo {author} {\bibfnamefont {S.}~\bibnamefont
  {{de Gironcoli}}}, \bibinfo {author} {\bibfnamefont {P.}~\bibnamefont
  {Delugas}}, \bibinfo {author} {\bibfnamefont {R.~A.}\ \bibnamefont {{DiStasio
  Jr.}}}, \bibinfo {author} {\bibfnamefont {A.}~\bibnamefont {Ferretti}},
  \bibinfo {author} {\bibfnamefont {A.}~\bibnamefont {Floris}}, \bibinfo
  {author} {\bibfnamefont {G.}~\bibnamefont {Fratesi}}, \bibinfo {author}
  {\bibfnamefont {G.}~\bibnamefont {Fugallo}}, \bibinfo {author} {\bibfnamefont
  {R.}~\bibnamefont {Gebauer}}, \bibinfo {author} {\bibfnamefont
  {U.}~\bibnamefont {Gerstmann}}, \bibinfo {author} {\bibfnamefont
  {F.}~\bibnamefont {Giustino}}, \bibinfo {author} {\bibfnamefont
  {T.}~\bibnamefont {Gorni}}, \bibinfo {author} {\bibfnamefont
  {J.}~\bibnamefont {Jia}}, \bibinfo {author} {\bibfnamefont {M.}~\bibnamefont
  {Kawamura}}, \bibinfo {author} {\bibfnamefont {H.-Y.}\ \bibnamefont {Ko}},
  \bibinfo {author} {\bibfnamefont {A.}~\bibnamefont {Kokalj}}, \bibinfo
  {author} {\bibfnamefont {E.}~\bibnamefont {K{\"u}{\c c}{\"u}kbenli}},
  \bibinfo {author} {\bibfnamefont {M.}~\bibnamefont {Lazzeri}}, \bibinfo
  {author} {\bibfnamefont {M.}~\bibnamefont {Marsili}}, \bibinfo {author}
  {\bibfnamefont {N.}~\bibnamefont {Marzari}}, \bibinfo {author} {\bibfnamefont
  {F.}~\bibnamefont {Mauri}}, \bibinfo {author} {\bibfnamefont {N.~L.}\
  \bibnamefont {Nguyen}}, \bibinfo {author} {\bibfnamefont {H.-V.}\
  \bibnamefont {Nguyen}}, \bibinfo {author} {\bibfnamefont {A.}~\bibnamefont
  {{Otero-de-la-Roza}}}, \bibinfo {author} {\bibfnamefont {L.}~\bibnamefont
  {Paulatto}}, \bibinfo {author} {\bibfnamefont {S.}~\bibnamefont {Ponc{\'e}}},
  \bibinfo {author} {\bibfnamefont {D.}~\bibnamefont {Rocca}}, \bibinfo
  {author} {\bibfnamefont {R.}~\bibnamefont {Sabatini}}, \bibinfo {author}
  {\bibfnamefont {B.}~\bibnamefont {Santra}}, \bibinfo {author} {\bibfnamefont
  {M.}~\bibnamefont {Schlipf}}, \bibinfo {author} {\bibfnamefont {A.~P.}\
  \bibnamefont {Seitsonen}}, \bibinfo {author} {\bibfnamefont {A.}~\bibnamefont
  {Smogunov}}, \bibinfo {author} {\bibfnamefont {I.}~\bibnamefont {Timrov}},
  \bibinfo {author} {\bibfnamefont {T.}~\bibnamefont {Thonhauser}}, \bibinfo
  {author} {\bibfnamefont {P.}~\bibnamefont {Umari}}, \bibinfo {author}
  {\bibfnamefont {N.}~\bibnamefont {Vast}}, \bibinfo {author} {\bibfnamefont
  {X.}~\bibnamefont {Wu}}, \ and\ \bibinfo {author} {\bibfnamefont
  {S.}~\bibnamefont {Baroni}},\ }\bibfield  {title} {\enquote {\bibinfo {title}
  {Advanced capabilities for materials modelling with \texttt{Quantum
  ESPRESSO}},}\ }\href@noop {} {\bibfield  {journal} {\bibinfo  {journal} {J.
  Phys.: Condens. Matter}\ }\textbf {\bibinfo {volume} {29}},\ \bibinfo {pages}
  {465901} (\bibinfo {year} {2017})}\BibitemShut {NoStop}%
\bibitem [{\citenamefont {Sharma}\ \emph {et~al.}(2003)\citenamefont {Sharma},
  \citenamefont {Wu},\ and\ \citenamefont {Car}}]{sharma_ab_2003}%
  \BibitemOpen
  \bibfield  {author} {\bibinfo {author} {\bibfnamefont {M.}~\bibnamefont
  {Sharma}}, \bibinfo {author} {\bibfnamefont {Y.}~\bibnamefont {Wu}}, \ and\
  \bibinfo {author} {\bibfnamefont {R.}~\bibnamefont {Car}},\ }\bibfield
  {title} {\enquote {\bibinfo {title} {\textit{Ab initio} molecular dynamics
  with maximally localized {Wannier} functions},}\ }\href@noop {} {\bibfield
  {journal} {\bibinfo  {journal} {Int. J. Quantum Chem.}\ }\textbf {\bibinfo
  {volume} {95}},\ \bibinfo {pages} {821--829} (\bibinfo {year}
  {2003})}\BibitemShut {NoStop}%
\bibitem [{\citenamefont {Wu}\ \emph {et~al.}(2009{\natexlab{b}})\citenamefont
  {Wu}, \citenamefont {Walter}, \citenamefont {Rappe}, \citenamefont {Car},\
  and\ \citenamefont {Selloni}}]{wu_hybrid_2009}%
  \BibitemOpen
  \bibfield  {author} {\bibinfo {author} {\bibfnamefont {X.}~\bibnamefont
  {Wu}}, \bibinfo {author} {\bibfnamefont {E.~J.}\ \bibnamefont {Walter}},
  \bibinfo {author} {\bibfnamefont {A.~M.}\ \bibnamefont {Rappe}}, \bibinfo
  {author} {\bibfnamefont {R.}~\bibnamefont {Car}}, \ and\ \bibinfo {author}
  {\bibfnamefont {A.}~\bibnamefont {Selloni}},\ }\bibfield  {title} {\enquote
  {\bibinfo {title} {Hybrid density functional calculations of the band gap of
  Ga$_{x}$In$_{1-x}$N},}\ }\href@noop {} {\bibfield  {journal} {\bibinfo
  {journal} {Phys. Rev. B}\ }\textbf {\bibinfo {volume} {80}},\ \bibinfo
  {pages} {115201} (\bibinfo {year} {2009}{\natexlab{b}})}\BibitemShut
  {NoStop}%
\bibitem [{\citenamefont {Chen}\ \emph {et~al.}(2011)\citenamefont {Chen},
  \citenamefont {Wu},\ and\ \citenamefont {Selloni}}]{chen_electronic_2011}%
  \BibitemOpen
  \bibfield  {author} {\bibinfo {author} {\bibfnamefont {J.}~\bibnamefont
  {Chen}}, \bibinfo {author} {\bibfnamefont {X.}~\bibnamefont {Wu}}, \ and\
  \bibinfo {author} {\bibfnamefont {A.}~\bibnamefont {Selloni}},\ }\bibfield
  {title} {\enquote {\bibinfo {title} {Electronic structure and bonding
  properties of cobalt oxide in the spinel structure},}\ }\href@noop {}
  {\bibfield  {journal} {\bibinfo  {journal} {Phys. Rev. B}\ }\textbf {\bibinfo
  {volume} {83}},\ \bibinfo {pages} {245204} (\bibinfo {year}
  {2011})}\BibitemShut {NoStop}%
\bibitem [{\citenamefont {Santra}\ \emph {et~al.}(2015)\citenamefont {Santra},
  \citenamefont {{DiStasio Jr.}}, \citenamefont {Martelli},\ and\ \citenamefont
  {Car}}]{santra_local_2015}%
  \BibitemOpen
  \bibfield  {author} {\bibinfo {author} {\bibfnamefont {B.}~\bibnamefont
  {Santra}}, \bibinfo {author} {\bibfnamefont {R.~A.}\ \bibnamefont {{DiStasio
  Jr.}}}, \bibinfo {author} {\bibfnamefont {F.}~\bibnamefont {Martelli}}, \
  and\ \bibinfo {author} {\bibfnamefont {R.}~\bibnamefont {Car}},\ }\bibfield
  {title} {\enquote {\bibinfo {title} {Local structure analysis in \textit{ab
  initio} liquid water},}\ }\href@noop {} {\bibfield  {journal} {\bibinfo
  {journal} {Mol. Phys.}\ }\textbf {\bibinfo {volume} {113}},\ \bibinfo {pages}
  {2829--2841} (\bibinfo {year} {2015})}\BibitemShut {NoStop}%
\bibitem [{\citenamefont {Bankura}\ \emph {et~al.}(2015)\citenamefont
  {Bankura}, \citenamefont {Santra}, \citenamefont {{DiStasio Jr.}},
  \citenamefont {Swartz}, \citenamefont {Klein},\ and\ \citenamefont
  {Wu}}]{bankura_systematic_2015}%
  \BibitemOpen
  \bibfield  {author} {\bibinfo {author} {\bibfnamefont {A.}~\bibnamefont
  {Bankura}}, \bibinfo {author} {\bibfnamefont {B.}~\bibnamefont {Santra}},
  \bibinfo {author} {\bibfnamefont {R.~A.}\ \bibnamefont {{DiStasio Jr.}}},
  \bibinfo {author} {\bibfnamefont {C.~W.}\ \bibnamefont {Swartz}}, \bibinfo
  {author} {\bibfnamefont {M.~L.}\ \bibnamefont {Klein}}, \ and\ \bibinfo
  {author} {\bibfnamefont {X.}~\bibnamefont {Wu}},\ }\bibfield  {title}
  {\enquote {\bibinfo {title} {A systematic study of chloride ion solvation in
  water using van der {Waals} inclusive hybrid density functional theory},}\
  }\href@noop {} {\bibfield  {journal} {\bibinfo  {journal} {Mol. Phys.}\
  }\textbf {\bibinfo {volume} {113}},\ \bibinfo {pages} {2842--2854} (\bibinfo
  {year} {2015})}\BibitemShut {NoStop}%
\bibitem [{\citenamefont {Chen}\ \emph {et~al.}(2018)\citenamefont {Chen},
  \citenamefont {Zheng}, \citenamefont {Santra}, \citenamefont {Ko},
  \citenamefont {{DiStasio Jr.}}, \citenamefont {Klein}, \citenamefont {Car},\
  and\ \citenamefont {Wu}}]{chen_hydroxide_2018}%
  \BibitemOpen
  \bibfield  {author} {\bibinfo {author} {\bibfnamefont {M.}~\bibnamefont
  {Chen}}, \bibinfo {author} {\bibfnamefont {L.}~\bibnamefont {Zheng}},
  \bibinfo {author} {\bibfnamefont {B.}~\bibnamefont {Santra}}, \bibinfo
  {author} {\bibfnamefont {H.-Y.}\ \bibnamefont {Ko}}, \bibinfo {author}
  {\bibfnamefont {R.~A.}\ \bibnamefont {{DiStasio Jr.}}}, \bibinfo {author}
  {\bibfnamefont {M.~L.}\ \bibnamefont {Klein}}, \bibinfo {author}
  {\bibfnamefont {R.}~\bibnamefont {Car}}, \ and\ \bibinfo {author}
  {\bibfnamefont {X.}~\bibnamefont {Wu}},\ }\bibfield  {title} {\enquote
  {\bibinfo {title} {Hydroxide diffuses slower than hydronium in water because
  its solvated structure inhibits correlated proton transfer},}\ }\href@noop {}
  {\bibfield  {journal} {\bibinfo  {journal} {Nat. Chem.}\ }\textbf {\bibinfo
  {volume} {10}},\ \bibinfo {pages} {413--419} (\bibinfo {year}
  {2018})}\BibitemShut {NoStop}%
\bibitem [{\citenamefont {Ko}\ \emph {et~al.}(2018)\citenamefont {Ko},
  \citenamefont {{DiStasio Jr.}}, \citenamefont {Santra},\ and\ \citenamefont
  {Car}}]{ko_thermal_2018}%
  \BibitemOpen
  \bibfield  {author} {\bibinfo {author} {\bibfnamefont {H.-Y.}\ \bibnamefont
  {Ko}}, \bibinfo {author} {\bibfnamefont {R.~A.}\ \bibnamefont {{DiStasio
  Jr.}}}, \bibinfo {author} {\bibfnamefont {B.}~\bibnamefont {Santra}}, \ and\
  \bibinfo {author} {\bibfnamefont {R.}~\bibnamefont {Car}},\ }\bibfield
  {title} {\enquote {\bibinfo {title} {Thermal expansion in dispersion-bound
  molecular crystals},}\ }\href@noop {} {\bibfield  {journal} {\bibinfo
  {journal} {Phys. Rev. Materials}\ }\textbf {\bibinfo {volume} {2}},\ \bibinfo
  {pages} {055603} (\bibinfo {year} {2018})}\BibitemShut {NoStop}%
\bibitem [{\citenamefont {Ko}\ \emph {et~al.}(2019)\citenamefont {Ko},
  \citenamefont {Zhang}, \citenamefont {Santra}, \citenamefont {Wang},
  \citenamefont {E}, \citenamefont {{DiStasio Jr.}},\ and\ \citenamefont
  {Car}}]{ko_isotope_2019}%
  \BibitemOpen
  \bibfield  {author} {\bibinfo {author} {\bibfnamefont {H.-Y.}\ \bibnamefont
  {Ko}}, \bibinfo {author} {\bibfnamefont {L.}~\bibnamefont {Zhang}}, \bibinfo
  {author} {\bibfnamefont {B.}~\bibnamefont {Santra}}, \bibinfo {author}
  {\bibfnamefont {H.}~\bibnamefont {Wang}}, \bibinfo {author} {\bibfnamefont
  {W.}~\bibnamefont {E}}, \bibinfo {author} {\bibfnamefont {R.~A.}\
  \bibnamefont {{DiStasio Jr.}}}, \ and\ \bibinfo {author} {\bibfnamefont
  {R.}~\bibnamefont {Car}},\ }\bibfield  {title} {\enquote {\bibinfo {title}
  {Isotope effects in liquid water via deep potential molecular dynamics},}\
  }\href@noop {} {\bibfield  {journal} {\bibinfo  {journal} {Mol. Phys.}\
  }\textbf {\bibinfo {volume} {117}},\ \bibinfo {pages} {3269--3281} (\bibinfo
  {year} {2019})}\BibitemShut {NoStop}%
\bibitem [{\citenamefont {Car}\ and\ \citenamefont
  {Parrinello}(1985)}]{car_unified_1985}%
  \BibitemOpen
  \bibfield  {author} {\bibinfo {author} {\bibfnamefont {R.}~\bibnamefont
  {Car}}\ and\ \bibinfo {author} {\bibfnamefont {M.}~\bibnamefont
  {Parrinello}},\ }\bibfield  {title} {\enquote {\bibinfo {title} {Unified
  approach for molecular dynamics and density-functional theory},}\ }\href@noop
  {} {\bibfield  {journal} {\bibinfo  {journal} {Phys. Rev. Lett.}\ }\textbf
  {\bibinfo {volume} {55}},\ \bibinfo {pages} {2471--2474} (\bibinfo {year}
  {1985})}\BibitemShut {NoStop}%
\bibitem [{\citenamefont {Sparrow}\ \emph {et~al.}(prep)\citenamefont
  {Sparrow}, \citenamefont {Ko}, \citenamefont {Zhang},\ and\ \citenamefont
  {{DiStasio Jr.}}}]{paper3}%
  \BibitemOpen
  \bibfield  {author} {\bibinfo {author} {\bibfnamefont {Z.~M.}\ \bibnamefont
  {Sparrow}}, \bibinfo {author} {\bibfnamefont {H.-Y.}\ \bibnamefont {Ko}},
  \bibinfo {author} {\bibfnamefont {J.}\ \bibnamefont {Zhang}}, \ and\
  \bibinfo {author} {\bibfnamefont {R.~A.}\ \bibnamefont {{DiStasio Jr.}}},\
  }\bibfield  {title} {\enquote {\bibinfo {title} {Enabling large-scale
  condensed-phase hybrid density functional theory-based \textit{ab initio}
  molecular dynamics. 3. {Extensions} to heterogeneous systems},}\ }\href@noop
  {} {\  (\bibinfo {year} {in prep})}\BibitemShut {NoStop}%
\bibitem [{\citenamefont {Tuckerman}\ \emph {et~al.}(1992)\citenamefont
  {Tuckerman}, \citenamefont {Berne},\ and\ \citenamefont
  {Martyna}}]{tuckerman_reversible_1992}%
  \BibitemOpen
  \bibfield  {author} {\bibinfo {author} {\bibfnamefont {M.}~\bibnamefont
  {Tuckerman}}, \bibinfo {author} {\bibfnamefont {B.~J.}\ \bibnamefont
  {Berne}}, \ and\ \bibinfo {author} {\bibfnamefont {G.~J.}\ \bibnamefont
  {Martyna}},\ }\bibfield  {title} {\enquote {\bibinfo {title} {Reversible
  multiple time scale molecular dynamics},}\ }\href@noop {} {\bibfield
  {journal} {\bibinfo  {journal} {J. Chem. Phys.}\ }\textbf {\bibinfo {volume}
  {97}},\ \bibinfo {pages} {1990--2001} (\bibinfo {year} {1992})}\BibitemShut
  {NoStop}%
\bibitem [{\citenamefont {Perdew}\ \emph
  {et~al.}(1996{\natexlab{a}})\citenamefont {Perdew}, \citenamefont
  {Ernzerhof},\ and\ \citenamefont {Burke}}]{perdew_rationale_1996}%
  \BibitemOpen
  \bibfield  {author} {\bibinfo {author} {\bibfnamefont {J.~P.}\ \bibnamefont
  {Perdew}}, \bibinfo {author} {\bibfnamefont {M.}~\bibnamefont {Ernzerhof}}, \
  and\ \bibinfo {author} {\bibfnamefont {K.}~\bibnamefont {Burke}},\ }\bibfield
   {title} {\enquote {\bibinfo {title} {Rationale for mixing exact exchange
  with density functional approximations},}\ }\href@noop {} {\bibfield
  {journal} {\bibinfo  {journal} {J. Chem. Phys.}\ }\textbf {\bibinfo {volume}
  {105}},\ \bibinfo {pages} {9982--9985} (\bibinfo {year}
  {1996}{\natexlab{a}})}\BibitemShut {NoStop}%
\bibitem [{\citenamefont {Adamo}\ and\ \citenamefont
  {Barone}(1999)}]{adamo_toward_1999}%
  \BibitemOpen
  \bibfield  {author} {\bibinfo {author} {\bibfnamefont {C.}~\bibnamefont
  {Adamo}}\ and\ \bibinfo {author} {\bibfnamefont {V.}~\bibnamefont {Barone}},\
  }\bibfield  {title} {\enquote {\bibinfo {title} {Toward reliable density
  functional methods without adjustable parameters: The {PBE0} model},}\
  }\href@noop {} {\bibfield  {journal} {\bibinfo  {journal} {J. Chem. Phys.}\
  }\textbf {\bibinfo {volume} {110}},\ \bibinfo {pages} {6158--6170} (\bibinfo
  {year} {1999})}\BibitemShut {NoStop}%
\bibitem [{\citenamefont {Damle}\ \emph {et~al.}(2017)\citenamefont {Damle},
  \citenamefont {Lin},\ and\ \citenamefont {Ying}}]{damle_scdm-k:_2017}%
  \BibitemOpen
  \bibfield  {author} {\bibinfo {author} {\bibfnamefont {A.}~\bibnamefont
  {Damle}}, \bibinfo {author} {\bibfnamefont {L.}~\bibnamefont {Lin}}, \ and\
  \bibinfo {author} {\bibfnamefont {L.}~\bibnamefont {Ying}},\ }\bibfield
  {title} {\enquote {\bibinfo {title} {{SCDM-\textit{k}}: {Localized} orbitals
  for solids via selected columns of the density matrix},}\ }\href@noop {}
  {\bibfield  {journal} {\bibinfo  {journal} {J. Comput. Phys.}\ }\textbf
  {\bibinfo {volume} {334}},\ \bibinfo {pages} {1--15} (\bibinfo {year}
  {2017})}\BibitemShut {NoStop}%
\bibitem [{\citenamefont {Kohn}(1996)}]{kohn_density_1996}%
  \BibitemOpen
  \bibfield  {author} {\bibinfo {author} {\bibfnamefont {W.}~\bibnamefont
  {Kohn}},\ }\bibfield  {title} {\enquote {\bibinfo {title} {Density functional
  and density matrix method scaling linearly with the number of atoms},}\
  }\href@noop {} {\bibfield  {journal} {\bibinfo  {journal} {Phys. Rev. Lett.}\
  }\textbf {\bibinfo {volume} {76}},\ \bibinfo {pages} {3168--3171} (\bibinfo
  {year} {1996})}\BibitemShut {NoStop}%
\bibitem [{\citenamefont {Prodan}\ and\ \citenamefont
  {Kohn}(2005)}]{prodan_nearsightedness_2005}%
  \BibitemOpen
  \bibfield  {author} {\bibinfo {author} {\bibfnamefont {E.}~\bibnamefont
  {Prodan}}\ and\ \bibinfo {author} {\bibfnamefont {W.}~\bibnamefont {Kohn}},\
  }\bibfield  {title} {\enquote {\bibinfo {title} {Nearsightedness of
  electronic matter},}\ }\href@noop {} {\bibfield  {journal} {\bibinfo
  {journal} {Proc. Natl. Acad. Sci. U.S.A.}\ }\textbf {\bibinfo {volume}
  {102}},\ \bibinfo {pages} {11635--11638} (\bibinfo {year}
  {2005})}\BibitemShut {NoStop}%
\bibitem [{\citenamefont {Kronik}\ \emph {et~al.}(2006)\citenamefont {Kronik},
  \citenamefont {Makmal}, \citenamefont {Tiago}, \citenamefont {Alemany},
  \citenamefont {Jain}, \citenamefont {Huang}, \citenamefont {Saad},\ and\
  \citenamefont {Chelikowsky}}]{kronik_parsec_2006}%
  \BibitemOpen
  \bibfield  {author} {\bibinfo {author} {\bibfnamefont {L.}~\bibnamefont
  {Kronik}}, \bibinfo {author} {\bibfnamefont {A.}~\bibnamefont {Makmal}},
  \bibinfo {author} {\bibfnamefont {M.~L.}\ \bibnamefont {Tiago}}, \bibinfo
  {author} {\bibfnamefont {M.~M.~G.}\ \bibnamefont {Alemany}}, \bibinfo
  {author} {\bibfnamefont {M.}~\bibnamefont {Jain}}, \bibinfo {author}
  {\bibfnamefont {X.}~\bibnamefont {Huang}}, \bibinfo {author} {\bibfnamefont
  {Y.}~\bibnamefont {Saad}}, \ and\ \bibinfo {author} {\bibfnamefont {J.~R.}\
  \bibnamefont {Chelikowsky}},\ }\bibfield  {title} {\enquote {\bibinfo {title}
  {{PARSEC}---{The} pseudopotential algorithm for real-space electronic
  structure calculations: {Recent} advances and novel applications to
  nano-structures},}\ }\href@noop {} {\bibfield  {journal} {\bibinfo  {journal}
  {Phys. Status Solidi B}\ }\textbf {\bibinfo {volume} {243}},\ \bibinfo
  {pages} {1063} (\bibinfo {year} {2006})}\BibitemShut {NoStop}%
\bibitem [{\citenamefont {Zhang}\ \emph {et~al.}(2019)\citenamefont {Zhang},
  \citenamefont {Lin}, \citenamefont {Wang}, \citenamefont {Car},\ and\
  \citenamefont {E}}]{zhang_active_2019}%
  \BibitemOpen
  \bibfield  {author} {\bibinfo {author} {\bibfnamefont {L.}~\bibnamefont
  {Zhang}}, \bibinfo {author} {\bibfnamefont {D.-Y.}\ \bibnamefont {Lin}},
  \bibinfo {author} {\bibfnamefont {H.}~\bibnamefont {Wang}}, \bibinfo {author}
  {\bibfnamefont {R.}~\bibnamefont {Car}}, \ and\ \bibinfo {author}
  {\bibfnamefont {W.}~\bibnamefont {E}},\ }\bibfield  {title} {\enquote
  {\bibinfo {title} {Active learning of uniformly accurate interatomic
  potentials for materials simulation},}\ }\href@noop {} {\bibfield  {journal}
  {\bibinfo  {journal} {Phys. Rev. Materials}\ }\textbf {\bibinfo {volume}
  {3}},\ \bibinfo {pages} {023804} (\bibinfo {year} {2019})}\BibitemShut
  {NoStop}%
\bibitem [{\citenamefont {Zhang}\ \emph {et~al.}(2020)\citenamefont {Zhang},
  \citenamefont {Wang}, \citenamefont {Chen}, \citenamefont {Zeng},
  \citenamefont {Zhang}, \citenamefont {Wang},\ and\ \citenamefont
  {E}}]{zhang_dp-gen:_2020}%
  \BibitemOpen
  \bibfield  {author} {\bibinfo {author} {\bibfnamefont {Y.}~\bibnamefont
  {Zhang}}, \bibinfo {author} {\bibfnamefont {H.}~\bibnamefont {Wang}},
  \bibinfo {author} {\bibfnamefont {W.}~\bibnamefont {Chen}}, \bibinfo {author}
  {\bibfnamefont {J.}~\bibnamefont {Zeng}}, \bibinfo {author} {\bibfnamefont
  {L.}~\bibnamefont {Zhang}}, \bibinfo {author} {\bibfnamefont
  {H.}~\bibnamefont {Wang}}, \ and\ \bibinfo {author} {\bibfnamefont
  {W.}~\bibnamefont {E}},\ }\bibfield  {title} {\enquote {\bibinfo {title}
  {{DP}-{GEN}: {A} concurrent learning platform for the generation of reliable
  deep learning based potential energy models},}\ }\href@noop {} {\bibfield
  {journal} {\bibinfo  {journal} {Comput. Phys. Commun.}\ }\textbf {\bibinfo
  {volume} {253}},\ \bibinfo {pages} {107206} (\bibinfo {year}
  {2020})}\BibitemShut {NoStop}%
\bibitem [{\citenamefont {Sun}\ \emph {et~al.}(2015)\citenamefont {Sun},
  \citenamefont {Ruzsinszky},\ and\ \citenamefont
  {Perdew}}]{sun_strongly_2015}%
  \BibitemOpen
  \bibfield  {author} {\bibinfo {author} {\bibfnamefont {J.}~\bibnamefont
  {Sun}}, \bibinfo {author} {\bibfnamefont {A.}~\bibnamefont {Ruzsinszky}}, \
  and\ \bibinfo {author} {\bibfnamefont {J.~P.}\ \bibnamefont {Perdew}},\
  }\bibfield  {title} {\enquote {\bibinfo {title} {Strongly constrained and
  appropriately normed semilocal density functional},}\ }\href@noop {}
  {\bibfield  {journal} {\bibinfo  {journal} {Phys. Rev. Lett.}\ }\textbf
  {\bibinfo {volume} {115}},\ \bibinfo {pages} {036402} (\bibinfo {year}
  {2015})}\BibitemShut {NoStop}%
\bibitem [{\citenamefont {Sun}\ \emph {et~al.}(2016)\citenamefont {Sun},
  \citenamefont {Remsing}, \citenamefont {Zhang}, \citenamefont {Sun},
  \citenamefont {Ruzsinszky}, \citenamefont {Peng}, \citenamefont {Yang},
  \citenamefont {Paul}, \citenamefont {Waghmare}, \citenamefont {Wu},
  \citenamefont {Klein},\ and\ \citenamefont {Perdew}}]{sun_accurate_2016}%
  \BibitemOpen
  \bibfield  {author} {\bibinfo {author} {\bibfnamefont {J.}~\bibnamefont
  {Sun}}, \bibinfo {author} {\bibfnamefont {R.~C.}\ \bibnamefont {Remsing}},
  \bibinfo {author} {\bibfnamefont {Y.}~\bibnamefont {Zhang}}, \bibinfo
  {author} {\bibfnamefont {Z.}~\bibnamefont {Sun}}, \bibinfo {author}
  {\bibfnamefont {A.}~\bibnamefont {Ruzsinszky}}, \bibinfo {author}
  {\bibfnamefont {H.}~\bibnamefont {Peng}}, \bibinfo {author} {\bibfnamefont
  {Z.}~\bibnamefont {Yang}}, \bibinfo {author} {\bibfnamefont {A.}~\bibnamefont
  {Paul}}, \bibinfo {author} {\bibfnamefont {U.}~\bibnamefont {Waghmare}},
  \bibinfo {author} {\bibfnamefont {X.}~\bibnamefont {Wu}}, \bibinfo {author}
  {\bibfnamefont {M.~L.}\ \bibnamefont {Klein}}, \ and\ \bibinfo {author}
  {\bibfnamefont {J.~P.}\ \bibnamefont {Perdew}},\ }\bibfield  {title}
  {\enquote {\bibinfo {title} {Accurate first-principles structures and
  energies of diversely bonded systems from an efficient density functional},}\
  }\href@noop {} {\bibfield  {journal} {\bibinfo  {journal} {Nat Chem}\
  }\textbf {\bibinfo {volume} {8}},\ \bibinfo {pages} {831--836} (\bibinfo
  {year} {2016})}\BibitemShut {NoStop}%
\bibitem [{\citenamefont {Chen}\ \emph {et~al.}(2017)\citenamefont {Chen},
  \citenamefont {Ko}, \citenamefont {Remsing}, \citenamefont {Andrade},
  \citenamefont {Santra}, \citenamefont {Sun}, \citenamefont {Selloni},
  \citenamefont {Car}, \citenamefont {Klein}, \citenamefont {Perdew},\ and\
  \citenamefont {Wu}}]{chen_ab_2017}%
  \BibitemOpen
  \bibfield  {author} {\bibinfo {author} {\bibfnamefont {M.}~\bibnamefont
  {Chen}}, \bibinfo {author} {\bibfnamefont {H.-Y.}\ \bibnamefont {Ko}},
  \bibinfo {author} {\bibfnamefont {R.~C.}\ \bibnamefont {Remsing}}, \bibinfo
  {author} {\bibfnamefont {M.~F.~C.}\ \bibnamefont {Andrade}}, \bibinfo
  {author} {\bibfnamefont {B.}~\bibnamefont {Santra}}, \bibinfo {author}
  {\bibfnamefont {Z.}~\bibnamefont {Sun}}, \bibinfo {author} {\bibfnamefont
  {A.}~\bibnamefont {Selloni}}, \bibinfo {author} {\bibfnamefont
  {R.}~\bibnamefont {Car}}, \bibinfo {author} {\bibfnamefont {M.~L.}\
  \bibnamefont {Klein}}, \bibinfo {author} {\bibfnamefont {J.~P.}\ \bibnamefont
  {Perdew}}, \ and\ \bibinfo {author} {\bibfnamefont {X.}~\bibnamefont {Wu}},\
  }\bibfield  {title} {\enquote {\bibinfo {title} {\textit{Ab initio} theory
  and modeling of water},}\ }\href@noop {} {\bibfield  {journal} {\bibinfo
  {journal} {Proc. Natl. Acad. Sci. U.S.A.}\ }\textbf {\bibinfo {volume}
  {114}},\ \bibinfo {pages} {10846--10851} (\bibinfo {year}
  {2017})}\BibitemShut {NoStop}%
\bibitem [{\citenamefont {Zheng}\ \emph {et~al.}(2018)\citenamefont {Zheng},
  \citenamefont {Chen}, \citenamefont {Sun}, \citenamefont {Ko}, \citenamefont
  {Santra}, \citenamefont {Dhuvad},\ and\ \citenamefont
  {Wu}}]{zheng_structural_2018}%
  \BibitemOpen
  \bibfield  {author} {\bibinfo {author} {\bibfnamefont {L.}~\bibnamefont
  {Zheng}}, \bibinfo {author} {\bibfnamefont {M.}~\bibnamefont {Chen}},
  \bibinfo {author} {\bibfnamefont {Z.}~\bibnamefont {Sun}}, \bibinfo {author}
  {\bibfnamefont {H.-Y.}\ \bibnamefont {Ko}}, \bibinfo {author} {\bibfnamefont
  {B.}~\bibnamefont {Santra}}, \bibinfo {author} {\bibfnamefont
  {P.}~\bibnamefont {Dhuvad}}, \ and\ \bibinfo {author} {\bibfnamefont
  {X.}~\bibnamefont {Wu}},\ }\bibfield  {title} {\enquote {\bibinfo {title}
  {Structural, electronic, and dynamical properties of liquid water by ab
  initio molecular dynamics based on {SCAN} functional within the canonical
  ensemble},}\ }\href@noop {} {\bibfield  {journal} {\bibinfo  {journal} {J.
  Chem. Phys.}\ }\textbf {\bibinfo {volume} {148}},\ \bibinfo {pages} {164505}
  (\bibinfo {year} {2018})}\BibitemShut {NoStop}%
\bibitem [{\citenamefont {Andreani}\ \emph {et~al.}(2020)\citenamefont
  {Andreani}, \citenamefont {Romanelli}, \citenamefont {Parmentier},
  \citenamefont {Senesi}, \citenamefont {Kolesnikov}, \citenamefont {Ko},
  \citenamefont {Calegari~Andrade},\ and\ \citenamefont
  {Car}}]{andreani_hydrogen_2020}%
  \BibitemOpen
  \bibfield  {author} {\bibinfo {author} {\bibfnamefont {C.}~\bibnamefont
  {Andreani}}, \bibinfo {author} {\bibfnamefont {G.}~\bibnamefont {Romanelli}},
  \bibinfo {author} {\bibfnamefont {A.}~\bibnamefont {Parmentier}}, \bibinfo
  {author} {\bibfnamefont {R.}~\bibnamefont {Senesi}}, \bibinfo {author}
  {\bibfnamefont {A.~I.}\ \bibnamefont {Kolesnikov}}, \bibinfo {author}
  {\bibfnamefont {H.-Y.}\ \bibnamefont {Ko}}, \bibinfo {author} {\bibfnamefont
  {M.~F.}\ \bibnamefont {Calegari~Andrade}}, \ and\ \bibinfo {author}
  {\bibfnamefont {R.}~\bibnamefont {Car}},\ }\bibfield  {title} {\enquote
  {\bibinfo {title} {Hydrogen dynamics in supercritical water probed by
  neutron scattering and computer simulations},}\ }\href@noop {} {\bibfield
  {journal} {\bibinfo  {journal} {J. Phys. Chem. Lett.}\ }\textbf {\bibinfo
  {volume} {11}},\ \bibinfo {pages} {9461--9467} (\bibinfo {year}
  {2020})}\BibitemShut {NoStop}%
\bibitem [{\citenamefont {Calegari~Andrade}\ \emph {et~al.}(2018)\citenamefont
  {Calegari~Andrade}, \citenamefont {Ko}, \citenamefont {Car},\ and\
  \citenamefont {Selloni}}]{calegari_andrade_structure_2018}%
  \BibitemOpen
  \bibfield  {author} {\bibinfo {author} {\bibfnamefont {M.~F.}\ \bibnamefont
  {Calegari~Andrade}}, \bibinfo {author} {\bibfnamefont {H.-Y.}\ \bibnamefont
  {Ko}}, \bibinfo {author} {\bibfnamefont {R.}~\bibnamefont {Car}}, \ and\
  \bibinfo {author} {\bibfnamefont {A.}~\bibnamefont {Selloni}},\ }\bibfield
  {title} {\enquote {\bibinfo {title} {Structure, polarization, and sum
  frequency generation spectrum of interfacial water on anatase {TiO}$_2$},}\
  }\href@noop {} {\bibfield  {journal} {\bibinfo  {journal} {J. Phys. Chem.
  Lett.}\ }\textbf {\bibinfo {volume} {9}},\ \bibinfo {pages} {6716--6721}
  (\bibinfo {year} {2018})}\BibitemShut {NoStop}%
\bibitem [{\citenamefont {Calegari~Andrade}\ \emph {et~al.}(2020)\citenamefont
  {Calegari~Andrade}, \citenamefont {Ko}, \citenamefont {Zhang}, \citenamefont
  {Car},\ and\ \citenamefont {Selloni}}]{calegari_andrade_free_2020}%
  \BibitemOpen
  \bibfield  {author} {\bibinfo {author} {\bibfnamefont {M.~F.}\ \bibnamefont
  {Calegari~Andrade}}, \bibinfo {author} {\bibfnamefont {H.-Y.}\ \bibnamefont
  {Ko}}, \bibinfo {author} {\bibfnamefont {L.}~\bibnamefont {Zhang}}, \bibinfo
  {author} {\bibfnamefont {R.}~\bibnamefont {Car}}, \ and\ \bibinfo {author}
  {\bibfnamefont {A.}~\bibnamefont {Selloni}},\ }\bibfield  {title} {\enquote
  {\bibinfo {title} {Free energy of proton transfer at the water--{TiO}$_2$
  interface from \textit{ab initio} deep potential molecular dynamics},}\
  }\href@noop {} {\bibfield  {journal} {\bibinfo  {journal} {Chem. Sci.}\
  }\textbf {\bibinfo {volume} {11}},\ \bibinfo {pages} {2335--2341} (\bibinfo
  {year} {2020})}\BibitemShut {NoStop}%
\bibitem [{\citenamefont {Zhang}\ \emph {et~al.}(2021)\citenamefont {Zhang},
  \citenamefont {Wang}, \citenamefont {Car},\ and\ \citenamefont
  {E}}]{zhang_phase_2021}%
  \BibitemOpen
  \bibfield  {author} {\bibinfo {author} {\bibfnamefont {L.}~\bibnamefont
  {Zhang}}, \bibinfo {author} {\bibfnamefont {H.}~\bibnamefont {Wang}},
  \bibinfo {author} {\bibfnamefont {R.}~\bibnamefont {Car}}, \ and\ \bibinfo
  {author} {\bibfnamefont {W.}~\bibnamefont {E}},\ }\bibfield  {title}
  {\enquote {\bibinfo {title} {Phase diagram of a deep potential water
  model},}\ }\href@noop {} {\bibfield  {journal} {\bibinfo  {journal} {Phys.
  Rev. Lett.}\ }\textbf {\bibinfo {volume} {126}},\ \bibinfo {pages} {236001}
  (\bibinfo {year} {2021})}\BibitemShut {NoStop}%
\bibitem [{\citenamefont {Nguyen}\ and\ \citenamefont
  {de~Gironcoli}(2009)}]{nguyen_efficient_2009}%
  \BibitemOpen
  \bibfield  {author} {\bibinfo {author} {\bibfnamefont {H.-V.}\ \bibnamefont
  {Nguyen}}\ and\ \bibinfo {author} {\bibfnamefont {S.}~\bibnamefont
  {de~Gironcoli}},\ }\bibfield  {title} {\enquote {\bibinfo {title} {Efficient
  calculation of exact exchange and {RPA} correlation energies in the
  adiabatic-connection fluctuation-dissipation theory},}\ }\href@noop {}
  {\bibfield  {journal} {\bibinfo  {journal} {Phys. Rev. B}\ }\textbf {\bibinfo
  {volume} {79}},\ \bibinfo {pages} {205114} (\bibinfo {year}
  {2009})}\BibitemShut {NoStop}%
\bibitem [{\citenamefont {Perdew}\ \emph
  {et~al.}(1996{\natexlab{b}})\citenamefont {Perdew}, \citenamefont {Burke},\
  and\ \citenamefont {Ernzerhof}}]{perdew_generalized_1996}%
  \BibitemOpen
  \bibfield  {author} {\bibinfo {author} {\bibfnamefont {J.~P.}\ \bibnamefont
  {Perdew}}, \bibinfo {author} {\bibfnamefont {K.}~\bibnamefont {Burke}}, \
  and\ \bibinfo {author} {\bibfnamefont {M.}~\bibnamefont {Ernzerhof}},\
  }\bibfield  {title} {\enquote {\bibinfo {title} {Generalized gradient
  approximation made simple},}\ }\href@noop {} {\bibfield  {journal} {\bibinfo
  {journal} {Phys. Rev. Lett.}\ }\textbf {\bibinfo {volume} {77}},\ \bibinfo
  {pages} {3865--3868} (\bibinfo {year} {1996}{\natexlab{b}})}\BibitemShut
  {NoStop}%
\bibitem [{foo()}]{footnote_sea_performance_density}%
  \BibitemOpen
  \href@noop {} {}\bibinfo {note} {The performance of \texttt{SeA} is expected
  to have a semi-systematic dependence on the system density since the degree
  of sparsity in a given configuration is also determined by the structural
  details of the underlying ionic framework.}\BibitemShut {Stop}%
\bibitem [{\citenamefont {Zhang}\ \emph
  {et~al.}(2018{\natexlab{a}})\citenamefont {Zhang}, \citenamefont {Han},
  \citenamefont {Wang}, \citenamefont {Car},\ and\ \citenamefont
  {E}}]{zhang_deep_2018}%
  \BibitemOpen
  \bibfield  {author} {\bibinfo {author} {\bibfnamefont {L.}~\bibnamefont
  {Zhang}}, \bibinfo {author} {\bibfnamefont {J.}~\bibnamefont {Han}}, \bibinfo
  {author} {\bibfnamefont {H.}~\bibnamefont {Wang}}, \bibinfo {author}
  {\bibfnamefont {R.}~\bibnamefont {Car}}, \ and\ \bibinfo {author}
  {\bibfnamefont {W.}~\bibnamefont {E}},\ }\bibfield  {title} {\enquote
  {\bibinfo {title} {Deep potential molecular dynamics: A scalable model with
  the accuracy of quantum mechanics},}\ }\href@noop {} {\bibfield  {journal}
  {\bibinfo  {journal} {Phys. Rev. Lett.}\ }\textbf {\bibinfo {volume} {120}},\
  \bibinfo {pages} {143001} (\bibinfo {year} {2018}{\natexlab{a}})}\BibitemShut
  {NoStop}%
\bibitem [{\citenamefont {Zhang}\ \emph
  {et~al.}(2018{\natexlab{b}})\citenamefont {Zhang}, \citenamefont {Han},
  \citenamefont {Wang}, \citenamefont {Saidi}, \citenamefont {Car},\ and\
  \citenamefont {E}}]{zhang_end--end_2018}%
  \BibitemOpen
  \bibfield  {author} {\bibinfo {author} {\bibfnamefont {L.}~\bibnamefont
  {Zhang}}, \bibinfo {author} {\bibfnamefont {J.}~\bibnamefont {Han}}, \bibinfo
  {author} {\bibfnamefont {H.}~\bibnamefont {Wang}}, \bibinfo {author}
  {\bibfnamefont {W.}~\bibnamefont {Saidi}}, \bibinfo {author} {\bibfnamefont
  {R.}~\bibnamefont {Car}}, \ and\ \bibinfo {author} {\bibfnamefont
  {W.}~\bibnamefont {E}},\ }\bibfield  {title} {\enquote {\bibinfo {title}
  {End-to-end symmetry preserving inter-atomic potential energy model for
  finite and extended systems},}\ }in\ \href@noop {} {\emph {\bibinfo
  {booktitle} {Advances in Neural Information Processing Systems 31}}},\
  \bibinfo {editor} {edited by\ \bibinfo {editor} {\bibfnamefont
  {S.}~\bibnamefont {Bengio}}, \bibinfo {editor} {\bibfnamefont
  {H.}~\bibnamefont {Wallach}}, \bibinfo {editor} {\bibfnamefont
  {H.}~\bibnamefont {Larochelle}}, \bibinfo {editor} {\bibfnamefont
  {K.}~\bibnamefont {Grauman}}, \bibinfo {editor} {\bibfnamefont
  {N.}~\bibnamefont {Cesa-Bianchi}}, \ and\ \bibinfo {editor} {\bibfnamefont
  {R.}~\bibnamefont {Garnett}}}\ (\bibinfo  {publisher} {Curran Associates},\
  \bibinfo {address} {Red Hook},\ \bibinfo {year} {2018})\ pp.\ \bibinfo
  {pages} {4436--4446}\BibitemShut {NoStop}%
\bibitem [{\citenamefont {Heyd}\ \emph {et~al.}(2003)\citenamefont {Heyd},
  \citenamefont {Scuseria},\ and\ \citenamefont
  {Ernzerhof}}]{heyd_hybrid_2003}%
  \BibitemOpen
  \bibfield  {author} {\bibinfo {author} {\bibfnamefont {J.}~\bibnamefont
  {Heyd}}, \bibinfo {author} {\bibfnamefont {G.~E.}\ \bibnamefont {Scuseria}},
  \ and\ \bibinfo {author} {\bibfnamefont {M.}~\bibnamefont {Ernzerhof}},\
  }\bibfield  {title} {\enquote {\bibinfo {title} {Hybrid functionals based on
  a screened {Coulomb} potential},}\ }\href@noop {} {\bibfield  {journal}
  {\bibinfo  {journal} {J. Chem. Phys.}\ }\textbf {\bibinfo {volume} {118}},\
  \bibinfo {pages} {8207--8215} (\bibinfo {year} {2003})}\BibitemShut {NoStop}%
\bibitem [{\citenamefont {Gerber}\ and\ \citenamefont
  {{\'A}ngy{\'a}n}(2005)}]{gerber_hybrid_2005}%
  \BibitemOpen
  \bibfield  {author} {\bibinfo {author} {\bibfnamefont {I.~C.}\ \bibnamefont
  {Gerber}}\ and\ \bibinfo {author} {\bibfnamefont {J.~G.}\ \bibnamefont
  {{\'A}ngy{\'a}n}},\ }\bibfield  {title} {\enquote {\bibinfo {title} {Hybrid
  functional with separated range},}\ }\href@noop {} {\bibfield  {journal}
  {\bibinfo  {journal} {Chem. Phys. Lett.}\ }\textbf {\bibinfo {volume}
  {415}},\ \bibinfo {pages} {100--105} (\bibinfo {year} {2005})}\BibitemShut
  {NoStop}%
\bibitem [{\citenamefont {Vydrov}\ and\ \citenamefont
  {Scuseria}(2006)}]{vydrov_assessment_2006}%
  \BibitemOpen
  \bibfield  {author} {\bibinfo {author} {\bibfnamefont {O.~A.}\ \bibnamefont
  {Vydrov}}\ and\ \bibinfo {author} {\bibfnamefont {G.~E.}\ \bibnamefont
  {Scuseria}},\ }\bibfield  {title} {\enquote {\bibinfo {title} {Assessment of
  a long-range corrected hybrid functional},}\ }\href@noop {} {\bibfield
  {journal} {\bibinfo  {journal} {J. Chem. Phys.}\ }\textbf {\bibinfo {volume}
  {125}},\ \bibinfo {pages} {234109} (\bibinfo {year} {2006})}\BibitemShut
  {NoStop}%
\bibitem [{\citenamefont {Janesko}\ \emph {et~al.}(2009)\citenamefont
  {Janesko}, \citenamefont {Henderson},\ and\ \citenamefont
  {Scuseria}}]{janesko_screened_2009}%
  \BibitemOpen
  \bibfield  {author} {\bibinfo {author} {\bibfnamefont {B.~G.}\ \bibnamefont
  {Janesko}}, \bibinfo {author} {\bibfnamefont {T.~M.}\ \bibnamefont
  {Henderson}}, \ and\ \bibinfo {author} {\bibfnamefont {G.~E.}\ \bibnamefont
  {Scuseria}},\ }\bibfield  {title} {\enquote {\bibinfo {title} {Screened
  hybrid density functionals for solid-state chemistry and physics},}\
  }\href@noop {} {\bibfield  {journal} {\bibinfo  {journal} {Phys. Chem. Chem.
  Phys.}\ }\textbf {\bibinfo {volume} {11}},\ \bibinfo {pages} {443--454}
  (\bibinfo {year} {2009})}\BibitemShut {NoStop}%
\bibitem [{\citenamefont {Baer}\ \emph {et~al.}(2010)\citenamefont {Baer},
  \citenamefont {Livshits},\ and\ \citenamefont {Salzner}}]{baer_tuned_2010}%
  \BibitemOpen
  \bibfield  {author} {\bibinfo {author} {\bibfnamefont {R.}~\bibnamefont
  {Baer}}, \bibinfo {author} {\bibfnamefont {E.}~\bibnamefont {Livshits}}, \
  and\ \bibinfo {author} {\bibfnamefont {U.}~\bibnamefont {Salzner}},\
  }\bibfield  {title} {\enquote {\bibinfo {title} {Tuned range-separated
  hybrids in density functional theory},}\ }\href@noop {} {\bibfield  {journal}
  {\bibinfo  {journal} {Annu. Rev. Phys. Chem.}\ }\textbf {\bibinfo {volume}
  {61}},\ \bibinfo {pages} {85--109} (\bibinfo {year} {2010})}\BibitemShut
  {NoStop}%
\bibitem [{\citenamefont {Kronik}\ \emph {et~al.}(2012)\citenamefont {Kronik},
  \citenamefont {Stein}, \citenamefont {Refaely-Abramson},\ and\ \citenamefont
  {Baer}}]{kronik_excitation_2012}%
  \BibitemOpen
  \bibfield  {author} {\bibinfo {author} {\bibfnamefont {L.}~\bibnamefont
  {Kronik}}, \bibinfo {author} {\bibfnamefont {T.}~\bibnamefont {Stein}},
  \bibinfo {author} {\bibfnamefont {S.}~\bibnamefont {Refaely-Abramson}}, \
  and\ \bibinfo {author} {\bibfnamefont {R.}~\bibnamefont {Baer}},\ }\bibfield
  {title} {\enquote {\bibinfo {title} {Excitation gaps of finite-sized systems
  from optimally tuned range-separated hybrid functionals},}\ }\href@noop {}
  {\bibfield  {journal} {\bibinfo  {journal} {J. Chem. Theory Comput.}\
  }\textbf {\bibinfo {volume} {8}},\ \bibinfo {pages} {1515--1531} (\bibinfo
  {year} {2012})}\BibitemShut {NoStop}%
\bibitem [{\citenamefont {Karolewski}\ \emph {et~al.}(2013)\citenamefont
  {Karolewski}, \citenamefont {Kronik},\ and\ \citenamefont
  {K{\"u}mmel}}]{karolewski_using_2013}%
  \BibitemOpen
  \bibfield  {author} {\bibinfo {author} {\bibfnamefont {A.}~\bibnamefont
  {Karolewski}}, \bibinfo {author} {\bibfnamefont {L.}~\bibnamefont {Kronik}},
  \ and\ \bibinfo {author} {\bibfnamefont {S.}~\bibnamefont {K{\"u}mmel}},\
  }\bibfield  {title} {\enquote {\bibinfo {title} {Using optimally tuned range
  separated hybrid functionals in ground-state calculations: Consequences and
  caveats},}\ }\href@noop {} {\bibfield  {journal} {\bibinfo  {journal} {J.
  Chem. Phys.}\ }\textbf {\bibinfo {volume} {138}},\ \bibinfo {pages} {204115}
  (\bibinfo {year} {2013})}\BibitemShut {NoStop}%
\bibitem [{\citenamefont {Tuckerman}\ and\ \citenamefont
  {Parrinello}(1994)}]{tuckerman_integrating_1994}%
  \BibitemOpen
  \bibfield  {author} {\bibinfo {author} {\bibfnamefont {M.~E.}\ \bibnamefont
  {Tuckerman}}\ and\ \bibinfo {author} {\bibfnamefont {M.}~\bibnamefont
  {Parrinello}},\ }\bibfield  {title} {\enquote {\bibinfo {title} {Integrating
  the {Car}{\textendash}{Parrinello} equations. {II}. {Multiple} time scale
  techniques},}\ }\href@noop {} {\bibfield  {journal} {\bibinfo  {journal} {J.
  Chem. Phys.}\ }\textbf {\bibinfo {volume} {101}},\ \bibinfo {pages}
  {1316--1329} (\bibinfo {year} {1994})}\BibitemShut {NoStop}%
\bibitem [{\citenamefont {Mandal}\ \emph {et~al.}(2022)\citenamefont {Mandal},
  \citenamefont {Kar}, \citenamefont {Kl{\"o}ffel}, \citenamefont {Meyer},\
  and\ \citenamefont {Nair}}]{mandal_improving_2022}%
  \BibitemOpen
  \bibfield  {author} {\bibinfo {author} {\bibfnamefont {S.}~\bibnamefont
  {Mandal}}, \bibinfo {author} {\bibfnamefont {R.}~\bibnamefont {Kar}},
  \bibinfo {author} {\bibfnamefont {T.}~\bibnamefont {Kl{\"o}ffel}}, \bibinfo
  {author} {\bibfnamefont {B.}~\bibnamefont {Meyer}}, \ and\ \bibinfo {author}
  {\bibfnamefont {N.~N.}\ \bibnamefont {Nair}},\ }\bibfield  {title} {\enquote
  {\bibinfo {title} {Improving the scaling and performance of multiple time
  stepping-based molecular dynamics with hybrid density functionals},}\
  }\href@noop {} {\bibfield  {journal} {\bibinfo  {journal} {J. Comput. Chem.}\
  }\textbf {\bibinfo {volume} {43}},\ \bibinfo {pages} {588--597} (\bibinfo
  {year} {2022})}\BibitemShut {NoStop}%
\bibitem [{\citenamefont {Thomas}\ \emph {et~al.}(2004)\citenamefont {Thomas},
  \citenamefont {Iftimie},\ and\ \citenamefont
  {Tuckerman}}]{thomas_field_2004}%
  \BibitemOpen
  \bibfield  {author} {\bibinfo {author} {\bibfnamefont {J.~W.}\ \bibnamefont
  {Thomas}}, \bibinfo {author} {\bibfnamefont {R.}~\bibnamefont {Iftimie}}, \
  and\ \bibinfo {author} {\bibfnamefont {M.~E.}\ \bibnamefont {Tuckerman}},\
  }\bibfield  {title} {\enquote {\bibinfo {title} {Field theoretic approach to
  dynamical orbital localization in \textit{ab initio} molecular dynamics},}\
  }\href@noop {} {\bibfield  {journal} {\bibinfo  {journal} {Phys. Rev. B}\
  }\textbf {\bibinfo {volume} {69}},\ \bibinfo {pages} {125105} (\bibinfo
  {year} {2004})}\BibitemShut {NoStop}%
\bibitem [{\citenamefont {Hamann}\ \emph {et~al.}(1979)\citenamefont {Hamann},
  \citenamefont {Schl{\"u}ter},\ and\ \citenamefont
  {Chiang}}]{hamann_norm-conserving_1979}%
  \BibitemOpen
  \bibfield  {author} {\bibinfo {author} {\bibfnamefont {D.~R.}\ \bibnamefont
  {Hamann}}, \bibinfo {author} {\bibfnamefont {M.}~\bibnamefont
  {Schl{\"u}ter}}, \ and\ \bibinfo {author} {\bibfnamefont {C.}~\bibnamefont
  {Chiang}},\ }\bibfield  {title} {\enquote {\bibinfo {title} {Norm-conserving
  pseudopotentials},}\ }\href@noop {} {\bibfield  {journal} {\bibinfo
  {journal} {Phys. Rev. Lett.}\ }\textbf {\bibinfo {volume} {43}},\ \bibinfo
  {pages} {1494--1497} (\bibinfo {year} {1979})}\BibitemShut {NoStop}%
\bibitem [{\citenamefont {Vanderbilt}(1985)}]{vanderbilt_optimally_1985}%
  \BibitemOpen
  \bibfield  {author} {\bibinfo {author} {\bibfnamefont {D.}~\bibnamefont
  {Vanderbilt}},\ }\bibfield  {title} {\enquote {\bibinfo {title} {Optimally
  smooth norm-conserving pseudopotentials},}\ }\href@noop {} {\bibfield
  {journal} {\bibinfo  {journal} {Phys. Rev. B}\ }\textbf {\bibinfo {volume}
  {32}},\ \bibinfo {pages} {8412--8415} (\bibinfo {year} {1985})}\BibitemShut
  {NoStop}%
\bibitem [{\citenamefont {Gygi}(2008)}]{gygi_architecture_2008}%
  \BibitemOpen
  \bibfield  {author} {\bibinfo {author} {\bibfnamefont {F.}~\bibnamefont
  {Gygi}},\ }\bibfield  {title} {\enquote {\bibinfo {title} {Architecture of
  \texttt{Qbox}: A scalable first-principles molecular dynamics code},}\
  }\href@noop {} {\bibfield  {journal} {\bibinfo  {journal} {IBM J. Res. Dev.}\
  }\textbf {\bibinfo {volume} {52}},\ \bibinfo {pages} {137--144} (\bibinfo
  {year} {2008})}\BibitemShut {NoStop}%
\bibitem [{\citenamefont {Wang}\ \emph {et~al.}(2018)\citenamefont {Wang},
  \citenamefont {Zhang}, \citenamefont {Han},\ and\ \citenamefont
  {E}}]{wang_deepmd-kit:_2018}%
  \BibitemOpen
  \bibfield  {author} {\bibinfo {author} {\bibfnamefont {H.}~\bibnamefont
  {Wang}}, \bibinfo {author} {\bibfnamefont {L.}~\bibnamefont {Zhang}},
  \bibinfo {author} {\bibfnamefont {J.}~\bibnamefont {Han}}, \ and\ \bibinfo
  {author} {\bibfnamefont {W.}~\bibnamefont {E}},\ }\bibfield  {title}
  {\enquote {\bibinfo {title} {{DeePMD}-kit: {A} deep learning package for
  many-body potential energy representation and molecular dynamics},}\
  }\href@noop {} {\bibfield  {journal} {\bibinfo  {journal} {Comput. Phys.
  Commun.}\ }\textbf {\bibinfo {volume} {228}},\ \bibinfo {pages} {178--184}
  (\bibinfo {year} {2018})}\BibitemShut {NoStop}%
\bibitem [{\citenamefont {Abadi}\ \emph {et~al.}(2016)\citenamefont {Abadi},
  \citenamefont {Barham}, \citenamefont {Chen}, \citenamefont {Chen},
  \citenamefont {Davis}, \citenamefont {Dean}, \citenamefont {Devin},
  \citenamefont {Ghemawat}, \citenamefont {Irving}, \citenamefont {Isard},
  \citenamefont {Kudlur}, \citenamefont {Levenberg}, \citenamefont {Monga},
  \citenamefont {Moore}, \citenamefont {Murray}, \citenamefont {Steiner},
  \citenamefont {Tucker}, \citenamefont {Vasudevan}, \citenamefont {Warden},
  \citenamefont {Wicke}, \citenamefont {Yu},\ and\ \citenamefont
  {Zheng}}]{tensorflow2015-whitepaper}%
  \BibitemOpen
  \bibfield  {author} {\bibinfo {author} {\bibfnamefont {M.}~\bibnamefont
  {Abadi}}, \bibinfo {author} {\bibfnamefont {P.}~\bibnamefont {Barham}},
  \bibinfo {author} {\bibfnamefont {J.}~\bibnamefont {Chen}}, \bibinfo {author}
  {\bibfnamefont {Z.}~\bibnamefont {Chen}}, \bibinfo {author} {\bibfnamefont
  {A.}~\bibnamefont {Davis}}, \bibinfo {author} {\bibfnamefont
  {J.}~\bibnamefont {Dean}}, \bibinfo {author} {\bibfnamefont {M.}~\bibnamefont
  {Devin}}, \bibinfo {author} {\bibfnamefont {S.}~\bibnamefont {Ghemawat}},
  \bibinfo {author} {\bibfnamefont {G.}~\bibnamefont {Irving}}, \bibinfo
  {author} {\bibfnamefont {M.}~\bibnamefont {Isard}}, \bibinfo {author}
  {\bibfnamefont {M.}~\bibnamefont {Kudlur}}, \bibinfo {author} {\bibfnamefont
  {J.}~\bibnamefont {Levenberg}}, \bibinfo {author} {\bibfnamefont
  {R.}~\bibnamefont {Monga}}, \bibinfo {author} {\bibfnamefont
  {S.}~\bibnamefont {Moore}}, \bibinfo {author} {\bibfnamefont {D.~G.}\
  \bibnamefont {Murray}}, \bibinfo {author} {\bibfnamefont {B.}~\bibnamefont
  {Steiner}}, \bibinfo {author} {\bibfnamefont {P.}~\bibnamefont {Tucker}},
  \bibinfo {author} {\bibfnamefont {V.}~\bibnamefont {Vasudevan}}, \bibinfo
  {author} {\bibfnamefont {P.}~\bibnamefont {Warden}}, \bibinfo {author}
  {\bibfnamefont {M.}~\bibnamefont {Wicke}}, \bibinfo {author} {\bibfnamefont
  {Y.}~\bibnamefont {Yu}}, \ and\ \bibinfo {author} {\bibfnamefont
  {X.}~\bibnamefont {Zheng}},\ }\bibfield  {title} {\enquote {\bibinfo {title}
  {Tensorflow: A system for large-scale machine learning},}\ }in\ \href@noop {}
  {\emph {\bibinfo {booktitle} {12th USENIX Symposium on Operating Systems
  Design and Implementation (OSDI 16)}}}\ (\bibinfo {year} {2016})\ pp.\
  \bibinfo {pages} {265--283}\BibitemShut {NoStop}%
\bibitem [{\citenamefont {Kingma}\ and\ \citenamefont
  {Ba}(2015)}]{Kingma2015adam}%
  \BibitemOpen
  \bibfield  {author} {\bibinfo {author} {\bibfnamefont {D.}~\bibnamefont
  {Kingma}}\ and\ \bibinfo {author} {\bibfnamefont {J.}~\bibnamefont {Ba}},\
  }\bibfield  {title} {\enquote {\bibinfo {title} {Adam: a method for
  stochastic optimization},}\ }in\ \href@noop {} {\emph {\bibinfo {booktitle}
  {Proceedings of the International Conference on Learning Representations
  (ICLR)}}}\ (\bibinfo {year} {2015})\BibitemShut {NoStop}%
\bibitem [{\citenamefont {Plimpton}(1995)}]{plimpton_fast_1995}%
  \BibitemOpen
  \bibfield  {author} {\bibinfo {author} {\bibfnamefont {S.}~\bibnamefont
  {Plimpton}},\ }\bibfield  {title} {\enquote {\bibinfo {title} {Fast parallel
  algorithms for short-range molecular dynamics},}\ }\href@noop {} {\bibfield
  {journal} {\bibinfo  {journal} {J. Comput. Phys.}\ }\textbf {\bibinfo
  {volume} {117}},\ \bibinfo {pages} {1--19} (\bibinfo {year}
  {1995})}\BibitemShut {NoStop}%
\bibitem [{\citenamefont {Tuckerman}\ \emph {et~al.}(2006)\citenamefont
  {Tuckerman}, \citenamefont {Alejandre}, \citenamefont {L{\'o}pez-Rend{\'o}n},
  \citenamefont {Jochim},\ and\ \citenamefont
  {Martyna}}]{tuckerman_liouville-operator_2006}%
  \BibitemOpen
  \bibfield  {author} {\bibinfo {author} {\bibfnamefont {M.~E.}\ \bibnamefont
  {Tuckerman}}, \bibinfo {author} {\bibfnamefont {J.}~\bibnamefont
  {Alejandre}}, \bibinfo {author} {\bibfnamefont {R.}~\bibnamefont
  {L{\'o}pez-Rend{\'o}n}}, \bibinfo {author} {\bibfnamefont {A.~L.}\
  \bibnamefont {Jochim}}, \ and\ \bibinfo {author} {\bibfnamefont {G.~J.}\
  \bibnamefont {Martyna}},\ }\bibfield  {title} {\enquote {\bibinfo {title} {A
  {Liouville}-operator derived measure-preserving integrator for molecular
  dynamics simulations in the isothermal--isobaric ensemble},}\ }\href@noop {}
  {\bibfield  {journal} {\bibinfo  {journal} {J. Phys. A: Math. Gen.}\ }\textbf
  {\bibinfo {volume} {39}},\ \bibinfo {pages} {5629--5651} (\bibinfo {year}
  {2006})}\BibitemShut {NoStop}%
\end{thebibliography}
\end{document}